\documentclass[12pt]{article}
\usepackage{amsfonts,amssymb,epsfig,amsmath}
\usepackage{color}

\renewcommand{\baselinestretch}{1.2}

\newcommand{\be}{\begin{eqnarray}}
\newcommand{\ee}{\end{eqnarray}}

\newcommand{\bn}{\begin{enumerate}}
\newcommand{\en}{\end{enumerate}}

\begin{document}

\makeatletter \@addtoreset{equation}{section} \makeatother
\renewcommand{\theequation}{\thesection.\arabic{equation}}
\renewcommand{\thefootnote}{\alph{footnote}}

\begin{titlepage}

\begin{center}
\hfill {\tt SNUTP16-002}\\

\vspace{2cm}


{\Large\bf 6d strings from new chiral gauge theories}

\vspace{2cm}

\renewcommand{\thefootnote}{\alph{footnote}}

{\large Hee-Cheol Kim$^1$, Seok Kim$^2$, Jaemo Park$^3$}

\vspace{0.7cm}

\textit{$^1$Perimeter Institute for Theoretical Physics,
Waterloo, Ontario N2L 2Y5, Canada.}\\

\vspace{0.2cm}

\textit{$^2$Department of Physics and Astronomy \& Center for
Theoretical Physics,\\
Seoul National University, Seoul 151-747, Korea.}\\

\vspace{0.2cm}

\textit{$^3$Department of Physics, Postech, Pohang 790-784, Korea.}\\

\vspace{0.7cm}

E-mails: {\tt hkim@perimeterinstitute.ca, skim@phya.snu.ac.kr, jaemo@postech.ac.kr}

\end{center}

\vspace{1cm}

\begin{abstract}

We study the 6d $\mathcal{N}=(1,0)$ superconformal field theory
with smallest non-Higgsable gauge symmetry $SU(3)$. In particular, we propose new
2d gauge theory descriptions of its self-dual strings in the tensor branch.
We use our gauge theories to compute the elliptic genera of the self-dual strings,
which completely agree with the partial data known from topological strings. We
further study the strings of the $(E_6,E_6)$ conformal matter by generalizing our
2d gauge theories. We also show that anomalies of all our gauge theories agree with
the self-dual string anomalies computed by inflows from 6d.

\end{abstract}

\end{titlepage}

\renewcommand{\thefootnote}{\arabic{footnote}}

\setcounter{footnote}{0}

\renewcommand{\baselinestretch}{1}

\tableofcontents

\renewcommand{\baselinestretch}{1.2}

\section{Introduction}

6d SCFTs are extensively studied recently. In particular,
plenty of $\mathcal{N}=(1,0)$ theories was discovered from F-theory
\cite{Heckman:2013pva,DelZotto:2014hpa,Heckman:2015bfa}.
A simple class of them \cite{Morrison:1996na,Morrison:1996pp,Witten:1996qb},
called 6d minimal SCFTs \cite{Haghighat:2014vxa}, play important roles.
Minimal SCFTs are engineered by putting F-theory on elliptic Calabi-Yau
3-folds with the bases
given by the Hirzebruch surfaces $\mathbb{F}_n$, with $n=1,\cdots,12$.
They also have $E_8\times E_8$ heterotic dual descriptions on
$K3$ with instanton numbers $(12+n,12-n)$. By definition, they have 1d
tensor branches and non-Higgsable gauge symmetries.\footnote{At
$n=9,10,11$, the heterotic setting contains extra `small
instantons,' thus making CFTs with higher dimensional tensor branches.}
They are atomic building blocks of 6d SCFTs \cite{Heckman:2013pva,Heckman:2015bfa}.
So it is important to understand their physics better.

6d SCFTs have tensionless strings coupled to the self-dual tensor
fields. These self-dual strings acquire tensions in the tensor
branch, when the tensor multiplet scalars have nonzero VEV. The strings host
2d $\mathcal{N}=(0,4)$ SCFTs on their worldsheets, which are the main
subject of this paper. They may be studied by engineering
2d gauge theories which are weakly coupled in UV, flowing to the desired CFTs
in the IR. Gauge theories on the self-dual strings have been
studied for some 6d SCFTs
\cite{Haghighat:2013gba,Haghighat:2013tka,Kim:2014dza,Haghighat:2014vxa}.
We find and study the self-dual string gauge theories for the minimal SCFT with
$SU(3)$ gauge group. The $SU(3)$ SCFT is interesting in itself, but is also a
building block of many other interesting 6d SCFTs.

\begin{table}[t!]
\begin{center}
\begin{tabular}{c||c|c|c|c|c|c|c|c|c
    }
	\hline
	$n$ & $1$ & $2$ & $3$ & $4$ & $5$ & $6$ & $7$ & $8$
    & $12$ \\
    \hline
    gauge symmetry & - & - & $SU(3)$ & $SO(8)$ & $F_4$ & $E_6$ & $E_7$ & $E_7$
    & $E_8$ \\
    \hline
    global symmetry &$ E_8$ & $SO(5)_R$ & - & - & - & - & - & - & 
    - \\
    \hline matters &&&-&-&-&-&$\frac{1}{2}{\bf 56}$&-&
    - \\
	\hline
\end{tabular}
\caption{Symmetries/matters of minimal SCFTs}\label{minimal}
\vskip 0.5cm
\begin{tabular}{c||c|c|c}
	\hline
	base & $3,2$ & $3,2,2$ & $2,3,2$\\
    \hline
    gauge symmetry & $G_2\times SU(2)$ & $G_2\times Sp(1)$ &
    $SU(2)\times SO(7)\times SU(2)$ \\
    \hline matters &$\frac{1}{2}({\bf 7}+{\bf 1},{\bf 2})$
    & $\frac{1}{2}({\bf 7}+{\bf 1},{\bf 2})$ &
    $\frac{1}{2}({\bf 2},{\bf 8},{\bf 1})+
    \frac{1}{2}({\bf 1},{\bf 8},{\bf 2})$ \\
	\hline
\end{tabular}
\caption{Symmetries/matters of other `atomic' non-Higgsable SCFTs}\label{other}
\end{center}
\end{table}
We first explain some aspects of the minimal SCFTs \cite{Morrison:1996na,Morrison:1996pp,Witten:1996qb} and their strings.
In the F-theory setting, there is a compact 2-cycle $\mathbb{P}^1$ in the
$\mathbb{F}_n$ base, whose volume is the VEV of the tensor
multiplet scalar. $n$ is the self intersection number of $\mathbb{P}^1$. The 6d CFT
is localized at the singularity of $\mathbb{F}_n$ when $\mathbb{P}^1$ degenerates.
By taking a non-compact limit of $\mathbb{F}_n$ to the $O(-n)\rightarrow\mathbb{P}^1$
bundle, one can decouple gravity and get the 6d CFT at low energy.
The elliptic fiber of CY$_3$ can degenerate on
the base $\mathbb{F}_n$, whose loci define 7-branes.
7-branes wrapping $\mathbb{P}^1$ yield 6d gauge symmetries.
Table \ref{minimal} shows the symmetries and matters
of the minimal SCFTs.\footnote{In Table \ref{minimal}, matters narrowly mean
charged hypermultiplets in 6d Yang-Mills. This loses meaning at $n=1,2$,
but they also have matters in a sense, as they
have Higgs branch ($n=1$) or $\mathcal{N}=(2,0)$ tensor branch ($n=2$).}
The self-dual strings are D3-branes wrapping $\mathbb{P}^1$. The CFTs
in Table \ref{minimal} are `atoms' of the recent F-theory construction
\cite{Heckman:2013pva,Heckman:2015bfa}, together with $3$ more
atomic building blocks listed in Table \ref{other} \cite{Morrison:2012np}.

The 2d gauge theories on the self-dual strings were found for minimal
SCFTs at $n=1,2,4$ so far. In these cases, there are D-brane
engineerings of the strings, whose open string dynamics at low energy
guarantees the existence of 2d gauge theories.
See \cite{Kim:2014dza,Kim:2015fxa} for the strings at $n=1$,
called `E-strings' \cite{Klemm:1996hh,Minahan:1998vr};
\cite{Haghighat:2013gba,Haghighat:2013tka} for the strings
at $n=2$, called `M-strings' for the 6d $(2,0)$ theory of $A_1$ type;
\cite{Haghighat:2014vxa} for the strings at $n=4$.
\cite{Haghighat:2013gba,Haghighat:2013tka,Gadde:2015tra,Kim:2015fxa}
studied the strings of 6d SCFTs with higher dimensional tensor branches,
made with the atoms at $n=1,2,4$.

The models in Table \ref{minimal} with $n\geq 5$ have
exceptional gauge symmetries. To understand what
it means to the constructions of 2d gauge theories for the strings,
we consider the self-dual strings from 6d effective Yang-Mills descriptions
in the tensor branch.
When the 6d SCFT has gauge symmetry, the self-dual
strings are instanton string solitons of the 6d Yang-Mills theory.
Namely, the Yang-Mills theory has soliton solutions of
the self-duality equation $F=\star_4 F$ on $\mathbb{R}^4$, being strings in 6d.
The low energy gauge theories on the worldvolume of instanton solitons are
well known for classical gauge groups, which is closely related to the so-called
ADHM construction of instantons \cite{Atiyah:1978ri}.
Such gauge theory descriptions are unknown for exceptional instantons.
So constructing UV gauge theories for the strings at $n\geq 5$ means that
one finds novel ADHM-like descriptions of exceptional instantons,
which sounds very challenging. On the other hand, the cases with $n=3,4$ apparently
look simple, since $SU(3)$ and $SO(8)$ instantons admit ADHM constructions.
Indeed, the 2d gauge theories at $n=4$ studied in
\cite{Haghighat:2014vxa} are the ADHM gauge theories
for the $SO(8)$ instanton strings. However, for $SU(3)$, it turns out that the naive
ADHM gauge theories suffer from 2d gauge anomalies, failing to provide good descriptions.
In fact, the 6d $SU(3)$ gauge fields emerge from various light $(p,q)$
string junctions suspended between $7$-branes, and not just
fundamental strings \cite{Grassi:2014sda}.\footnote{In a suitable $SL(2,\mathbb{Z})$
frame, two A-branes (i.e. D7-branes) and two C-branes make $SU(3)$ \cite{Grassi:2014sda}.} So there are no reasons to expect the naive ADHM gauge theories to work.
It is more suitable to regard non-Higgsable $SU(3)$
as the simplest exceptional gauge group. For instance, as reviewed
in section 2, the $SU(3)$ minimal SCFT cannot be unHiggsed to an
infinite sequence of classical gauge theories, and only admits finite unHiggsing
sequences to exceptional gauge theories.

We construct anomaly-free 2d gauge theories which describe these $SU(3)$ strings.
More precisely, our UV gauge theories are made of $\mathcal{N}=(0,2)$
supermultiplets, with novel non-holomorphic interactions which only
preserve $\mathcal{N}=(0,1)$ supersymmetry. We propose that these gauge theories
flow to the $\mathcal{N}=(0,4)$ SCFTs on the self-dual strings in IR. We first
find that the classical moduli space is given by the $SU(3)$ instanton moduli spaces,
and make some consistency checks on quantum corrections. We then show that the
elliptic genera computed from our gauge theories completely agree with
the data known from topological strings \cite{Haghighat:2014vxa}. Also, our
gauge theories reduced to 1d should be identical to the $SU(3)$ ADHM quantum
mechanics in IR. We support this by showing that our
gauge theories (reduced to 1d) and the standard $SU(3)$
ADHM quantum mechanics have same Witten indices, the instanton partition
functions \cite{Nekrasov:2002qd}.
Finally, our gauge theories show the correct 2d anomalies for the self-dual
strings, which we can independently compute from anomaly inflows from 6d.

There are many interesting extensions of our $SU(3)$ strings' gauge theories,
since $SU(3)$ minimal SCFT is a building block of novel
6d CFTs. The $(E_6,E_6)$ conformal matter called $131$ \cite{DelZotto:2014hpa}
is an example, constructed by gluing an $SU(3)$ minimal SCFT with two
E-string theories. In M-theory, it is engineered by putting an M5-brane
on an $E_6$ ALE singularity, after which a novel fragmentation of M5-brane happens.
We construct the gauge theories for their strings, and make nontrivial tests
from the elliptic genera and anomalies.

Other interesting strings, such as the strings of the SCFTs in
Table \ref{other}, can also be studied. Note that the theories in Table \ref{other}
either include exceptional gauge group $G_2$, or $SO(7)$ hypermultiplets in
the spinor representation ${\bf 8}$. Before forming quivers as shown in
Table \ref{other}, these parts are related to our $SU(3)$ theory by Higgsing:
see (\ref{Higgsing-SU3}). They are cases in which the ADHM
descriptions are unknown. We find that our new descriptions of $SU(3)$ instantons
can be easily extended to describe some aspects of $G_2$
instantons or $SO(7)$ instantons with hypermultiplets in ${\bf 8}$.
So one can study the strings of all 6d CFTs listed
in Table \ref{other}. The theories in Table \ref{other} are important building
blocks of interesting 6d CFTs, e.g. the $(E_7,E_7)$ conformal matter
$12321$ \cite{DelZotto:2014hpa}. What we mentioned in this paragraph will be reported
in a separate publication \cite{kkkp}.

The rest of this paper is organized as follows. In section 2, we review some
useful facts about 6d SCFTs in the tensor branch, from the Yang-Mills theory
viewpoint. In section 3, we explain aspects of 2d $\mathcal{N}=(0,2)$
gauge theories and novel interactions.
We propose our 2d gauge theories for the $SU(3)$ self-dual strings.
We show that they have correct moduli spaces and elliptic genera.
In section 4, we glue our $SU(3)$ string gauge theory with
the E-string gauge theories, to describe the strings of the $(E_6,E_6)$ conformal
matter. In section 5, we show that the 2d chiral anomalies of our gauge theories
cancel with the 6d anomaly inflow calculations. Section 6 discusses
various open issues and future directions.

\section{6d SCFTs and Yang-Mills theories}

We first review some aspects of 6d SCFTs with gauge symmetries, and their 6d Yang-Mills
effective descriptions in the tensor branch. For simplicity, we only consider the
cases with one dimensional tensor branch. One has a
tensor multiplet, consisting of a 2-form $B_{\mu\nu}$
(whose 3-form flux satisfies the self-duality condition), a real scalar $\Phi$,
and fermions. When the 6d theory has a
gauge symmetry $G$, its vector multiplet consists of the gauge field
$A_\mu$ and fermions. There may be hypermultiplet matters in
the representation ${\bf R}$ of $G$. In the tensor branch, this system
admits an effective field theory description in which the VEV $\langle\Phi\rangle>0$
sets the inverse gauge coupling $\frac{1}{g_{YM}^2}$ of the 6d Yang-Mills theory.
The bosonic part of the tensor/vector multiplet action is given by
\begin{equation}
  S^{\rm bos}_{\rm vector+tensor}=\int\left[\frac{1}{2}d\Phi\wedge \star d\Phi
  +\frac{1}{2}H\wedge \star H\right]+\sqrt{c}
  \int\left[-\Phi {\rm tr}(F\wedge \star F)
  +B\wedge{\rm tr}(F\wedge F)\right]
\end{equation}
with certain $c>0$ that depends on the theory, where
\begin{equation}
  H\equiv dB+\sqrt{c}\ {\rm tr}\left(AdA-\frac{2i}{3}A^3\right)\ .
\end{equation}
This action should be understood as providing the equation of motion by varying
the action with $B_{\mu\nu}$, after which the self-duality constraint $H=\star H$ is
imposed by hand. The hypermultiplet part of the action is standard, which we do
not explain here.

The equation of motion for $B$ is given by
\begin{equation}
  d\star H\left(=dH\right)=\sqrt{c}\ {\rm tr}(F\wedge F)\ .
\end{equation}
Self-dual string solutions should have their tensions proportional
to $\langle\Phi\rangle$, and source nonzero $H=\star H$. The configurations
with nonzero ${\rm tr}(F\wedge F)$, namely the instanton string solitons,
provide such sources. They are extended along
$\mathbb{R}^{1,1}\subset\mathbb{R}^{5,1}$ and satisfy
\begin{equation}\label{self-dual}
  F=\pm\star_4 F
\end{equation}
on the transverse $\mathbb{R}^4$, where the instanton number
$k=\frac{1}{8\pi^2}\int_{\mathbb{R}^4}{\rm tr}(F\wedge F)$ is quantized.
BPS self-dual strings further satisfy $H=\mp\star_4 d\Phi$.
Here, the upper/lower signs correspond to $k>0$ and $k<0$, respectively.
We shall consider self-dual instantons with $k>0$.

Let us explain various consistency conditions for the 6d SCFTs,
from the Yang-Mills description. We first discuss the gauge anomaly cancelation.
Gauge anomalies come both from tree and 1-loop levels, which should cancel each
other via the Green-Schwarz mechanism \cite{Green:1984bx,Sagnotti:1992qw}.
Under the following gauge transformation
\begin{equation}
  \delta A_\mu=D_\mu\epsilon\ ,\ \
  \delta B_{\mu\nu}=-\sqrt{c}\ {\rm tr}(\epsilon F_{\mu\nu})\ ,
\end{equation}
the action suffers from the classical anomaly
\begin{equation}
  \delta S=\delta\left[\sqrt{c}\int B\wedge{\rm tr}(F\wedge F)\right]
  =-c\int{\rm tr}(\epsilon F)\wedge{\rm tr}(F\wedge F)\ ,
\end{equation}
contributing a term proportional to $c\ {\rm tr}(F^2)^2$
to the anomaly polynomial. The 1-loop anomaly from the box diagrams,
where fermions in the vector and hypermultiplets run through the loop,
contributes terms proportional to ${\rm tr}_{\rm adj}(F^4)$
and ${\rm tr}_{\bf R}(F^4)$. For the net anomaly to cancel, the combination
of the quartic Casimirs ${\rm tr}_{\rm adj}(F^4)$ and ${\rm tr}_{\bf R}(F^4)$
appearing at 1-loop should factorize to a square of quadratic Casimirs.
The factorization condition of the quartic Casimir severely constrains
possible gauge groups and matters.
Firstly, the factorization can happen when $G$ does not have
independent quartic Casimirs. Among simple groups, this is true for $G=SU(2)$,
$SU(3)$, $G_2$, $F_4$, $E_6$, $E_7$, $E_8$. So for these $G$, one can introduce
hypermultiplets in any representation ${\bf R}$, if it is compatible with $c>0$.
Exceptionally, $G=SO(8)$ may also yield anomaly-free systems. This is because
${\rm tr}_{\rm adj}(F^4)$ can be factorized as ${\rm tr}(F^2)^2$ for $SO(8)$. So
the 6d $SO(8)$ super-Yang-Mills without matters can be made anomaly free.
If ${\bf R}$ is taken to be a suitable representation of $SO(8)$ whose
quatic Casimir factorizes, such hypermultiplets can also be introduced.
An example is ${\bf R}=n({\bf 8}_v\oplus{\bf 8}_s\oplus{\bf 8}_c)$. There are
other possible choices of $G$ and ${\bf R}$, such as $G=SU(N)$ and $N_f=2N$
fundamental hypermultiplets.

The global anomalies
\cite{Bershadsky:1997sb} further constrain possible ${\bf R}$.
For instance, among the gauge groups mentioned in the previous paragraph,
the cases with $G=SU(2)$, $G_2$ suffer from global anomalies without matters.
So if one restricts oneself to the non-Higgsable
theories with void ${\bf R}$ or just with one half hypermultiplet (which
cannot be used to Higgs the system), one recovers the gauge groups listed
in Table \ref{minimal}. More generally, the numbers of some simple
hypermultiplets are constrained as \cite{Bershadsky:1997sb}
\begin{eqnarray}\label{list}
  SU(2)&:&n_{\bf 2}=4,10,\cdots\hspace{1cm}\in 4+6\mathbb{Z}_{\geq 0}\nonumber\\
  SU(3)&:&n_{\bf 3}=0,6,12,\cdots\hspace{.6cm}\in 6\mathbb{Z}_{\geq 0}\nonumber\\
  G_2&:&n_{\bf 7}=1,4,7,\cdots\hspace{.8cm}\in 1+3\mathbb{Z}_{\geq 0}\ ,
\end{eqnarray}
where $\mathbb{Z}_{\geq 0}=\{0,1,2,3,\cdots\}$, to avoid global anomalies.

Going back to the ordinary gauge anomaly, suppose that the 1-loop anomaly
factorizes. Then, the coefficient $c$ appearing in the classical action has to
be tuned to have the classical anomaly to cancel the factorized 1-loop
anomaly. One has to make sure that the value of $c$ that ensures
anomaly cancelation is positive. If this condition is not met, one fails to have
a conformal field theory decoupled from gravity. Here, we note that the signs in front
of the 1-loop anomaly contributions are different for vector multiplet and
hypermultiplets. Anomaly
from vector multiplet tends to increase $c$, while those from hypermultiplets tend to
decrease $c$. So to have $c>0$, ${\bf R}$ should not be too big \cite{Seiberg:1996qx}.
For instance, in the list (\ref{list}) illustrated above, $c>0$ is satisfied
at $n_{\bf 2}=4,10$ for $SU(2)$, at $n_{\bf 6}=0,6,12$ for $SU(3)$, and at
$n_{\bf 7}=1,4,7$ for $G_2$.

These are the consistency conditions for 6d SCFTs from Yang-Mills viewpoint.
It will be helpful to know all consistent 6d super-Yang-Mills-Higgs systems
coupled to single tensor multiplet (with a simple gauge group $G$)
\cite{Bershadsky:1997sb,Heckman:2015bfa}, without imposing the non-Higgsability
condition which leads to Table \ref{minimal}. These theories form various
sequences, related to other theories by Higgsing/unHiggsing.
All possible Higgsing sequences finish by one of the minimal SCFTs listed in Table
\ref{minimal}. Of course the procedure of unHiggsing is not unique, which means
that two sequences can merge. The possible
unHiggsing sequences can be classified into `classical gauge theories'
and `exceptional gauge theories,' depending on whether the sequence is
infinite or finite.

Firstly, the E-string theory at $n=1$ has two classical sequences
\begin{equation}\label{E-string-classical}
  (n=1)\leftarrow(SU(2),n_{\bf 2}=10)\leftarrow
  \left\{\begin{array}{l}(Sp(N),n_{\bf 2N}=8+2N)\ \ \ \ (N\geq 2)\\
  (SU(3),n_{\bf 3}=12)\leftarrow(SU(N),n_{\bf N}=8+N,n_{\bf anti}=1)
  \ \ \ \ (N\geq 4)\end{array}\right.
\end{equation}
and three exceptional sequences,
\begin{eqnarray}\label{n=1-exceptional}
  (n=1)&\leftarrow&(SU(2),n_{\bf 2}=10)\leftarrow(SU(3),n_{\bf 3}=12)
  \leftarrow(SU(4),n_{\bf 4}=12,n_{\bf 6}=1)\\
  &&\hspace*{.7cm}\leftarrow(SU(5),n_{\bf 5}=13,n_{\bf 10}=1)\leftarrow
  (SU(6),n_{\bf 6}=15,n_{\frac{1}{2}{\bf 20}}=1)\nonumber\\
  (n=1)&\leftarrow&(SU(2),n_{\bf 2}=10)\leftarrow(SU(3),n_{\bf 3}=12)
  \leftarrow(G_2,n_{\bf 7}=7)\nonumber\\
  &&\hspace*{.7cm}\leftarrow(SO(7),n_{\bf 7}=2,n_{\bf 8}=6)
  \leftarrow(SO(8),n_{{\bf 8}_v}\!=\!n_{{\bf 8}_s}\!=\!n_{{\bf 8}_c}\!=\!3)
  \nonumber\\
  &&\hspace*{2cm}\leftarrow \left\{\begin{array}{lll}
  (SO(N),n_{\bf N}=N-5,n_{\bf S}=\frac{48}{d_{\bf S}})_{N=9,\cdots,12}\\
  (F_4,n_{\bf 26}=4)\leftarrow(E_6,n_{\bf 27}=5)
  \leftarrow(E_7,n_{\frac{1}{2}\bf 56}=7)\leftarrow(E_8,n_{\rm inst}=11)
  \end{array}
  \right.\ .\nonumber
\end{eqnarray}
Here, $d_{\bf S}=2^{\left[\frac{N-1}{2}\right]}$, and $n_{\bf S}$ denotes the
number of hypermultiplets in spinor representation for odd $N$, and
twice the number of hypermultiplets in chiral and anti-chiral spinor
representations, respectively, for even $N$. (Namely, there are
$\frac{n_{\bf S}}{2}$ hypermultiplets
in chiral and anti-chiral spinor representations, respectively). The notation
$(G,n_{\bf R})$ means a gauge theory with gauge group $G$ and $n_{\bf R}$
hypermultiplets in representation ${\bf R}$. `${\bf anti}$' denotes rank $2$
antisymmetric representation of $SU(N)$, of dimension $\frac{N^2-N}{2}$, and
${\bf 20}$ for $SU(6)$ is the rank $3$ antisymmetric representation.
The arrows $\leftarrow$ mean Higgsings, and whenever two sequences
meet like $\leftarrow\left\{\frac{}{}\!\!\right.\ \ $, it means that two
Higgsing sequences merge to a single one.
When a sequence contains the parameter $N$, the sequence of Higgsings proceeds
by reducing $N$ by $1$. Note also that,
although the first sequence of (\ref{n=1-exceptional}) is uplifted to $SO(N)$
gauge theories, their rank $N$ cannot be made indefinitely large because this
sequence contains matters in spinor representations. Namely, since the dimension
and the quartic/quadratic Casimirs of spinor representations grow exponentially
in $N$, one cannot have anomaly-free gauge theories with spinor
matters at too large $N$, meaning that the sequence should terminate. So
$SO(N)$ gauge theories with matters in spinor representation should be
regarded as exceptional gauge theories. This is natural also because such matters
cannot be engineered by D-branes and open fundamental strings only. Same
comment applies to the first exceptional unHiggsing sequence, which terminates
by the $SU(6)$ theory with a half hyperultiplet in ${\bf 20}$.

Before proceeding, we explain what $(E_8,n_{\rm inst})$ means,
as this is not well defined within Yang-Mills theory.
It is the $E_8$ gauge theory with one tensor multiplet, coupled to extra
$n_{\rm inst}$ `small instantons' in the heterotic string picture.
In field theoretic terms, `$n_{\rm inst}$ small instantons' means the rank
$n_{\rm inst}$ E-string theory, with $n_{\rm inst}$ dimensional tensor branch and
also a Higgs branch given by the moduli space of $n_{\rm inst}$ $E_8$ instantons.
In M-theory realization, this is $n_{\rm inst}$ M5-branes probing an M9-plane.
So this rank $n_{\rm inst}$ E-string theory has $E_8$ global symmetry, which
acts nonlinearly on the Higgs branch. In the $(E_8,n_{\rm inst})$ theory, the
$E_8$ of the rank $n_{\rm inst}$ E-string theory is gauged, by identifying it
with the gauge symmetry of the $E_8$ minimal SCFT (at $n=12$ in Table \ref{minimal}).
This means that the Higgs
branch moduli space is reduced from that of the E-string
theory, and sometimes even absent if $n_{\rm inst}$ is too small.
(More comments about these theories are given at the end of this section.)

The $A_1$ $(2,0)$ theory at $n=2$ has one classical unHiggsing sequence
\begin{equation}
  (n=2)\leftarrow(SU(N),n_{\bf N}=2N)\ \ \ \ (N\geq 2)
\end{equation}
and two exceptional unHiggsing sequences
\begin{eqnarray}
  (n=2)&\leftarrow&(SU(2),n_{\bf 2}=4)\leftarrow(SU(3),n_{\bf 3}=6)\leftarrow
  (G_2,n_{\bf 7}=4)\\
  &&\hspace*{.7cm}\leftarrow(SO(7),n_{\bf 7}=1,n_{\bf 8}=4)
  \leftarrow(SO(8),n_{{\bf 8}_v}\!=\!n_{{\bf 8}_s}\!=\!n_{{\bf 8}_c}\!=\!2)
  \nonumber\\
  &&\hspace*{2cm}\leftarrow \left\{\begin{array}{lll}
  (SO(N),n_{\bf N}=N-6,n_{\bf S}=\frac{32}{d_{\bf S}})_{N=9,\cdots,12}\\
  (F_4,n_{\bf 26}=3)\leftarrow(E_6,n_{\bf 27}=4)
  \leftarrow(E_7,n_{\frac{1}{2}\bf 56}=6)\leftarrow(E_8,n_{\rm inst}=10)
  \end{array}
  \right.\ .\nonumber
\end{eqnarray}
The next two cases with $n=3,4$, with non-Higgsable gauge groups
$SU(3)$, $SO(8)$, can be unHiggsed as follows. The case with
$n=3$ only has two exceptional unHiggsing sequences:
\begin{eqnarray}\label{Higgsing-SU3}
  (SU(3))&\leftarrow&
  (G_2,n_{\bf 7}=1)\leftarrow(SO(7),n_{\bf 7}=0,n_{\bf 8}=2)
  \leftarrow(SO(8),n_{{\bf 8}_v}=n_{{\bf 8}_s}=n_{{\bf 8}_c}=1)\nonumber\\
  &&\leftarrow \left\{\begin{array}{lll}
  (SO(N),n_{\bf N}=N-7,n_{\bf S}=\frac{16}{d_{\bf S}})_{N=9,\cdots,12}\\
  (F_4,n_{\bf 26}=2)\leftarrow(E_6,n_{\bf 27}=3)
  \leftarrow(E_7,n_{\frac{1}{2}\bf 56}=5)\leftarrow(E_8,n_{\rm inst}=9)
  \end{array}
  \right.\ \ \ \ .
\end{eqnarray}
The case with $n=4$ has one
classical and one exceptional sequence:
\begin{equation}
  (SO(8))\leftarrow \left\{\begin{array}{lll}(SO(N),n_{\bf N}=N-8)
  \ \ \ \ N\geq 9\\
  (F_4,n_{\bf 26}=1)\leftarrow(E_6,n_{\bf 27}=2)
  \leftarrow(E_7,n_{\frac{1}{2}\bf 56}=4)\leftarrow(E_8,n_{\rm inst}=8)
  \end{array}
  \right.\ .
\end{equation}
It is curious to note that the $SU(3)$ minimal SCFT, although its gauge group
is classical in group theoretical sense, can never be unHiggsed into an
infinite sequence of classical gauge theories. In this sense, non-Higgsable
$SU(3)$ theory should be regarded as an exceptional gauge theory.

The other minimal SCFTs with $n\geq 5$ have non-Higgsable exceptional gauge
groups, and they only have finite unHiggsing sequences of
exceptional gauge theories. Each minimal SCFT at $n\geq 5$ has only
one unHiggsing sequence, given by
\begin{eqnarray}
  n=5&:&(F_4)\leftarrow(E_6,n_{\bf 27}=1)\leftarrow
  (E_7,n_{\frac{1}{2}{\bf 56}}=3)\leftarrow(E_8,n_{\rm inst}=7)\nonumber\\
  n=6&:&(E_6)\leftarrow(E_7,n_{\frac{1}{2}{\bf 56}}=2)\leftarrow(E_8,n_{\rm inst}=6)
  \nonumber\\
  n=7&:&(E_7,n_{\frac{1}{2}{\bf 56}}=1)\leftarrow(E_8,n_{\rm inst}=5)\nonumber\\
  n=8&:&(E_7)\leftarrow(E_8,n_{\rm inst}=4)\nonumber\\
  n=9,10,11,12&:&(E_8,n_{\rm inst}=12-n)\ .
\end{eqnarray}

In many exceptional sequences listed above, the final unHiggsed theory is often the
$(E_8,n_{\rm inst})$ theory with $n_{\rm inst}=0,\cdots,11$. This theory can be Higgsed
to $(E_7,n_{\frac{1}{2}{\bf 56}}=n_{\rm inst}-4)$ for $n_{\rm inst}\geq 4$,
but is non-Higgsable for $n_{\rm inst}<4$, which can
be understood as follows. Recall that the $(E_8,n_{\rm inst})$ theory has the Higgs
branch obtained by first taking the moduli space of $n_{\rm inst}$ $E_8$ instantons,
which is further reduced by
gauging the $E_8$ symmetry. The Higgsing to the $E_7$ theory is possible if
one can go into the reduced Higgs branch, Higgsing the gauge symmetry as
$E_8\rightarrow E_7$. It is easy to see when such Higgsings are possible.
For instance, at $n_{\rm inst}=1$, the ungauged moduli space of single $E_8$ instanton
is the cone over the coset $E_8/E_7$ (times a decoupled factor $\mathbb{R}^4$).
So after gauging the $E_8$, the Higgs branch
does not remain. More generally, to study when one can go to the Higgs branch
and break $E_8$ into $E_7$, it is useful to first consider the ungauged moduli
space of $n_{\rm inst}$ $E_8$ instantons. Away from the origin of the ungauged
Higgs branch, at most the $E_7\subset E_8$ global symmetry can be preserved.
At such points preserving $E_7$, one locally finds $n_{\rm inst}$ copies of
single instanton moduli space. This is described by $n_{\rm inst}$ copies of
hypermultiplets in $\frac{1}{2}{\bf 56}+{\bf 1}+{\bf 1}$ representation of $E_7$.
To see if going into the Higgs branch (preserving $E_7\subset E_8$) is
compatible with gauging $E_8$, one should check if the ungauged moduli space
can provide the necessary fields to make the Higgsed part of $E_8$ gauge
fields massive. Since the adjoint of
$E_8$ decomposes in $E_7$ to ${\bf 248}\rightarrow {\bf 133}+2\times {\bf 56}+3\times {\bf 1}$, one needs at least $2\times{\bf 56}$ in the moduli space of $n_{\rm inst}$
$E_8$ instantons after going to an $E_7$ preserving point.
So the Higgsing cannot happen for $n_{\rm inst}<4$, since the Higgs branch fields
only have $\frac{n_{\rm inst}}{2}{\bf 56}$. This explains why
$(E_8,n_{\rm inst})$ theories can be Higgsed to
$(E_7,n_{\frac{1}{2}{\bf 56}}=n_{\rm inst}-4)$ only for $n_{\rm inst}\geq 4$.

The cases with $n\geq 5$ and also
the case with $n=3$ (with non-Higgsable $SU(3)$ gauge group) should be regarded as
exceptional gauge theories. On the other hand, the four classical Higgsing
sequences at $n=1,2,4$ all have D-brane realizations in weakly coupled (massless
or massive) type IIA string
theory \cite{Hanany:1997gh,Brunner:1997gf}, using NS5-branes,
D6- and D8-branes, optionally an O8$^\pm$-plane, and in the second sequence of
(\ref{E-string-classical}), also a half NS5-brane stuck to O8 \cite{Kim:2015fxa}.

\section{6d $SU(3)$ self-dual strings}

We shall study the 2d supersymmetric gauge theories of
the $SU(3)$ self-dual strings. The IR SCFT is expected
to preserve $(0,4)$ supersymmetry. The natural candidate gauge theories motivated
by the ADHM construction have this amount of supersymmetry, but will suffer from
2d gauge anomalies. Improving these bad theories, we shall
find anomaly-free gauge theories.
Our final gauge theories will basically take the form of an $\mathcal{N}=(0,2)$
supersymmetric gauge theory, broken to $\mathcal{N}=(0,1)$ by certain interactions.
So we start by reviewing in section 3.1 the basic aspects of $\mathcal{N}=(0,2)$ gauge
theories, and one necessary ingredient for $\mathcal{N}=(0,1)$ theory as well.
We shall further present the gauge theories for our $SU(3)$ strings, whose
physics will be studied in sections 3.2 and 3.3.

\subsection{2d gauge theories for the $SU(3)$ strings}

We first explain the generalities of $\mathcal{N}=(0,2)$ gauge theories,
following \cite{Witten:1993yc}. First consider the vector
multiplet with gauge symmetry $G$. The vector superfield $V$ is given by
\begin{equation}\label{vector}
  V=A_0-A_1-2i\theta^+\bar\lambda_-
  -2i\bar\theta^+\lambda_-+2\theta^+\bar\theta^+ D\ .
\end{equation}
$\pm$ subscripts on spinors denote right/left chiral components,
respectively, where the supercharges are right handed.
The superspace covariant derivatives are defined by
\begin{eqnarray}
  &&\mathcal{D}_0-\mathcal{D}_1=\partial_0-\partial_1+iV\ \ ,\ \ \
  \mathcal{D}_0+\mathcal{D}_1\ =\ \partial_0+\partial_1+i(A_0+A_1)\nonumber\\
  &&\mathcal{D}_+=\frac{\partial}{\partial\theta^+}-i\bar\theta^+
  (\mathcal{D}_0+\mathcal{D}_1)\ \ ,\ \ \
  \overline{\mathcal{D}}_+\ =\ -\frac{\partial}{\partial\bar\theta^+}
  +i\theta^+(\mathcal{D}_0+\mathcal{D}_1)\ ,
\end{eqnarray}
where derivatives in $\mathcal{D}_\mu$ are defined with respect
to $y^\mu$, given by $y^0+y^1=x^0+x^1$ and
$y^0-y^1=x^0-x^1-2i\theta^+\bar\theta^+$.
$\mathcal{D}_0+\mathcal{D}_1$ is simply the usual gauge covariant derivative
$D_0+D_1$ using $x^\mu$. The gaugino superfield is given by
\begin{equation}
  \Upsilon=-\frac{1}{2}[\overline{\mathcal{D}}_+,\mathcal{D}_0-\mathcal{D}_1]
  =\lambda_--\theta^+\left(F_{01}+iD\right)
  -i\theta^+\bar\theta^+ (D_0+D_1)\lambda_-\ .
\end{equation}
The supersymmetric action for the vector multiplet is given by
\begin{equation}
  S_{V}=\frac{1}{2e^2}
  \int d^2yd\theta^+d\bar\theta^+\ {\rm tr}\overline\Upsilon\Upsilon
  =\frac{1}{e^2}\int d^2x\left[\frac{1}{2}F_{01}^2
  +i\bar\lambda_-(D_0+D_1)\lambda_-+\frac{1}{2}D^2\right]\ .
\end{equation}
The chiral superfield $\Phi$ is given by
\begin{equation}
  \Phi=\phi+\sqrt{2}\theta^+\psi_{+}-i\theta^+\bar\theta^+
  (D_0+D_1)\phi\ ,
\end{equation}
where $D_\mu$ is the covariant derivative containing the gauge field $A_\mu$.
The chiral multiplet kinetic terms (with gauge couplings) are given by
\begin{eqnarray}
  S_{\rm ch}&=&-\frac{i}{2}\int d^2yd^2\theta\
  \overline\Phi(\mathcal{D}_0-\mathcal{D}_1)\Phi\\
  &=&\int d^2x\left[-|D_\mu\phi|^2+i\bar\psi_{+}(D_0-D_1)\psi_{+}
  +Q_\Phi D\bar\phi\phi-\sqrt{2}iQ_\Phi\bar\phi\lambda_-\psi_{+}
  +\sqrt{2}iQ_\Phi\bar\psi_{+}\bar\lambda_-\phi\right]\nonumber
\end{eqnarray}
where $Q_\Phi$ is the charge of $\Phi$ in $V$ (which can be straightforwardly
extended to non-Abelian gauge group $G$ as well). The Fermi multiplet $\Lambda$
is given by
\begin{equation}\label{fermi}
  \Lambda=\lambda_--\sqrt{2}\theta^+ G_\Lambda
  -i\theta^+\bar\theta^+(D_0+D_1)\lambda_--\sqrt{2}\bar\theta^+E_\Lambda(\Phi_i)\ .
\end{equation}
$E_\Lambda(\Phi_i)$ is a composite chiral superfield which is holomorphic in
$\Phi_i$, expanded as
\begin{equation}
  E_\Lambda(\Phi_i)=E_\Lambda(\phi_i)+\sqrt{2}\theta^+
  \frac{\partial E_\Lambda}{\partial\phi_i}\psi_{i+}-i\theta^+\bar\theta^+
  (D_0+D_1)E_\Lambda(\phi_i)\ .
\end{equation}
The Fermi superfield obeys $\bar{\mathcal{D}}_+\Lambda_-=\sqrt{2}E_\Lambda(\Phi_i)$.
The Fermi multiplet action is given by
\begin{equation}
  S_F=\frac{1}{2}\int d^2y d^2\theta \bar\Lambda\Lambda=
  \int d^2x\left[i\bar\lambda_-(D_0+D_1)\lambda_-+|G_\Lambda|^2-|E_\Lambda(\phi_i)|^2
  -\left(\bar\lambda_-\frac{\partial E_\Lambda}{\partial\phi_i}\psi_{i+}+\frac{
  \partial\bar{E_\Lambda}}{\partial\bar\phi_i}\bar\psi_{i+}\lambda_-\right)\right]\ .
\end{equation}
One can introduce other interactions associated with the Fermi multiplets.
Introduce holomorphic functions $J_a(\Phi_i)$ for each Fermi multiplet $\Lambda_a$.
Then one can add the following interaction
\begin{equation}
  S_J=\frac{1}{\sqrt{2}}\int d^2y d\theta^+ \left.\Lambda_a J_a
  \right|_{\bar\theta^+=0}+c.c.=
  -\int d^2x\left(G_a J_a(\phi_i)+\lambda_{a-}\psi_{i+}
  \frac{\partial J_a}{\partial\phi_i}\right)+c.c.\ ,
\end{equation}
which preserves SUSY provided that $\sum_a E_aJ_a=0$ condition is met.
Note that, the gaugino superfield $\Upsilon$ is a Fermi multiplet, whose
associated $E$ function is zero.

So far we reviewed the standard features of $\mathcal{N}=(0,2)$ gauge theories.
However, it will turn out that holomorphic potentials $J_\Psi(\Phi)$ and
$E_\Psi(\Phi)$ associated with Fermi multiplets are not enough to realize
the physics for the $SU(3)$ self-dual strings. Namely, later in this subsection,
after presenting the $\mathcal{N}=(0,2)$ supersymmetric field contents
whose gauge anomalies cancel, we shall turn on various potentials
which guarantees the correct global symmetries and the moduli spaces.
It will turn out the holomorphic potentials $E_\Psi$ and $J_\Psi$ are
not sufficient to constrain the symmetries and moduli spaces. To achieve
our goal, we introduce certain non-holomorphic potentials for
some Fermi multiplet fields, which respect $\mathcal{N}=(0,1)$ supersymmetry
only.

The $\mathcal{N}=(0,1)$ supersymmetry of our interest is obtained from
the $\mathcal{N}=(0,2)$ supersymmetry transformation that we explained so far
by restricting the $(0,2)$ SUSY transformation
$\delta=\varepsilon Q +\bar{\varepsilon}\bar{Q}$ by $\varepsilon=\bar{\varepsilon}$.
Then, we shall introduce non-holomorphic potentials $F(\Phi,\bar\Phi)$ associated
with certain Fermi multiplets, that preserve only one real supercharge
$\mathcal{Q}\equiv Q+\bar{Q}$. $\mathcal{N}=(0,1)$ superfields are obtained by
introducing only one real coordinate $\theta$ ($=\bar\theta$) on the superspace.
The superspace realizations of the supersymmetry $\mathcal{Q}$
and the superderivative $\mathcal{D}$ are given by
\begin{equation}
  \mathcal{Q}=\frac{\partial}{\partial\theta}+2i\theta(D_0+D_1)
  \ ,\ \ \mathcal{D}=\frac{\partial}{\partial\theta}-2i\theta(D_0+D_1)\ .
\end{equation}
An $\mathcal{N}=(0,2)$ chiral multiplet $\Phi^i\sim (\phi^i,\psi^i)$ can
be written as an $\mathcal{N}=(0,1)$ complex superfield of the form
\begin{equation}
  \Phi^i=\phi^i+\sqrt{2}\theta\psi^i\ .
\end{equation}
An $\mathcal{N}=(0,2)$ Fermi multiplet $\Lambda\sim(\lambda,G)$ can be written as an
$\mathcal{N}=(0,1)$ complex superfield of the form
\begin{equation}
  \Lambda=\lambda-\sqrt{2}\theta G\ .
\end{equation}
In the $\mathcal{N}=(0,1)$ formalism, one can turn on the following
interaction $\mathcal{L}_{\rm int}^{\mathcal{N}=(0,1)}$,
\begin{equation}
  \mathcal{L}_{\rm int}^{\mathcal{N}=(0,1)}=\frac{1}{\sqrt{2}}\int d\theta
  \Lambda F(\Phi,\bar\Phi)+c.c.=-GF(\phi,\bar\phi)+
  \left(\frac{\partial F}{\partial\phi^i}\psi^i-\frac{\partial F}{\partial\bar\phi_i}
  \bar\psi_i\right)\lambda+c.c.
\end{equation}
using a non-holomorphic function $F(\Phi,\bar\Phi)$ associated with $\Lambda$.
Combined with the kinetic term for $\Lambda$, $|G|^2+i\bar\lambda D_+\lambda$,
one obtains the potential energy $-|F(\phi,\bar\phi)|^2$ after integrating
out $G$. This non-holomorphic potential will play important roles in constructing
the gauge theories for the $SU(3)$ self-dual strings (explained later in this
subsection and section 3.2).
Also, once all the fields are combined into $\mathcal{N}=(0,2)$ supermultiplets,
it will suffice to have one right-moving Hermitian supercharge $\mathcal{Q}$ in
section 3.3 to define and compute the elliptic genus.

With these backgrounds on SUSY gauge theories, we shall now propose our
models for the 6d $SU(3)$ self-dual strings.

We first review some aspects of the ADHM construction for $SU(N)$
multi-instantons, emphasizing the associated worldvolume gauge theories on
instantons. The ADHM construction was originally discovered as an ansatz for solving
the self-duality equation (\ref{self-dual}) on $\mathbb{R}^4$ \cite{Atiyah:1978ri}.
The parameters appearing in this ansatz are called ADHM data, which have to satisfy
certain algebraic equations for the ansatz to solve (\ref{self-dual}). This
construction can be promoted to gauge theory descriptions on the
worldvolumes of these solitons. The gauge theories are UV uplifts of the moduli space
approximation \cite{Manton:1981mp} of instanton solitons given by $\mathcal{N}=(0,4)$
non-linear sigma models. The ADHM data are the zero modes of the scalar fields in the
UV gauge theory, and the algebraic constraints are conditions for the scalars to
minimize the bosonic potential to zero. In our context,
the gauge theories and sigma models live on $\mathbb{R}^{1+1}$. The non-linear
sigma model description is incomplete, because the moduli space metric of instantons
has small instanton singularities. So the gauge theories provide UV
completions of nonlinear sigma models on singular target spaces.

Let us consider the classical aspects of the
$\mathcal{N}=(0,4)$ supersymmetric ADHM gauge theories, for $k$
instanton strings in $SU(N)$ Yang-Mills theory.  We first explain symmetries.
The transverse space
$\mathbb{R}^4$ to the strings has $SO(4)=SU(2)_l\times SU(2)_r$ symmetry, which
descends to 2d internal symmetries. We  denote by $\alpha$, $\dot\alpha$ the
doublet indices of $SU(2)_l$, $SU(2)_r$, respectively. The 6d superconformal
R-symmetry $SU(2)_R$ is also realized as 2d symmetry, whose doublet
index we denote by $A=1,2$. For the self-dual instantons on $\mathbb{R}^4$,
the preserved $(0,4)$ supercharges can be written as $Q_{+\dot\alpha A}$,
with $\dot\alpha=1,2$, $A=1,2$. For $SU(N)$ $k$ instantons, the standard
ADHM gauge theory has $U(k)$ gauge symmetry. The fields are given by
\begin{eqnarray}\label{ADHM-(0,4)}
  (A_\mu,\lambda_{-\dot\alpha A})&:&U(k)\textrm{ vector multiplet}\nonumber\\
  (q_{\dot\alpha},\psi_{+A})&:&\textrm{ hypermultiplets in }({\bf k},\overline{\bf N})
  \nonumber\\
  (a_{\alpha\dot\beta},\Psi_{+\alpha A})&:&\textrm{ hypermultiplets in }
  ({\bf adj},{\bf 1})\ .
\end{eqnarray}
We follow the notations of \cite{Dorey:2002ik,Kim:2011mv,Hwang:2014uwa}.
On the right side of each line, we denoted the supermultiplet type and the
representation in $U(k)\times SU(N)$. The last adjoint hypermultiplet fields
satisfy a reality condition, so that
$a_m\sim a_{\alpha\dot\beta}(\bar\sigma_m)^{\dot\beta\alpha}$ become $4$
Hermitian matrices. $\pm$ subscripts denote the right/left chiral fermions,
respectively, as in the previous subsection. These are on-shell field contents.
To explain the action and SUSY, we shall use the $\mathcal{N}=(0,2)$
off-shell formalism that we already explained, following \cite{Tong:2014yna}.
A vector multiplet decomposes into a $(0,2)$ vector multiplet and
an adjoint Fermi multiplet. A hypermultiplet decomposes into
a pair of chiral multiplets in conjugate representations. Thus we obtain
\begin{eqnarray}\label{ADHM-(0,2)}
  (A_\mu,\lambda_{-\dot\alpha A})&\rightarrow&
  (A_\mu,\lambda_0,D)+(\lambda,G_\lambda)_{(R,J)=(1,-1)}\nonumber\\
  (q_{\dot\alpha},\psi_{+A})&\rightarrow&
  (q,\psi_+)_{(R,J)=(0,\frac{1}{2})}
  +(\tilde{q},\tilde\psi_+)_{(R,J)=(0,\frac{1}{2})}
  \nonumber\\
  (a_{\alpha\dot\beta},\Psi_{+\alpha A})&\rightarrow&
  (a,\Psi)_{(R,J,J_l)=(0,\frac{1}{2},\frac{1}{2})}
  +(\tilde{a},\tilde\Psi)_{(R,J,J_l)=(0,\frac{1}{2},-\frac{1}{2})}\ .
\end{eqnarray}
Here, the $(0,2)$ R-charge is given by $R=2J_{R}$, where $J_R$ is the Cartan
of $SU(2)_R$, so that $R[Q]=-1$ for
$Q\equiv Q^{\dot{1}2}\sim Q_{\dot{2}1}$. $J\equiv J_r+J_R$ and $J_l$ are
treated as flavor symmetries in the $(0,2)$ setting, where $J_l$ and $J_r$ are
the Cartans of $SU(2)_l$ and $SU(2)_r$, respectively.
The Lagrangian in the $(0,2)$ formalism is determined by specifying the holomorphic
$E_\Psi(\Phi)$, $J_\Psi(\Phi)$ for each Fermi field $\Psi$, as aleady explained.
The choice which yields enhanced $\mathcal{N}=(0,4)$ supersymmetry is \cite{Tong:2014yna}
\begin{equation}
  J_\lambda=\sqrt{2}\left(q\tilde{q}+[a,\tilde{a}]\right)\ ,
\end{equation}
and $E_\lambda=0$.
The bosonic potential is given by
\begin{equation}
  V={\rm tr}\left[\frac{1}{2}D^2+\sum_{\Psi}\left(|J_\Psi|^2+|E_\Psi|^2\right)\right]
\end{equation}
where $D$ is the on-shell value of the D-term field
\begin{equation}
  D=qq^\dag-\tilde{q}^\dag\tilde{q}+[a,a^\dag]+[\tilde{a},\tilde{a}^\dag]\ .
\end{equation}
This potential shows enhanced $SO(4)_R=SU(2)_r\times SU(2)_R$ R-symmetry,
which is a consequence of the enhanced $\mathcal{N}=(0,4)$ SUSY.
The minima of the bosonic potential are solutions of $D=0$, $J_\lambda=0$.
These are the so-called ADHM constraints
for $q_{\dot\alpha}$, $a_{\alpha\dot\beta}$ \cite{Atiyah:1978ri}. The resulting
Higgs branch is identified
as the moduli space of $k$ $SU(N)$ instantons. Had this gauge theory been a good
quantum theory, it would have yielded at low energy the $\mathcal{N}=(0,4)$
non-linear sigma model whose target space is given by the instanton 
moduli space.

However, since the theory is chiral, one should worry about the $U(k)$
gauge anomaly. We separately consider the $SU(k)$ and $U(1)$ parts of the
anomalies. It turns out that this quiver is anomalous. The $SU(k)$ anomaly from
a fermion in the representation ${\bf R}$ is
proportional to the index $D_{\bf R}$, defined by
${\rm tr}_{\bf R}(T^aT^b)=D_{\bf R}\delta^{ab}$. With $D_{\bf k}=1$ and
$D_{\bf adj}=2k$, the $SU(k)$ anomaly at $N=3$ is proportional to
\begin{equation}\label{ADHM-anomaly}
  2\cdot 3\cdot 1+2\cdot 2k-2\cdot 2k=6\neq 0\ .
\end{equation}
The three terms come from $\psi_{+A}$, $\Psi_{+A}$,
$\lambda_{-\dot\alpha A}$, respectively, where we count the numbers
of complex fermions. The overall $U(1)$ part of $U(k)$ is also anomalous.
So the naive ADHM quivers fail to provide consistent
UV theories.

Now we present the modified 2d quiver for the $SU(3)$ instanton strings.
We first explain the fields.
Take $G=U(k)$ for $k$ self-dual strings and first keep all the ADHM fields
given by (\ref{ADHM-(0,4)}), or equivalently (\ref{ADHM-(0,2)}), that we copy
here for $N=3$:
\begin{eqnarray}\label{original-fields}
  (A_\mu,\lambda_0,\bar\lambda_0,D)+(\lambda,G_\lambda)&:&U(k)\textrm{ vector multiplet}
  +\textrm{complex adjoint Fermi multiplet}
  \nonumber\\
  (q,\psi_+)+(\tilde{q},\tilde\psi_+)
  &:&\textrm{ chiral multiplets in }({\bf k},\overline{\bf 3})
  +(\overline{\bf k},{\bf 3})\nonumber\\
  (a,\Psi_+)+(\tilde{a},\tilde\Psi_+)&:&\textrm{chiral multiplets in }
  ({\bf adj},{\bf 1})+({\bf adj},{\bf 1})\ .
\end{eqnarray}
Then we add the following $\mathcal{N}=(0,2)$ supermultiplets:
\begin{eqnarray}\label{new-fields}
  (\phi,\chi)&:&\textrm{chiral multiplet in }
  (\overline{\bf k},\overline{\bf 3})\nonumber\\
  (b,\xi)+(\tilde{b},\tilde\xi)&:&
  \textrm{two chiral multiplet in }(\overline{\bf anti},{\bf 1})\nonumber\\
  (\hat\lambda,\hat{G})&:&\textrm{complex Fermi multiplet
  in }({\bf sym},{\bf 1})\nonumber\\
  (\check\lambda,\check{G})&:&\textrm{complex Fermi multiplet in }
  ({\bf sym},{\bf 1})\nonumber\\
  (\zeta,G_\zeta)&:&\textrm{complex Fermi multiplet in }
  (\overline{\bf k},{\bf 1})\ .
\end{eqnarray}
Here, `${\bf anti}$' and `${\bf sym}$' denote rank $2$ antisymmetric and symmetric
representations of $U(k)$, respectively. Finally, we also introduce the following pair
of chiral and Fermi multiplets, which will also play some roles later,
\begin{eqnarray}\label{new-fields-2}
  (\tilde\phi,\tilde\chi)&:&\textrm{chiral multiplet in }({\bf k},{\bf 1})\nonumber\\
  (\eta,G_\eta)&:&\textrm{complex Fermi multiplet in }(\bar{\bf k},{\bf 1})\ .
\end{eqnarray}
The pair of multiplets in (\ref{new-fields-2}) will not be directly relevant
to the IR physics, but will play a subtle role in section 3.2,
to make the quantum moduli space to have to right structures.

Before explaining the interactions, we show that the above fields
define anomaly-free gauge theories. We start by considering the $SU(k)$ anomaly.
Recall that the anomaly from (\ref{original-fields}) is proportional to
$+6$, as computed in (\ref{ADHM-anomaly}). We compute the $SU(k)$ anomaly of
the fermions in the new multiplets (\ref{new-fields}). With
$D_{\bf sym}=k+2$, $D_{\bf anti}=k-2$, one obtains
\begin{equation}
  +3\cdot 1+2(k-2)-(k+2)-(k+2)-1=-6\ ,
\end{equation}
where the five terms come from $\chi$, ($\xi,\tilde\xi$), $\hat\lambda$, $\check\lambda$,
$\zeta$, respectively. This perfectly cancels (\ref{ADHM-anomaly}). As for
(\ref{new-fields-2}), the anomalies of $\tilde\chi$ and $\eta$ mutually cancel.
The overall $U(1)$ anomaly, proportional to $\pm Q^2$ for a right/left chiral
fermion with $U(1)$ charge $Q$, is given by
\begin{equation}
  +3\cdot 2 \cdot 1^2\cdot k+3\cdot 1^2\cdot k+2\cdot 2^2\cdot
  \frac{k^2-k}{2}-2^2\cdot\frac{k^2+k}{2}-2^2\cdot\frac{k^2+k}{2}-1^2\cdot k
  =0\ ,
\end{equation}
where the six terms come from $(\psi_+,\tilde\psi_+)$, $\chi_+$,
($\xi_{+}$, $\tilde\xi_+$), $\hat\lambda_-$, $\check\lambda_-$, $\zeta$, respectively.
Again the anomalies from (\ref{new-fields-2}) mutually cancel.
So the $U(1)$ gauge anomaly also cancels.

Now we write the action. We
only need to specify $J, E$  or $F(\Phi,\bar\Phi)$ for each Fermi multiplet.
We first take
\begin{eqnarray}
  &&J_\lambda=e\left[q\tilde{q}+[a,\tilde{a}]\right]\ ,\ \ E_\lambda=0\nonumber\\
  &&J_{\check\lambda}=m_{\check\lambda}\left[
  (\phi\tilde{q})_S+(\tilde{b}a-b\tilde{a})_S\right]
  \ ,\ \ E_{\check\lambda}=0\ ,
\end{eqnarray}
where $(\cdots)_S$ denotes symmetrizing the two $\bar{\bf k}$ indices. Note that
$\epsilon^{\alpha\beta}a_\alpha b_\beta=a\tilde{b}-\tilde{a}b$ with
$a_\alpha\equiv (a,\tilde{a})$, $b_\alpha\equiv(b,\tilde{b})$ is an $SU(2)_l$
invariant. Furthermore, we introduce the interaction $\mathcal{L}_{\rm int}^{\mathcal{N}=(0,1)}$
given by
\begin{eqnarray}\label{SU(3)-non-holomorphic}
  \mathcal{L}_{\rm int}^{\mathcal{N}=(0,1)}&=&\frac{1}{\sqrt{2}}\int d\theta
  \left[\hat{\lambda}^{IJ}\left(\alpha\phi_I q^\dag_J+\beta
  (b_{IK}(a^{\dag})^K_{\ \ J}+\tilde{b}_{IK}(\tilde{a}^\dag)^K_{\ \ J})
  \right)\right.\\
  &&\hspace*{2.5cm}\left.\frac{}{}\!\!+
  \zeta_I\left(\gamma_1\epsilon^{ijk}q_i^I\phi_{jJ}(\tilde{q}^\dag)_k^J+
  \gamma_2\epsilon^{ijk}q_i^J\phi_{jJ}(\tilde{q}^\dag)_k^I\right)\right]
  +c.c.\ ,\nonumber
\end{eqnarray}
where upper/lower $I,J,K=1,\cdots,k$ indices denote fundamental/anti-fundamental
indices of $U(k)$. Here, all the superfields are $\mathcal{N}=(0,1)$ superfields,
explained earlier in this subsection. 
Note that, although we took the relative coefficients of various terms
in a given $J$ potential to be $1$, $\mathcal{N}=(0,2)$ SUSY allows more general
relative coefficients. We do not know whether the IR SCFT (which we claim to have
enhanced $\mathcal{N}=(0,4)$ SUSY) is affected by these relative coefficients or not.
However, to clearly see a signal of $(0,4)$ SUSY enhancement at 1-loop order in
the next subsection, we have taken the mass scale for $J_\lambda$ to be the
gauge coupling $e$. The other parameters $m_{\check\lambda}$, 
$\alpha,\beta,\gamma_1,\gamma_2$ of mass dimensions 
will be taken to be much larger than $e$.
Note that, at $k=1$, the last two terms with coefficients $\gamma_1$ and $\gamma_2$
become indistinguishable. Finally, we turn on the following extra $\mathcal{N}=(0,1)$
interactions for the fields in (\ref{new-fields-2})
\begin{equation}\label{extra-non-holomorphic}
  \Delta\mathcal{L}^{\mathcal{N}=(0,1)}_{\rm int}=
  \frac{1}{\sqrt{2}}\int d\theta\eta\tilde\phi
  \left[\kappa_1|q|^{2a}|\tilde{q}|^{2b}+\kappa_2f(|q|^2,|\phi|^2)\right]
\end{equation}
where the exponents have to fixed to $a=1$, $b=1$.
$|q|^2$, $|\tilde{q}|^2$ etc. imply the $U(k)\times SU(3)$ singlet contractions.
$f$ is quite flexibly allowed to be a function of $|q|^2$ and $|\phi|^2$
that will not affect our IR physics too much, supposing that it meets
some conditions (to be addressed in section 3.2).
Again, we do not know how to fix the coefficients $\kappa_1,\kappa_2$.
The last interaction $\Delta\mathcal{L}^{\mathcal{N}=(0,1)}_{\rm int}$ will
only play important roles in section 3.2, to have the quantum correction to
the classical moduli space to be consistent with the self-dual string physics. Note
also that, the interaction $\Delta\mathcal{L}^{\mathcal{N}=(0,1)}_{\rm int}$
demands that the global symmeries of $\tilde\phi$ and $\eta$ are all frozen
to be mutually conjugate. So $\tilde\phi$ and
$\eta$ will completely cancel out in the elliptic genus of section 3.3.

\begin{table}[t!]
\begin{center}
$$
\begin{array}{c||c|c|c|c|c||c|c}
	\hline
    {\rm superfields}&U(k)&SU(3)&SU(2)_l&U(1)_R&U(1)_J&U(1)_\phi&U(1)_g\\
    \hline
	V&{\bf adj}&{\bf 1}&{\bf 1}&0&0&0&0\\
    \hline \lambda &{\bf adj}&{\bf 1}&{\bf 1}&1&-1&0&0\\
    \hline
    \hat{\lambda}&{\bf sym}&{\bf 1}&{\bf 1}&0&0&-1&0\\
    \hline \check\lambda &{\bf sym}&{\bf 1}&{\bf 1}&1&-1&-1&0\\
    \hline \zeta&\bar{\bf k}&{\bf 1}&{\bf 1}&0&-\frac{1}{2}&-1&-3\\
    \hline
    q&{\bf k}&\bar{\bf 3}&{\bf 1}&0&\frac{1}{2}&0&1\\
    \hline \tilde{q}&\bar{\bf k}&{\bf 3}&{\bf 1}&0&\frac{1}{2}&0&-1\\
    \hline
    \phi&\bar{\bf k}&\bar{\bf 3}&{\bf 1}&0&\frac{1}{2}&1&1\\
    \hline
    (a,\tilde{a})\equiv a_\alpha&{\bf adj}&{\bf 1}&{\bf 2}&0&\frac{1}{2}&0&0\\
    \hline (b,\tilde{b})\equiv b_\alpha
    &\overline{\bf anti}&{\bf 1}&{\bf 2}&0&\frac{1}{2}&1&0\\
    \hline
\end{array}
$$
\caption{The $\mathcal{N}=(0,2)$ superfields and global symmetries}\label{charges}
\end{center}
\end{table}
We study the global symmetries of our system. The flavor symmetries should be
compatible with the potentials that we turned on,
and also should not have mixed anomalies with the $U(1)\subset U(k)$ gauge
symmetry. Including the symmetries $U(k)\times SU(3)$, the full set of anomaly-free
global symmetries and charges/representations of fields are listed in Table
\ref{charges}. We first explain $U(1)_R$, $U(1)_J$.
We claim that this theory flows to a system with $\mathcal{N}=(0,4)$ SUSY in the IR,
which has $SO(4)=SU(2)_R\times SU(2)_r$ R-symmetry that rotates the supercharges
$Q_{\dot\alpha A}$. Their Cartans $J_R$, $J_r$ are identified with $R,J$ in our UV
theory by $R=2J_R$, $J=J_r+J_R$.
Had there been no $\mathcal{N}=(0,1)$ interactions, the $\mathcal{N}=(0,2)$ superspace
coordinates $\theta$, $\bar\theta$ would have carried $R[\theta]=R[\bar\theta]=-1$,
making $R$ to be the $\mathcal{N}=(0,2)$ R-symmetry. However, in our current setting,
$U(1)_R$ is regarded as a flavor symmetry.
We have omitted one flavor symmetry, $U(1)_{\tilde\phi}$, in Table \ref{charges},
which acts only
on $\tilde\phi$ and $\eta$ with charges $+1$ and $-1$, respectively.
The two fields rotate only under $U(1)_{\tilde\phi}$, and are neutral
under all the symmetries shown in Table \ref{charges}. $U(1)_{\tilde\phi}$
will play not role at all in the IR physics that we study. We expect
this $U(1)$ to decouple from the IR SCFT that we are interested in,
which is compatible with all the studies that we make in this paper.

We pay more attention to $U(1)_g$ and $U(1)_\phi$.
$U(1)_g$ is a linear combination of the global symmetry $U(1)_\phi$
and the overall $U(1)$ part of the $U(k)$ gauge symmetry. Namely, from Table
\ref{charges}, one finds $Q_g=2Q_\phi+Q$, where $Q$ is the $U(1)\subset U(k)$
charge, $Q_\phi$ is the $U(1)_\phi$ charge, and $Q_g$ is the $U(1)_g$ charge.
So $Q_g-2Q_\phi$ can be eaten up by the $U(1)\subset U(k)$ gauge transformation,
meaning that its values are zero for gauge invariant observables.
The remaining $U(1)_\phi$ in Table \ref{charges} is
not expected from the 6d $SU(3)$ SCFT.
Here, it is useful to know that $U(1)_g$ is naturally regarded as the overall
$U(1)\subset U(3)$, if we regard $SU(3)$ as embedded in $U(3)$,
and if we take $\zeta$ to be in the determinant representation of $\bar{\bf 3}$
in $U(3)$. Note also that, for $U(N)$ or $SU(N)$ instantons, the ADHM formalism always
has $U(N)$ symmetry as a default. One either kills the overall $U(1)$ part of $U(N)$
by hand, or sometimes it can be eaten up by the overall $U(1)\subset U(k)$
gauge symmetry. In our case, $U(1)_g$ is not completely
eaten up by gauge symmetry, but only a combination of $U(1)_g\times U(1)_\phi$
is. So for our 2d gauge theories describe the self-dual strings of this 6d CFT,
$U(1)_\phi$ should somehow decouples in IR.
In the next subsections, we shall claim (with evidences) that the IR
$\mathcal{N}=(0,4)$ SCFT does not see $U(1)_\phi$, by illustrating that our gauge
theories exhibit an IR decoupling into two sectors. The sector that we
are interested in is neutral under $U(1)_\phi$ rotation, so that the remaining
$U(1)_g$ is eaten up by the gauge transformation. Thus, we shall have the correct
bosonic global symmetries that we want for self-dual strings.

In the next two subsections, we provide evidences that
our gauge theories describe the physics of
the $SU(3)$ self-dual strings. We first show that the moduli space is the
$SU(3)$ instanton moduli space, at classical level and then at 1-loop quantum level
in the asymptotic region of the moduli space.
This provides (modest) evidences for our claim that the system flows to an
$\mathcal{N}=(0,4)$ theory in IR. We shall then study the
elliptic genera, showing that they reproduce
the results from topological strings and extend them to all genus sum.
We finally show that the zero momentum sectors of our elliptic genera
reproduce the Nekrasov's instanton partition functions, computed from the
standard ADHM quantum mechanics.

\subsection{Moduli spaces: classical and quantum aspects}

We study the low energy physics of our gauge theories in the Higgs branch, when
the energy scale $E$ is much lower than the 2d gauge coupling, $E\ll e$.
We shall obtain the $SU(3)$ instanton moduli space target space, in which only 
the ADHM fields $q,\tilde{q},a,\tilde{a}$ develop flat
directions. However, when the ADHM fields are not large, near the small instanton
singularity of the moduli space, other fields provide extra light degrees of freedom
supported near the small instanton region, and UV complete the non-linear sigma model.
There will also be an extra branch apart from the above instanton moduli space, which will
meet the first branch at a point classically. After a quantum consideration following
\cite{Melnikov:2012nm}, we shall argue that the two branches decouple
and also that the observables we shall discuss in this paper acquire contributions from
the first branch only.

We first make a classical analysis of the moduli space, by studying the zero loci
of the bosonic potential energy. They are given by the zeros of the D-term potential,
$J$ potentials, and also the non-holomorphic potentials $F$. They are given by
\begin{eqnarray}\label{moduli-condition}
  D&:&qq^\dag-\tilde{q}^\dag q-\phi\phi^\dag+[a,a^\dag]+[\tilde{a},\tilde{a}^\dag]
  -2b^\dag b-2\tilde{b}^\dag \tilde{b}=0\nonumber\\
  J_\lambda&:&q\tilde{q}+[a,\tilde{a}]=0\nonumber\\
  F_{\hat{\lambda}}&:&
  \alpha(\phi q^\dag)_S+\beta(ba^\dag+\tilde{b}\tilde{a}^\dag)=0\nonumber\\
  J_{\check\lambda}&:&(\phi\tilde{q})_S+(\tilde{b}a-b\tilde{a})_S=0\nonumber\\
  F_\zeta&:&\epsilon^{ijk}(\gamma_1q_i^I\phi_{jJ}(\tilde{q}^\dag)_k^J+
  \gamma_2q_i^J\phi_{iJ}(\tilde{q}^\dag)^I_k)=0\ ,
\end{eqnarray}
and additionally
\begin{equation}\label{extra-condition}
  F_\eta\ :\ \tilde\phi
  \left(\kappa_1|q|^{2a}|\tilde{q}|^{2b}+\kappa_2f(|q|^2,|\phi|^2)\right)=0
\end{equation}
with $a=1$, $b=1$.
We are interested in nonzero solutions to these equations, i.e. away from
the small instanton singularity where the nonlinear sigma model will be singular.
Away from the singularity, the expression inside the parenthesis of
(\ref{extra-condition}) will always be nonzero, so that one always finds $\tilde\phi=0$.
(This can be guaranteed by suitable choices of $f$.) From now on, we shall always set
$\tilde\phi=0$ in the classical analysis, coming back to its role later for quantum studies.

We shall mostly study the case with $k=1$, and will only briefly comment on the
cases with $k>1$ at the end of this subsection.
At $k=1$, $b,\tilde{b}$ fields are absent, and (\ref{moduli-condition})
reduces to
\begin{eqnarray}\label{k=1-constraints}
  &&q_i(q^\dag)^i-(\tilde{q}^\dag)_i \tilde{q}^i-\phi_i(\phi^\dag)^i=0\ ,\ \
  q_i\tilde{q}^i=0\nonumber\\
  &&\phi_i(q^\dag)^i=0\ ,\ \ \phi_i\tilde{q}^i=0\ ,\ \
  \epsilon^{ijk}q_i\phi_j(\tilde{q}^\dag)_k=0\ .
\end{eqnarray}
We view $q$, $\tilde{q}^\dag$, $\phi$ as complex vectors of $SU(3)$,
transforming in $\bar{\bf 3}$, and call them $v_1,v_2,v_3$ respectively. The
conditions (\ref{k=1-constraints}) can be rewritten as
\begin{eqnarray}
  &&|v_1|^2-|v_2|^2-|v_3|^2=0\ ,\ \ v_2^\dag \cdot v_1=0\ ,\nonumber\\
  &&v_1^\dag\cdot v_3=0\ ,\ \ v_2^\dag \cdot v_3=0\ ,\ \
  v_3^T\cdot(v_1\times v_2)=0
\end{eqnarray}
The
second line of (\ref{k=1-constraints}) requires that $v_3$ is orthogonal
to $v_1$, $v_2$ and $v_1^\ast\times v_2^\ast$, where $\times$ denotes
the exterior product of two 3d vectors. Orthogonality of two vectors $u,v$
are defined by $u^\ast\cdot v=0$. On the other hand, $q_i\tilde{q}^i=0$
requires $v_1,v_2$ to be orthogonal. So the three equations on the second
line demands that $v_3=\phi$ is zero if both $v_1$ and $v_2$ are nonzero.
This defines our first branch. Then the first line of (\ref{k=1-constraints}) is
the ADHM constraints for single $SU(3)$ instantons, realizing the classical
instanton moduli space.

There is an extra branch, in which some of $v_1,v_2$ are zero.
From $|v_1|^2=|v_2|^2+|v_3|^2$ on the first line, $v_1=0$ means that all
three vectors are zero, which is the small instanton singularity.
When $v_2=0$ with $v_1\neq 0$, one finds $|v_1|^2=|v_3|^2$ and
$v_1^\ast \cdot v_3=0$. This defines our second branch. Both branches
have to be modded out by the $U(1)$ gauge orbit, which achieves the Kahler
quotient. The two branches only meet at the origin $q=0$, $\tilde{q}=0$,
$\phi=0$.

A similar analysis can be made after turning on Fayet-Iliopoulos (FI) parameter
$r$, adding a term $-rD$ to the action. Then the D-term condition is
modified to $|q|^2-|\tilde{q}|^2-|\phi|^2=r$. In this case, the origins of
the two branches are resolved into $\mathbb{CP}^2$. When $r>0$, the two $\mathbb{CP}^2$
factors are both given by $|q|^2=r$, so the two branches meed at $\mathbb{CP}^2$.
When $r<0$, the two $\mathbb{CP}^2$ factors are given by
$|\tilde{q}|^2=-r$ and $|\phi|^2=-r$, respectively. In this case, the two
$\mathbb{CP}^2$ factors are different subspaces of a $\mathbb{CP}^4$ defined by
\begin{equation}
  |\tilde{q}|^2+|\phi|^2=-r\ \ ,\ \ \tilde{q}^i\phi_i=0\ \ ,\ \ q_i=0\ .
\end{equation}
So when $r<0$, the two branches meet by having the
$\mathbb{CP}^4$ to bridge them.

We now discuss the quantum moduli space. The classical picture of the moduli space will drastically change once we consider the quantum effects.
For instance, we shall find evidences that the two
branches are disconnected after including quantum corrections,
along the line of \cite{Melnikov:2012nm}.
Although some interactions in our gauge theories preserves only $\mathcal{N}=(0,1)$
supersymmetry, a crucial quantum analysis along the line of \cite{Witten:1993yc,Melnikov:2012nm} will go in a similar manner as
$\mathcal{N}=(2,2)$ or $\mathcal{N}=(0,2)$ gauge theories.
So we first review aspects of quantum corrections in $\mathcal{N}=(0,2)$ gauge
theories.

Following \cite{Melnikov:2012nm}, the idea is to take the massive parameters
$m_{\check\lambda}$ ,$\alpha,\beta,\gamma_1,\gamma_2$, $\kappa_1,\kappa_2$
to be much larger than $e$. In $\mathcal{N}=(0,2)$ theories, there can be quantum
corrections to the classical analysis in the following way.
We keep nonzero light fields $\Phi_L$, including those which form our classical 
moduli space, and integrate out heavy fields
$\Phi_H$ \cite{Melnikov:2012nm}
when the masses of heavy fields $\Phi_H$
are sufficiently large. We shall only study the leading
1-loop correction to the classical results, which will be justified when 
the values of nonzero fields forming the classical moduli space are not
too small. Thus the analysis of this section will be valid away from the small 
instanton singularity. One important quantum correction
is the field-dependent 1-loop renormalization of the FI parameter \cite{Witten:1993yc}, 
based on the log interactions \cite{Quigley:2012gq}.
\cite{Melnikov:2012nm} makes a careful integration over the massive modes
to derive other associated effects in the $\mathcal{N}=(0,2)$ effective action,
including the moduli space metric. We shall only pay attention
to the renormalization of the 1-loop FI parameter.
Since we study the case with $k=1$, we explain the 1-loop analysis for the
$U(1)$ gauge theory.
The 1-loop renormalization of the FI parameter $r$ can be computed by keeping
the auxiliary field $D$, and computing its 1-point function with nonzero
background field $\Phi_L$ \cite{Witten:1993yc,Melnikov:2012nm}.
D-term field classically couples to the scalars in the charged chiral multiplets,
\begin{equation}
  D\left(\frac{}{}\!\!\right.\sum_i Q_i|\Phi_i|^2-r\left.\frac{}{}\!\!\right)\ ,
\end{equation}
where $r$ is the bare FI parameter. So at 1-loop order, only the chiral multiplet
scalars $\phi_i$ contribute to the 1-point function of $D$. Since the calculation
includes scalars only, the analysis is the same in our gauge theories.
When some light fields $\Phi_L$ are nonzero,
the heavy fields $\Phi_H$ acquire nonzero masses $M_H$ which
depend on $\Phi_L$. Integrating out
the heavy fields yields the following 1-loop correction to the FI parameter
\cite{Witten:1993yc,Melnikov:2012nm}
\begin{equation}\label{1-loop-heavy}
  \Delta r=-\langle \sum_HQ_H\Phi_H(0)\Phi_H(0)^\dag\rangle=
  -\sum_H \frac{Q_H}{(2\pi)^2}\int \frac{d^2p}{p^2+M_H^2}=
  \sum_H \frac{Q_H}{2\pi}\log\left(\frac{M_H}{\Lambda}\right)\ .
\end{equation}
Here $\Lambda$ is the cut-off energy scale. This 1-loop analysis from the weakly-coupled
gauge theory is reliable when all $M_H$'s are large enough so that $e\ll M_H$.
The microscopic FI parameter is well defined if $\sum_iQ_i=0$, where $i$
runs over all chiral multiplets. This is not the case for our theory.
If $\sum_H Q_H=0$ which will also not true in our case, $r+\Delta r$
is the net field-dependent FI parameter at low energy. If $\sum_H Q_H\neq 0$,
the bare parameter $r$ can be absorbed into a redefinition of $\Lambda$,
which one may call renormalized FI parameter $r_{\rm ren}$.

Now we consider $\Delta r$ of our gauge theory, at $k=1$,
in the two classical branches. We first discuss the first branch, in which
$q_i$ and $\tilde{q}^i$ satisfying $q_i\tilde{q}^i=0$ are the low energy bosonic
degrees of freedom in $\Phi_L$. Without losing generality,
we can take them to be $q_i=(q,0,0)$ and $\tilde{q}^i=(0,\tilde{q},0)$
with $q,\tilde{q}\neq 0$ in a region of the 2d spacetime, which are subject to
very slow variations with wavelength $\lambda\gg e^{-1}\gg M_H^{-1}$.
To see the possibility of $\mathcal{N}=(0,4)$ enhancement more concisely, 
we also keep the auxiliary field $G_\lambda$ unintegrated, which we expect 
to form a triplet of D-term fields in the $(0,4)$ language.
Keeping $G_\lambda$ unintegrated, and taking other mass parameters of 
section 3.1 to be large, the massless fields in this background are 
$q_i$, $\tilde{q}^i$, with $i=1,2,3$. The high frequency modes of
these light fields at most provide field independent renormalization of
the FI parameter \cite{Melnikov:2012nm}, which is absorbed into $r_{\rm ren}$.
The heavy fields in this background, belonging to $\Phi_H$, are
$\phi_i$ for all $i=1,2,3$, and $\tilde\phi$.
Note that $\sum_HQ_H\neq 0$. Their masses computed from the potentials of section 3.1
are given by
\begin{equation}
  M_{\phi_1}\sim|q|\ ,\ \ M_{\phi_2}\sim|\tilde{q}|\ ,\ \
  M_{\phi_3}\sim|q||\tilde{q}|
\end{equation}
and
\begin{equation}
  M_{\tilde\phi}=|q|^{2a}|\tilde{q}|^{2b}
\end{equation}
with $a=b=1$. Collecting the 1-loop contributions (\ref{1-loop-heavy})
for all fields in $\Phi_H$, one obtains
\begin{eqnarray}
  r+\Delta r&=&\frac{1}{2\pi}\left[r-\log|q|
  -\log|\tilde{q}|-\log(|q||\tilde{q}|)+\log(|q|^{2a}|\tilde{q}|^{2b})+
  \textrm{const.}\right]\nonumber\\
  &=&\frac{1}{2\pi}\left[r_{\rm ren}+\log\left(|q|^{2a-2}|\tilde{q}|^{2b-2}\right)\right]
  \longrightarrow\frac{r_{\rm ren}}{2\pi}\ ,
\end{eqnarray}
where the last step holds for the choice $a=b=1$. Thus, we have integrated out 
the fields whose masses are much larger than $e$, and found that the 1-loop correction 
leaves the D-term unchanged.  So the triplet of D-term conditions under $SU(2)_r$
$D$, $G_\lambda$ remains unchanged. Since $SU(2)_r$ is 
part of the $\mathcal{N}=(0,4)$ R-symmetry, this can be regarded as an evidence 
for the IR SUSY enhancement.

So at least away from the small instanton singularity (i.e. 
when $|q|,|\tilde{q}|$ are not too small), we saw that 
the triplet of classical D-term constraints remains
unchanged. One slightly uncomfortable aspect is that 
the FI parameter $r_{\rm ren}$ is still allowed to deform the potential. 
We think this is in contrast to the 6d self-dual string perspective, due to the
following reason. From the viewpoint of instanton moduli space,
The FI parameter makes a non-commutative deformation of the spacetime
$\mathbb{R}^4$. However, it is unlikely
that such backgrounds would be allowed. This is because we do not know how to introduce
such non-commutative deformations in the 6d $SU(3)$ Yang-Mills theory,
unless we extend it to a $U(3)$ theory. In particular, having the Higgsing
sequence (\ref{Higgsing-SU3}) in mind, the 6d $SU(3)$ theory should come
from 6d exceptional gauge theories or $SO(N)$ gauge theories with matters,
for which one cannot turn on non-commutative deformations.
Perhaps the consistency of SUSY enhancement, or other physics, at 
the small instanton singularity might be requiring a further constraint 
$r_{\rm ren}=0$ that we cannot address with our techniques. 
It will be nice if one can formulate this problem in a quantitative way.

Next we study the 1-loop correction in the second branch, at large enough
$|q|$ and $|\phi|$. Now since $q$ and $\phi$ have no chance to combine into 
a doublet of $SU(2)_r$, unlike $q$ and $\tilde{q}$ in the first branch, 
there is no motivation
to keep $G_\lambda$ unintegrated. So we only keep the off-shell auxiliary 
field $D$, and compute its 1-point function. We shall be eventually interested 
in the low energy theory at energies much lower than all energy scales of 
our theory, including $e$, so that we restrict our studies to the 
asymptotic region $|q|,|\phi|\gg 1$. In this setting, we again rely on 
the 1-loop analysis.
We take $q_i=(q,0,0)$ and $\phi_i=(0,\phi,0)$ as our background, without losing
generality. Then, $q\equiv q_1$, $q_3$, $\phi\equiv\phi_2$, $\phi_3$ are massless
fields, and $\tilde{q}^i$ for all $i=1,2,3$, $\tilde\phi$ are
massive fields belonging to $\Phi_H$, with their masses given by
\begin{equation}
  M_{\tilde{q}^1}\sim|q|\ ,\ \ M_{\tilde{q}^2}\sim|\phi|\ ,\ \
  M_{\tilde{q}^3}\sim|q||\phi|\ ,
\end{equation}
while $M_{\tilde\phi}$ will depend on the choice of $f(|q|^2,|\phi|^2)$
in (\ref{extra-non-holomorphic}). For instance, we take
$f\sim |q|^{2c}|\phi|^{2d}$,
with $c$ and $d$ being positive integers.
Then one finds $M_{\tilde\phi}\sim|q|^{2c}|\phi|^{2d}$. 
The contribution of these massive fields to $\Delta r$ is 
$\frac{1}{2\pi}\log\left(|q|^{2c-2}|\phi|^{2d-2}\right)$.
Finally, $q_2$, $\phi_1$ become partly massive due to the potential 
$m_{\hat{m}}^2|\phi_i q^{\dag i}|^2=m_{\hat\lambda}^2|q\phi_1+\phi q^{\dag 2}|^2$. 
Going to a new basis of fields, diagonalizing the kinetic terms 
including this mass matrix, and integrating out both fields, one finds a 
complicated contribution to $\Delta r$. We do not need to know 
its form for our discussions. After all, one finds
\begin{equation}
  \Delta r=\frac{1}{2\pi}\left[{\rm const.}+\log\left(|q|^{2c-2}
  |\phi|^{2d-2}\right)+\cdots\right]\ ,
\end{equation}
where $\cdots$ stands for the contribution from $\phi_1$, $q_2$.
The 1-loop modified D-term condition is given by
\begin{equation}\label{1-loop-2nd}
  |q|^2-|\phi|^2=r+\Delta r\ .
\end{equation}
To make the physics simpler to understand, it is helpful 
to choose generic values of $c,d$, especially satisfying $c>1$, $d>1$. Then, 
as in \cite{Melnikov:2012nm}, neither $q$
nor $\phi$ can vanish if they are required to solve the D-term condition, 
since the right hand side diverges at $q=\phi=0$. This implies that, by quantum effects,
the second branch becomes disconnected from the first branch. 
This phenomenon has been studied in detail in \cite{Melnikov:2012nm}. Although 
our anlaysis can be trusted at large enough $|q|$, $|\phi|$, it is 
quite natural to believe that this phenomenon will happen exactly. 
We shall assume the disconnection and make further discussions.\footnote{Of course, 
observables like the elliptic genera are
insensitive to the choice of $c,d$. It is not clear to us what actually
happens at, say, $c=d=1$.}

The disconnection of the two branches implies that
our gauge theory flows to two decoupled QFTs in IR. The first branch is of our interst. 
Of course, the nonlinear sigma model is incomplete, since the target space has a small 
instanton singularity at $q=0$, $\tilde{q}=0$. So the RG flow of our gauge theory 
implicitly defines a UV completion of the nonlinear sigma model near the
singularity. Since our moduli space analysis is reliable away from the singularity,
it is a priori unclear whether the full SCFT on the first branch
has enhanced supersymmetry. It will be very desirable to set up a framework in which
one can test our claim much more nontrivially, beyond the consistency checks
that we made.

We revisit our previous statements about global symmetries of the QFT on
the first branch. Note that our UV gauge theory had one unwanted global
symmetry $U(1)_\phi$, which leaves the original ADHM fields
invariant, while rotating all the extra added fields. Since the nonlinear
sigma model on the first branch is made with the original ADHM fields only,
it is very likely that the first branch does not see $U(1)_\phi$ (unless the
UV completion at small instanton singularity spoils this fact). We shall
assume that this IR decoupling of $U(1)_\phi$ is an exact property,
even at small instanton singularity, which will be consistent with our elliptic
genus calculus in the next subsection. So with this understood, $U(1)_g$ can be
absorbed by $U(1)\subset U(k)$ gauge symmetry and we have exactly the desired
global symmetries in the first branch.

We also make a comment on one important property of the BPS states and their
$U(1)_\phi$ charges, that will be studied in more detail in the next
subsection. Namely, one can argue that the second branch will not contribute
to the elliptic genus, to be defined and computed in the next subsection.
This can be seen by turning on the chemical potential $m$ for $U(1)_\phi$ in
the index, which amounts to covariantizing the Euclidean action inside
the path integral with the background gauge field proportional to $m$,
$D_\tau\rightarrow D_\tau+imQ_\phi$, where $\tau$ is the compactified
Euclidean time for the index. At low energy, this covariantization will
yield mass terms for the fields charged in $U(1)_\phi$ in the second branch,
proportional to $m^2|\phi|^2$ at $k=1$. But since we already know that the
vanishing condition $D=0$ of the D-term potential forbids $\phi$ from vanishing,
the net potential energy after adding
$m^2|\phi|^2$ can not have a minimum at zero. Since the net potential cannot
vanish in the second branch,
it will not have supersymmetric saddle points so that the second branch
will not contribute to the index. It more abstractly implies that the states with
$U(1)_\phi$ charges will not appear in the index, meaning that the indices studied
in section 3.3 will be all independent of $m$. As we shall comment again in
the next subsection, we concretely checked the $m$ independence of the indices,
which is a very nontrivial support of our general argument here. More practically,
since the second branch will not contribute to the BPS states, one can ensure that
the indices computed in our UV gauge theories are identical to the indices of the
IR SCFTs in the first branch.

We close this subsection by commenting on the moduli spaces at $k>1$,
which is much more complicated than the case with $k=1$. We only sketch
some aspects. We first start by counting the number of fields and constraints.
We have $3k^2+8k$ complex bosonic fields, and $3k^2+2k$ complex constraints
from potentials. So
the moduli space has $6k$ complex, or $12k$ real dimensions, including
the position moduli of $k$ strings on $\mathbb{R}^4$. This is the right
dimension of $SU(3)$ instanton moduli space. An obvious class of solution
is obtained by setting $\phi=0$, $b=0$, $\tilde{b}=0$, and taking
$q,\tilde{q},a,\tilde{a}$ to satisfy the ADHM constraints
$qq^\dag-\tilde{q}^\dag\tilde{q}+[a,a^\dag]+[\tilde{a},\tilde{a}^\dag]=0$,
$q\tilde{q}+[a,\tilde{a}]=0$.
Since the moduli space has $6k$ complex dimension, at generic values of
$q,\tilde{q}$, $a,\tilde{a}$ satisfying the ADHM constraints, the other fields
$\phi,b,\tilde{b}$ will be massive. However, there is an extra branch
of classical moduli space which meets the instanton moduli space at
special subspaces of dimension lower than $6k$. The equations are very complicated
that we did not manage to make a general analysis like the case with $k=1$:
we only managed to find special solutions, to see some structures of the
possible extra branches and how they intersect with the instanton moduli space.
(For instance, we have studied the extra branch formed by nonzero $b,\tilde{b}$
which meets a subspace of the instanton moduli space at $a=\tilde{a}=0$.)
In all the special solutions that we found,
the intersection of the instanton moduli space with other branch
happens at small instanton singularities, where the nonlinear sigma model on the
first branch breaks down. (Note that for $k>1$, small instanton singularity is
not just a point, but forms a nontrivial subspace when any of the $k$ instantons
become small.)
It will be interesting to first solve the classical problem of identifying
the full set of extra branches for $k>1$ (for instance at
$k=2$). One can then check whether they meet
the instanton moduli space only at the small instanton singularity,
and further check if the 1-loop quantum corrections can make the
two branches disconnected. We do not carry out this analysis in this paper for
$k\geq 2$.

\subsection{Elliptic genera and Witten indices}

We investigate the BPS spectrum of our gauge theories from their elliptic genera.
Elliptic genus is a Witten index which captures the BPS spectrum of the circle
compactified theory. It is represented by a supersymmetric partition
function of Euclidean QFT on $T^2$. The definition of the elliptic genus, as well
as its formula for gauge theories, can be found in \cite{Benini:2013nda}.
The elliptic genus for $k$ strings
in the Hamiltonian picture is defined by \cite{Kim:2014dza}
\begin{equation}
  Z_k(\tau,\epsilon_{1,2},m_a)
  ={\rm Tr}\left[(-1)^Fe^{2\pi i\tau H_+}e^{2\pi i\bar\tau H_-}
  e^{2\pi i\epsilon_1(J_1+J_R)}e^{2\pi i\epsilon_2(J_2+J_R)}\cdot
  \prod_{a\in{\rm flavor}}e^{2\pi i m_aF_a}\right]\ ,
\end{equation}
where $H_\pm\equiv\frac{H\pm P}{2}$ with Hamiltonian $H$ and momentum $P$,
$J_1,J_2$ are two angular momenta in $SO(4)$ which rotate orthogonal 2-planes of
$\mathbb{R}^4$, $J_R$ is the Cartan of the $SU(2)_R$ symmetry, and $F_a$
are the other flavor charges.
For our $SU(3)$ theory, they are the Cartans of the 6d $SU(3)$
gauge symmetry, which are realized as global symmetries in 2d. The corresponding
fugacity factor is given by $\prod_{i=1,2,3}e^{2\pi i v_i F_i}$, with
the constraint $v_1+v_2+v_3=0$ from the traceless condition of $SU(3)\subset U(3)$.
In $\mathcal{N}=(0,2)$ theories, one finds $H_-\sim \{Q,\overline{Q}\}$.
$\bar\tau$ is the standard regulator of the Witten index and does not
appear in $Z_k$. One can also define and compute this partition function for theories
with $\mathcal{N}=(0,1)$ supersymmetry, if the theory has $\mathcal{N}=(0,2)$
supermultiplets in its fields and the symmetry is just broken by interactions.
$\bar\tau$ does not appear in $Z_k$ because $H_-\sim\mathcal{Q}^2$, where $\mathcal{Q}$
is the Hermitian $\mathcal{N}=(0,1)$ supercharge. Also, the calculation of $Z_k$ can
be done by localizing the partition function by $\mathcal{Q}$-exact terms,
following the methods of \cite{Benini:2013nda}. The result is also completely the
same as \cite{Benini:2013nda}, since the calculation can be done after turning off
$\mathcal{L}_{\rm int}^{\mathcal{N}=(0,1)}$.

We first summarize the result for the elliptic genera of gauge theories, following
\cite{Benini:2013nda}. The supersymmetric path integral of the
gauge theory on $T^2$ can be computed in the weak-coupling limit,
essentially by doing suitable Gaussian path integrals around saddle points.
The saddle points are given by the flat connections of $U(k)$ gauge fields on
$T^2$, which can be labeled (up to conjugation) by two commuting matrices with
$k$ eigenvalues,
\begin{equation}\label{zero-modes}
  A_1+\tau A_2={\rm diag}(u_1,u_2,\cdots,u_k)\ \ \ ,\ \ \ \
  u_I\in\mathbb{C}/(\mathbb{Z}+\tau\mathbb{Z})\ .
\end{equation}
Around a given saddle point, one integrates over all the massive modes,
given by a 1-loop determinant $Z_{\textrm{1-loop}}(u,\tau,z)$.
($z$ collectively denotes the chemical
potentials $\epsilon_{1,2},m_a$.) After evaluating the 1-loop integrals, one finally
integrates over the flat connections. This integral turns
out to be a residue sum, called the Jeffrey-Kirwan residue.
Each chiral multiplet and
complex Fermi multiplet contributes the following
factor to $Z_{\textrm{1-loop}}$,
\begin{equation}\label{1-loop}
  Z_{\textrm{chiral}}=\frac{i\eta}{\theta_1(Q\cdot u+\rho\cdot z)}\ ,\ \
  Z_{\textrm{Fermi}}=\frac{\theta_1(Q\cdot u+\rho\cdot z)}{i\eta}\ ,
\end{equation}
where $\rho$ is the global charge of the field which is conjugate to the chemical
potential $z$, and $Q$ denotes the gauge charge conjugate to $u$.
The contribution
to $Z_{\textrm{1-loop}}$ from the vector multiplet (in the adjoint representation
of the gauge group $G$) is given by
\begin{equation}
  Z_{\rm vec}=\prod_{I=1}^k\frac{2\pi\eta^2 du_I}{i}\cdot
  \prod_{\alpha\in{\rm root}}\frac{\theta_1(\alpha\cdot u)}{i\eta}\ ,
\end{equation}
where $\alpha$ runs over the roots of $G=U(k)$. After multiplying these factors
to form $Z_{\textrm{1-loop}}$, one integrates over $u$ to obtain the
elliptic genus:
\begin{equation}
  Z=\frac{1}{(2\pi i)^k}\oint\frac{1}{|W(G)|}Z_{\textrm{1-loop}}\ .
\end{equation}
$|W(G)|$ is the order of the Weyl group of $G$, which is $k!$ for $U(k)$.
The contour integral is given by the Jeffrey-Kirwan residue sum,
which we denote by JK-Res. See \cite{Benini:2013nda} for its definition.

We now study our gauge theories of section 3.1, using the elliptic genus.
The elliptic genus for $k$ $SU(3)$ strings is given by\footnote{The overall
factor $(-1)^{\frac{k^2-k}{2}}$ is obtained by first collecting all factors of
$i$'s appearing in $Z_{\textrm{1-loop}}$, and then putting an extra factor of
$(-1)^k$ by hand. This overall sign factor cannot be determined in 2d QFT,
and should be determined by considering the 6d physics. For instance, one
can fix the sign as (\ref{SU(3)-contour}) by comparing the 1d limit of our
result with the well-known 5d $SU(3)$ instanton partition function.}
\begin{eqnarray}\label{SU(3)-contour}
  \hspace*{-.5cm}Z_k^{SU(3)}&=&\frac{(-1)^{\frac{k^2-k}{2}}}{(2\pi i)^kk!}\oint \prod_{I=1}^k\left(2\pi\eta^2du_I\right)\cdot
  \frac{\theta_1(2\epsilon_+)^k}{\eta^k}\cdot\frac{\eta^{2k}}{\theta_1(\epsilon_{1,2})^k}
  \prod_{I<J}\frac{\theta_1(\pm u_{IJ})\theta_1(2\epsilon_+\!\mp \!u_{IJ})}{\eta^4}
  \nonumber\\
  &&\prod_{I<J}\frac{\eta^4}{\theta_1(\epsilon_{1,2}\pm u_{IJ})}
  \cdot\frac{\eta^2}{\theta_1(\epsilon_{1,2}-(u_I+u_J))}
  \cdot\frac{\theta_1(u_I\!+\!u_J)\theta_1(2\epsilon_+\!-\!(u_I\!+\!u_J))}{\eta^2}\\
  &&\cdot\prod_{I=1}^k\frac{\theta_1(2u_I)\theta_1(2\epsilon_+-2u_I)}{\eta^2}
  \cdot\frac{\eta^{9}}{\theta_1(\epsilon_+\pm(u_I-v_{1,2,3}))
  \theta_1(\epsilon_+-u_I-v_{1,2,3})}\cdot
  \frac{\theta_1(u_I+\epsilon_+)}{\eta}\ .\nonumber
\end{eqnarray}
Our notation is to multiply all the $\frac{\theta_1}{\eta}$ factors
for the repeated arguments in $\theta_1$, e.g.
$\theta_1(\epsilon_{1,2})\equiv\theta_1(\epsilon_1)\theta_1(\epsilon_2)$,
$\theta_1(\pm u_{IJ})\equiv\theta_1(u_{IJ})\theta_1(-u_{IJ})$, and so on.
Also, we have omitted the modular parameter $\tau$ in all functions,
$\eta\equiv\eta(\tau)$, $\theta_1(z)\equiv\theta_1(\tau|z)$.
As explained in the previous subsection, the QFT in the second branch
does not contribute to the elliptic genus. So we can trust the above UV index
as the index in the first branch of our interest. To check this fact more
concretely, one could have introduced one more chemical potential $m_\phi$ in the
above contour integral, for $U(1)_\phi$. After including this extra parameter
in doing the residue sums, we have checked that $m_\phi$
never appears in $Z_k$, either exactly in some sectors, or more generally
by making series expansions in $q\equiv e^{2\pi i \tau}$ to some high orders.

The result of the contour integral is given by a residue sum called
the Jeffrey-Kirwan residue \cite{Benini:2013nda}. To find an explicit expression,
we choose an auxiliary vector $\eta=(1,\cdots,1)$ as in \cite{Hwang:2014uwa}.
With our fields and integrand, the poles with nonzero JK-Res is classified by
the so-called colored Young diagrams. The colored Young diagrams first appeared
in the context of instanton countings in $U(N)$ gauge theories in \cite{Flume:2002az},
and was derived in the context of JK-Res calculus in \cite{Hwang:2014uwa}.
In particular, the discussions of \cite{Hwang:2014uwa} straightforwardly applies
to our problem (\ref{SU(3)-contour}). We simply summarize the result here.
Firstly, to parametrize all the poles with nonzero JK-Res, one introduces
a set of $3$ Young diagrams $\vec{Y}=(Y_1,Y_2,Y_3)$, whose box numbers sum up to be $k$.
Some Young diagrams may be empty, i.e. having no boxes. Let us label these $k$
boxes of $Y$ by $s=i,(m,n)$, where $i=1,2,3$
picks one of the three Young diagrams and $(m,n)$ are the coordinates of
the box $s$ in $i$'th Young diagram from the upper-left corner. Namely,
if the box $s$ is at the $m$'th column and $n$'th row of $Y_i$, we assign
the coordinate $(m,n)$. Each box of the Young diagram encodes
the information on the pole location for a contour integral variable
$u_I$, $I=1,\cdots,k$. The order of $u_I$ is irrelevant because of the permutation
symmetry and $\frac{1}{k!}$ factor in (\ref{SU(3)-contour}). Thus for
given $Y$, we relabel $u_I$ as $u(s)$. The pole location $u(s)$ is given by \cite{Flume:2002az,Hwang:2014uwa}
\begin{equation}
  u(s)=v_i-\epsilon_+-(m-1)\epsilon_1-(n-1)\epsilon_2\ .
\end{equation}
The residue sum is given (after cancelations of various $\theta_1$ functions) by
\cite{Flume:2002az}
\begin{eqnarray}
  Z_k^{SU(3)}&=&(-1)^{\frac{k^2-k}{2}}
  \eta^{6k}\sum_{\vec{Y};|\vec{Y}|=k}\prod_{i=1}^3\prod_{s\in Y_i}
  \frac{\theta_1(2u(s))\theta_1(2\epsilon_+-2u(s))\theta_1(\epsilon_++u(s))}
  {\prod_{j=1}^3\theta_1(E_{ij})\theta_1(E_{ij}-2\epsilon_+)
  \theta_1(\epsilon_+-u(s)-v_j)}\nonumber\\
  &&\times\prod_{i\leq j}^3\prod_{s_{i,j}\in Y_{i,j}; s_i<s_j}
  \frac{\theta_1(u(s_i)+u(s_j))\theta_1(2\epsilon_+-u(s_i)-u(s_j))}
  {\theta_1(\epsilon_{1,2}-u(s_i)-u(s_j))}
\end{eqnarray}
where
\begin{equation}
  E_{ij}=v_i-v_j-\epsilon_1 h_i(s)+\epsilon_2(v_j(s)+1)\ .
\end{equation}
Here,
$h_i(s)$ is the distance from the box $s\in Y_i$ to the edge of the Young diagram $Y_i$
reached by moving right. $v_j(s)$ is the distance from the box $s\in Y_i$ to the
edge of the \textit{$j$'th Young diagram} $Y_j$ reached by moving down. See
\cite{Flume:2002az,Hwang:2014uwa,Kim:2011mv} for more explanations with examples.

There are two important tests that we make about these elliptic genera.
Firstly, we take the modular parameter $\tau$ conjugate to the circle momentum
to be $\tau\rightarrow i\infty$, or equivalently $q=e^{2\pi i\tau}\rightarrow 0$.
In this limit, one keeps the lowest order terms of the elliptic genus in $q$.
Then this will yield the index of our gauge theories reduced on a small circle.
However, in the 1d limit, the naive $SU(3)$ ADHM gauge theories that were
anomalous in 2d define completely good quantum systems. Their Witten indices are
the well known instanton partition functions of 5d $SU(3)$ super-Yang-Mills theory.
We showed for some low $k$'s that the two expressions completely agree, and will also
provide an all order argument below that they should agree.
We can also test the 2d spectrum, i.e. higher order coefficients of
the elliptic genera in the $q$ expansion. This is because the elliptic genera in
certain limits have been computed using the topological string methods
\cite{Haghighat:2014vxa}.

\hspace*{-.6cm}{\bf \underline{One string}:}
The elliptic genus at $k=1$ is given by
\begin{equation}\label{k=1-general}
  Z_{k=1}^{SU(3)}(v,\epsilon_{1,2})=\frac{\eta^2}{\theta_1(\epsilon_{1,2})}
  \sum_{i=1}^3\frac{\eta^4\theta_1(2v_i-4\epsilon_+)\theta_1(v_i)}
  {\prod_{j(\neq i)}\theta_1(v_{ij})\theta_1(2\epsilon_+-v_{ij})
  \theta_1(2\epsilon_++v_j)}
\end{equation}
where $v_{ij}\equiv v_1-v_j$, and we used $v_1+v_2+v_3=0$.
Below, we shall also be interested in the limit $\epsilon_+=0$.
Furthermore, we shall often consider the `genus $0$'
contribution in the topological string language, to compare our results with
the data of \cite{Haghighat:2014vxa}. Namely, the genus $0$ contribution at $k=1$
is $z_1\equiv\lim_{\epsilon_1,\epsilon_2\rightarrow 0}
(-4\pi^2\epsilon_1\epsilon_2Z_{1}^{SU(3)})$. Using
$\theta_1(\epsilon_1)\theta_1(\epsilon_2)\approx 4\pi^2\epsilon_1\epsilon_2\eta^6$,
one obtains the following genus $0$ contribution
\begin{equation}\label{k=1-g=0}
  z_1\equiv-\left[\frac{\theta_1(2v_1)\theta_1(v_1)}
  {\theta_1(v_{12})^2\theta_1(v_{13})^2\theta_1(v_2)\theta_1(v_3)}+
  \frac{\theta_1(2v_2)\theta_1(v_2)}
  {\theta_1(v_{21})^2\theta_1(v_{23})^2\theta_1(v_3)\theta_1(v_1)}+
  \frac{\theta_1(2v_3)\theta_1(v_3)}
  {\theta_1(v_{31})^2\theta_1(v_{32})^2\theta_1(v_1)\theta_1(v_2)}\right]
\end{equation}
at $\epsilon_+=0$.

Let us first study the zero momentum modes in the $q$ expansion.
Here, as emphasized above, we
expect that the Witten index should agree with Nekrasov's $SU(3)$ single instanton
partition function. We have checked that the two expressions at $k=1$ completely agree
with each other. To illustrate how this happens, we explain in detail the two expressions
at $\epsilon_+=0$ in the `genus $0$' limit, for simplicity. Nekrasov's partition function
in this limit is given by \cite{Nekrasov:2002qd}
\begin{equation}\label{nekrasov-k=1}
  Z_{\rm 5d}=\sum_{i=1}^3\frac{1}{\prod_{j(\neq i)}\left[2\sin(\pi v_{ij})\right]^2}
\end{equation}
From our (\ref{k=1-g=0}), using
$\theta_1(z)\approx q^{\frac{1}{8}}\cdot 2\sin(\pi z)$ at small $q$,
one obtains
\begin{equation}\label{k=1,p=0}
  z_1\stackrel{q\rightarrow 0}{\longrightarrow}
  -\frac{1}{16}\left[\frac{\sin(2\pi v_1)\sin(\pi v_1)}
  {\sin^2(\pi v_{12})\sin^2(\pi v_{13})\sin(\pi v_2)\sinh(\pi v_3)}
  +({\rm cyclic})\right]\equiv \tilde{Z}_{\rm 5d}
\end{equation}
up to an overall factor $q^{-\frac{1}{2}}$, where (cyclic) denotes two
more terms obtained by replacing $v_1,v_2,v_3$ by $v_2,v_3,v_1$ and $v_3,v_1,v_2$,
respectively. Here, one can show that
\begin{equation}\label{trigonometric-identity}
  \sin(2\pi v_1)\sin(\pi v_1)=\sin(\pi v_{12})\sin(\pi v_{13})-\sin(\pi v_2)\sin(\pi v_3)\ ,
\end{equation}
etc. on the numerators, so that (\ref{k=1,p=0}) can be written as
\begin{equation}
  \tilde{Z}_{\rm 5d}=Z_{\rm 5d}-\sum_{i=1}^3\frac{1}{\prod_{j(\neq i)}2\sin(\pi v_{ij})
  \cdot 2\sinh(\pi v_j)}\ .
\end{equation}
One can show that the second term is zero at $v_1+v_2+v_3=0$, proving
$\tilde{Z}_{\rm 5d}=Z_{\rm 5d}$.
Of course, we have shown that the two results completely agree at general
$\epsilon_1,\epsilon_2$.

This exercise shows that Nekrasov's $SU(3)$ partition function is
elliptically uplifted in a very intriguing way. Naively, each $\frac{1}{2\sin(\pi z)}$
factor comes from a bosonic mode living on $S^1$. So it is natural to try
to uplift this factor as $\frac{\eta}{\theta_1(z)}$ for $SU(3)$ instanton strings.
However, such an uplift of (\ref{nekrasov-k=1}) would yield the following trial function,
\begin{equation}\label{wrong-elliptic-uplift}
  \sum_{i=1}^3\frac{\eta^2}{\prod_{j(\neq j)}\theta_1(v_{ij})}
  =\frac{\eta^2}{\theta_1(v_{12})\theta_1(v_{13})}+
  \frac{\eta^2}{\theta_1(v_{21})\theta_1(v_{23})}+
  \frac{\eta^2}{\theta_1(v_{31})\theta_1(v_{32})}
\end{equation}
which fails to provide a reasonable expression. To explain why, we first note that
the elliptic genus $Z(\tau,m_a)$ with nonzero flavor chemical potentials
$m_a$ suffers from a conformal anomaly under the modular transformation $\tau\rightarrow-\frac{1}{\tau}$ on $T^2$. This anomaly is given by \cite{Benini:2013nda}
\begin{equation}\label{modular-anomaly}
  Z\left(-\frac{1}{\tau},\frac{m_a}{\tau}\right)\sim
  \exp\left(-\frac{\pi i}{\tau}\mathcal{A}^{ab}m_am_b\right)Z(\tau,m_a)\ ,
\end{equation}
where $\mathcal{A}^{ab}\equiv\sum_{\rm fermions}\gamma_3 Q_a Q_b$ is the
anomaly matrix of the flavor symmetries with charges $Q_a$. The inconsistency of
(\ref{wrong-elliptic-uplift}) is that the three terms of (\ref{wrong-elliptic-uplift})
have different modular factors, inconsistent with (\ref{modular-anomaly}).
(\ref{wrong-elliptic-uplift}) also disagrees with the elliptic genus computed
from topological strings \cite{Haghighat:2014vxa}. On the other hand,
after rewriting (\ref{nekrasov-k=1}) into (\ref{k=1,p=0}) using the trigonometric
identity (\ref{trigonometric-identity}), the three terms of its elliptic uplift
(\ref{k=1-g=0}) transform in the same way under the modular transformation,
being consistent with (\ref{modular-anomaly}).

\begin{table}[t!]
\parbox{0.49\textwidth}
{
$$
\begin{array}{c|c|c|c|c|c|c|c}
  \hline d_1\setminus d_2&0&1&2&3&4&5&6\\
  \hline 0& {\color{red} 1}&{\color{red} 3}&{\color{red} 5}
  &{\color{red} 7}&{\color{red} 9}&{\color{red} 11}&13\\
  \hline 1& {\color{red} 3}&{\color{red} 4}&{\color{red} 8}
  &{\color{red} 12}&{\color{red} 16}&{\color{red} 20}&24\\
  \hline 2& {\color{red} 5}&{\color{red} 8}&{\color{red} 9}
  &{\color{red} 15}&{\color{red} 21}&{\color{red} 27}&33\\
  \hline 3& {\color{red} 7}&{\color{red} 12}&{\color{red} 15}
  &{\color{red} 16}&{\color{red} 24}&{\color{red} 32}&40\\
  \hline 4& {\color{red} 9}&{\color{red} 16}&{\color{red} 21}
  &{\color{red} 24}&{\color{red} 25}&{\color{red} 35}&45\\
  \hline 5& {\color{red} 11}&{\color{red} 20}&{\color{red} 27}
  &{\color{red} 32}&{\color{red} 35}&{\color{red} 36}&48\\
  \hline 6& 13&24&33&40&45&48&49\\
  \hline
\end{array}
$$
\caption{$N^{(1)}_{d_1,d_2,0}$}\label{p=0}
}
\parbox{0.56\textwidth}
{
$$
\begin{array}{c|c|c|c|c|c|c|c}
  \hline d_1\setminus d_2&0&1&2&3&4&5&6\\
  \hline 0& {\color{red} 3}&{\color{red} 4}&{\color{red} 8}
  &{\color{red} 12}&{\color{red} 16}&{\color{red} 20}&24\\
  \hline 1& {\color{red} 4}&{\color{red} 16}&{\color{red} 36}
  &{\color{red} 60}&{\color{red} 84}&{\color{red} 108}&132\\
  \hline 2& {\color{red} 8}&{\color{red} 36}&{\color{red} 56}
  &{\color{red} 96}&{\color{red} 144}&{\color{red} 192}&240\\
  \hline 3& {\color{red} 12}&{\color{red} 60}&{\color{red} 96}
  &{\color{red} 120}&{\color{red} 180}&{\color{red} 252}&324\\
  \hline 4& {\color{red} 16}&{\color{red} 84}&{\color{red} 144}
  &{\color{red} 180}&{\color{red} 208}&{\color{red} 288}&384\\
  \hline 5& {\color{red} 20}&{\color{red} 108}
  &{\color{red} 192}&{\color{red} 252}&{\color{red} 288}
  &{\color{red} 320}&420\\
  \hline 6& 24&132&240&324&384&420&456\\
  \hline
\end{array}
$$
\caption{$N^{(1)}_{d_1,d_2,1}$}\label{p=1}
}
\parbox{0.50\textwidth}
{\small
$$
\begin{array}{c|c|c|c|c|c|c|c}
  \hline d_1\setminus d_2&0&1&2&3&4&5&6\\
  \hline 0& {\color{red} 5}&{\color{red} 8}&{\color{red} 9}
  &{\color{red} 15}&{\color{red} 21}&{\color{red} 27}&33\\
  \hline 1& {\color{red}8}&{\color{red}36}&{\color{red} 56}&{\color{red} 96}
  &{\color{red} 144}&{\color{red} 192}&240\\
  \hline 2& {\color{red} 9}&{\color{red} 56}&{\color{red} 149}&{\color{red} 288}
  &{\color{red} 465}&{\color{red} 651}&837\\
  \hline 3& {\color{red} 15}&{\color{red} 96}&{\color{red} 288}&{\color{red} 456}
  &{\color{red} 735}&{\color{red} 1080}&1440\\
  \hline 4& {\color{red} 21}&{\color{red} 144}&{\color{red} 465}&{\color{red} 735}
  &{\color{red} 954}&{\color{red} 1371}&1890\\
  \hline 5& {\color{red} 27}&{\color{red} 192}&{\color{red} 651}
  &{\color{red} 1080}&{\color{red} 1371}&1632&2187\\
  \hline 6& 33&240&837&1440&1890&2187&2490\\
  \hline
\end{array}
$$
\caption{$N^{(1)}_{d_1,d_2,2}$}\label{p=2}
}
\parbox{0.55\textwidth}
{\small
$$
\begin{array}{c|c|c|c|c|c|c|c}
  \hline d_1\setminus d_2&0&1&2&3&4&5&6\\
  \hline 0& {\color{red} 7}&{\color{red} 12}&{\color{red} 15}&{\color{red} 16}
  &{\color{red} 24}&{\color{red} 32}&40\\
  \hline 1& {\color{red} 12}&{\color{red} 60}&{\color{red} 96}&{\color{red} 120}
  &{\color{red} 180}&{\color{red} 252}&324\\
  \hline 2& {\color{red} 15}&{\color{red} 96}&{\color{red} 288}&{\color{red} 456}
  &{\color{red} 735}&{\color{red} 1080}&1440\\
  \hline 3& {\color{red} 16}&{\color{red} 120}&{\color{red} 456}
  &{\color{red} 1012}&{\color{red} 1788}&2796&3892\\
  \hline 4& {\color{red} 24}&{\color{red} 180}&{\color{red} 735}
  &{\color{red} 1788}&2823&4356&6288\\
  \hline 5& {\color{red} 32}&{\color{red} 252}&{\color{red} 1080}&2796&4356&5760&8052\\
  \hline 6& 40&324&1440&3892&6288&8052&9760\\
  \hline
\end{array}
$$
\caption{$N^{(1)}_{d_1,d_2,3}$}\label{p=3}
}
\end{table}
Now we come back to the full elliptic genus (\ref{k=1-general}), (\ref{k=1-g=0}).
We focus on (\ref{k=1-g=0}) because partial data on it is known from topological
string calculus in the F-/M-theory setting of the 6d SCFT. Namely, we expand
(\ref{k=1-g=0}) in the $SU(3)$ fugacities as
\begin{equation}
  z_1=q^{-\frac{1}{2}}y_{13}
  \sum_{d_1,d_2,d_3=0}^\infty\left(qy_{31}\right)^{d_3} (y_{12})^{d_1}(y_{23})^{d_2}N^{(1)}_{d_1,d_2,d_3}
\end{equation}
where $y_{ij}=e^{2\pi iv_{ij}}$. The coefficients $N_{d_1,d_2,d_3}$ can be
computed in the M-theory dual setting. Namely,
note first that we compactified one direction of the 6d CFT on a circle.
We can T-dualize the F-theory to type IIA setting, in which the momentum conjugate
to $q$ maps to winding string number. Uplifting the system to M-theory, one obtains
an M-theory on exactly the same elliptic CY$_3$ as one had in the F-theory
side, but now having its elliptic fiber as part of the 11d geometry. The IIA
winding string charge (T-dual to momentum) maps to the M2-branes wrapping the $T^2$
fiber. So the BPS states captured by our elliptic genus map to the M2-branes in
M-theory wrapping various 2-cycles of CY$_3$.
From the last viewpoint, the Witten indices for
the single particle BPS states are computed in \cite{Haghighat:2014vxa}
for some $N_{d_1,d_2,d_3}$. We would like to compare these
results with ours. We summarize the results obtained by expanding our (\ref{k=1-g=0})
in Tables \ref{p=0}-\ref{p=3}.
Note that $N_{d_1,d_2,0}$ are simply the coefficients of Nekrasov's $SU(3)$
instanton partition function.
Many of these coefficients are computed in \cite{Haghighat:2014vxa} from
the topological string calculus. The first four tables on p.32 of
\cite{Haghighat:2014vxa} at $k=1$ completely agree with our results.
In our Tables \ref{p=0}-\ref{p=3}, and also other Tables below in this
subsection, red numbers are those shown in \cite{Haghighat:2014vxa}, or
deducible from the data of \cite{Haghighat:2014vxa} using the affine
$SU(3)$ symmetry which permutes $d_1,d_2,d_3$.

\hspace*{-.6cm}{\bf \underline{Two strings}:} At $k=2$,
we sum over the JK-Res at the following poles,
\begin{equation}
  (u_1,u_2)=(v_i-\epsilon_+,v_i-\epsilon_+-\epsilon_{1,2})\ ,\ \
  (v_i-\epsilon_+,v_j-\epsilon_+)\textrm{ with }i<j\ ,
\end{equation}
up to Weyl copies. The residue sum is given by
\begin{eqnarray}
  \hspace*{-1cm}&&\left[\frac{}{}\!\!\right.
  \frac{\eta^4}{\theta_1(\epsilon_1)\theta_1(\epsilon_2)
  \theta_1(2\epsilon_1)\theta_1(\epsilon_2\!-\!\epsilon_1)}\\
  \hspace*{-1cm}&&\times\sum_{i=1}^3\frac{\eta^8
  \theta_1(v_i)\theta_1(v_i-\epsilon_1)
  \theta_1(2v_i-4\epsilon_+-\epsilon_1)\theta_1(2v_i-4\epsilon_+-2\epsilon_1)}
  {\prod_{j(\neq i)}\theta_1(v_{ij})\theta_1(v_{ij}\!-\!2\epsilon_+)\theta_1(v_{ij}\!-\!\epsilon_1)
  \theta_1(v_{ij}\!-\!\epsilon_1\!-\!2\epsilon_+)
  \theta_1(2\epsilon_+\!+\!v_j)\theta_1(2\epsilon_+\!+\!\epsilon_1\!+\!v_j)}
  \left.\frac{}{}\!\!\right]
  +(\epsilon_1\leftrightarrow\epsilon_2)\frac{}{}\nonumber\\
  \hspace*{-1cm}&&+\frac{\eta^4}{\theta_1(\epsilon_{1,2})^2}\!\!\sum_{i<j;k(\neq i,j)}
  \!\!\frac{\eta^8\theta_1(2v_{i,j}-4\epsilon_+)\theta_1(v_{i,j})\theta_1(4\epsilon_++v_k)}
  {\theta_1(v_{i,j}\!-\!v_k)\theta_1(2\epsilon_+\!-\!(v_{i,j}\!-\!v_k))
  \theta_1(\epsilon_{1,2}\!\pm\! v_{ij})
  \theta_1(2\epsilon_+\!+\!v_{i,j})
  \theta_1(2\epsilon_+\!+\!\epsilon_{1,2}\!+\!v_k)\theta_1(2\epsilon_+\!+\!v_k)}
  \nonumber
\end{eqnarray}
where repeated subscripts like $\epsilon_{1,2}$ or
$v_{i,j}$ imply that both factors are included, e.g.
$\theta_1(\epsilon_{1,2})\equiv\theta_1(\epsilon_1)\theta_1(\epsilon_2)$. Also,
$\pm$ signs in the arguments also mean that both $\theta_1$ factors are multiplied.

First of all, we expand this quantity in $q$ and consider the leading
order coefficient at $q^{-1}$. This coefficient is obtained simply by replacing
all $\theta_1(x)$ functions by $2\sin(\pi x)$ functions and discarding all $\eta$
functions. We have shown that this is exactly the same as the 5d $SU(3)$ 2-instanton
partition function.

\begin{table}[t!]
$$
\begin{array}{c|c|c|c|c|c|c|c}
  \hline d_1\setminus d_2&0&1&2&3&4&5&6\\
  \hline 0& {\color{red} 0}&{\color{red} 0}&{\color{red} -6}
  &{\color{red} -32}&{\color{red} -110}&{\color{red} -288}&-644\\
  \hline 1& {\color{red} 0}&{\color{red} 0}&{\color{red} -10}&{\color{red} -70}
  &{\color{red} -270}&{\color{red} -770}&-1820\\
  \hline 2& {\color{red} -6}&{\color{red} -10}&{\color{red} -32}&{\color{red} -126}
  &{\color{red} -456}&{\color{red} -1330}&-3264\\
  \hline 3& {\color{red} -32}&{\color{red} -70}&{\color{red} -126}&{\color{red} -300}
  &{\color{red} -784}&{\color{red} -2052}&-4928\\
  \hline 4& {\color{red} -110}&{\color{red} -270}&{\color{red} -456}&{\color{red} -784}
  &{\color{red} -1584}&{\color{red} -3360}&-7260\\
  \hline 5& {\color{red} -288}&{\color{red} -770}&{\color{red} -1330}
  &{\color{red} -2052}&{\color{red} -3360}&{\color{red} -6076}&-11340\\
  \hline 6& -644&-1820&-3264&-4928&-7260&-11340&-18944\\
  \hline
\end{array}
$$
\caption{$N^{(2)}_{d_1,d_2,0}$}\label{k=2-p=0}
$$
\begin{array}{c|c|c|c|c|c|c|c}
  \hline d_1\setminus d_2&0&1&2&3&4&5&6\\
  \hline 0& {\color{red} 0}&{\color{red} 0}&{\color{red} -10}&{\color{red} -70}
  &{\color{red} -270}&{\color{red} -770}&-1820\\
  \hline 1& {\color{red} 0}&{\color{red} -8}&{\color{red} -60}&{\color{red} -360}
  &{\color{red} -1432}&{\color{red} -4280}&-10548\\
  \hline 2& {\color{red} -10}&{\color{red} -60}&{\color{red} -216}&{\color{red} -850}
  &{\color{red} -3164}&{\color{red} -9720}&-24970\\
  \hline 3& {\color{red} -70}&{\color{red} -360}&{\color{red} -850}&{\color{red} -2176}
  &{\color{red} -6084}&{\color{red} -16960}&-43100\\
  \hline 4& {\color{red} -270}&{\color{red} -1432}&{\color{red} -3164}
  &{\color{red} -6084}&{\color{red} -13000}&{\color{red} -29526}&-67878\\
  \hline 5& {\color{red} -770}&{\color{red} -4280}&{\color{red} -9720}
  &{\color{red} -16960}&{\color{red} -29526}&-55944&-110600\\
  \hline 6& -1820&-10548&-24970&-43100&-67878&-110600&-192080\\
  \hline
\end{array}
$$
\caption{$N^{(2)}_{d_1,d_2,1}$}\label{k=2-p=1}
$$
\begin{array}{c|c|c|c|c|c|c|c}
  \hline d_1\setminus d_2&0&1&2&3&4&5&6\\
  \hline 0& -6&-10&-32&-126&-456&-1330&-3264\\
  \hline 1& -10&-60&-216&-850&-3164&-9720&-24970\\
  \hline 2& -32&-216&-856&-3016&-10656&-33200&-88240\\
  \hline 3& -126&-850&-3016&-8604&-24780&-71232&-188244\\
  \hline 4& -456&-3164&-10656&-24780&-57128&-136944&-330976\\
  \hline 5& -1330&-9720&-33200&-71232&-136944&-274680&-572866\\
  \hline 6& -3264&-24970&-88240&-188244&-330976&-572866&-1041144\\
  \hline
\end{array}
$$
\caption{$N^{(2)}_{d_1,d_2,2}$}\label{k=2-p=2}
\end{table}
We can compare our elliptic genus with the topological string amplitudes.
The `BPS invariants' computed in \cite{Haghighat:2014vxa} captures the single
particle spectrum. So we first consider the single particle partition function
at $k=2$ sector, which is given by
\begin{equation}
  f_2^{SU(3)}\equiv Z_{2}^{SU(3)}-\frac{Z_1^{SU(3)}(\tau,\epsilon_{1,2},v_i)^2
  +Z_1^{SU(3)}(2\tau,2\epsilon_{1,2},2v_i)}{2}\ .
\end{equation}
Again we consider the `genus $0$' part of $f_2^{SU(3)}$,
\begin{equation}
  \lim_{\epsilon_{1,2}\rightarrow 0}\left(-4\pi^2\epsilon_1\epsilon_2f_2^{SU(3)}\right)
  =q^{-1}(y_{13})^2\sum_{d_1,d_2,d_3=0}^\infty
  (qy_{31})^{d_3}(y_{12})^{d_1}(y_{23})^{d_2}N^{(2)}_{d_1,d_2,d_3}\ .
\end{equation}
Some low order coefficients $N^{(2)}_{d_1,d_2,d_3}$ are given by Tables
\ref{k=2-p=0}-\ref{k=2-p=2}, which completely agrees with the result of
\cite{Haghighat:2014vxa} and extends it. Namely, the red numbers of our
Tables \ref{k=2-p=0}, \ref{k=2-p=1} are computed in \cite{Haghighat:2014vxa},
shown in their last table on p.32 and the first table on p.33, perfectly
agreeing with ours.

\begin{table}[t!]
$$
\begin{array}{c|c|c|c|c|c|c|c}
  \hline d_1\setminus d_2&0&1&2&3&4&5&6\\
  \hline 0& {\color{red} 0}&{\color{red} 0}&{\color{red} 0}&{\color{red} 27}
  &{\color{red} 286}&{\color{red} 1651}&6885\\
  \hline 1& {\color{red} 0}&{\color{red} 0}&{\color{red} 0}&{\color{red} 64}
  &{\color{red} 800}&{\color{red} 5184}&23520\\
  \hline 2& {\color{red} 0}&{\color{red} 0}&{\color{red} 25}&{\color{red} 266}
  &{\color{red} 1998}&{\color{red} 11473}&51506\\
  \hline 3& {\color{red} 27}&{\color{red} 64}&{\color{red} 266}&{\color{red} 1332}
  &{\color{red} 6260}&{\color{red} 26880}&104454\\
  \hline 4& {\color{red} 286}&{\color{red} 800}&{\color{red} 1998}
  &{\color{red} 6260}&{\color{red} 21070}&{\color{red} 70362}&226160\\
  \hline 5& {\color{red} 1651}&{\color{red} 5184}&{\color{red} 11473}
  &{\color{red} 26880}&{\color{red} 70362}&{\color{red} 191424}&521430\\
  \hline 6& 6885&23520&51506&104454&226160&521430&1231767\\
  \hline
\end{array}
$$
\caption{$N^{(3)}_{d_1,d_2,0}$}\label{k=3-p=0}
$$
\begin{array}{c|c|c|c|c|c|c|c}
  \hline d_1\setminus d_2&0&1&2&3&4&5&6\\
  \hline 0& 0&0&0&64&800&5184&23520\\
  \hline 1& 0&0&20&448&4752&30932&145288\\
  \hline 2& 0&20&224&2052&15088&88460&412128\\
  \hline 3& 64&448&2052&10200&50008&227880&939488\\
  \hline 4& 800&4752&15088&50008&176688&626800&2150960\\
  \hline 5& 5184&30932&88460&227880&626800&1797320&5185944\\
  \hline 6& 23520&145288&412128&939488&2150960&5185944&12858880\\
  \hline
\end{array}
$$
\caption{$N^{(3)}_{d_1,d_2,1}$}\label{k=3-p=1}
\end{table}
\hspace*{-.6cm}{\bf \underline{Three strings}:}
We finally report the results at $k=3$. Again taking into account
the $\frac{1}{3!}$ Weyl factor, there are $3\times 3+6\times 2+1=22$ distinct poles
at
\begin{eqnarray}
  (u_1,u_2,u_3)&:&(v_i-\epsilon_+,v_i-\epsilon_+-\epsilon_{1,2},
  v_i-\epsilon_+-2\epsilon_{1,2})\ ,\nonumber\\
  &&(v_i-\epsilon_+,v_i-\epsilon_+-\epsilon_1,v_i-\epsilon_+-\epsilon_2)\ ,\nonumber\\
  &&(v_i-\epsilon_+,v_i-\epsilon_+-\epsilon_{1,2},v_j-\epsilon_+)\ \ \
  \textrm{with } i\neq j\ ,\nonumber\\
  &&(v_1-\epsilon_+,v_2-\epsilon_+,v_3-\epsilon_+)\ .
\end{eqnarray}
The residue sum from the $6$ poles on the first line
is given by
\begin{eqnarray}\label{k=3-1}
  \hspace*{-1cm}&&
  \frac{\eta^6}{\theta_1(\epsilon_1)\theta_1(2\epsilon_1)\theta_1(3\epsilon_1)
  \theta_1(\epsilon_{2})
  \theta_1(\epsilon_2\!-\!\epsilon_1)\theta_1(\epsilon_2\!-\!2\epsilon_1)}\\
  \hspace*{-1cm}&&\times\sum_{i=1}^3
  \frac{\eta^{12}}
  {\prod_{j(\neq i)}\theta_1(v_{ij})\theta_1(v_{ij}-2\epsilon_+)
  \theta_1(v_{ij}-\epsilon_1)\theta_1(v_{ij}-2\epsilon_+-\epsilon_1)
  \theta_1(v_{ij}-2\epsilon_1)\theta_1(v_{ij}-2\epsilon_+-2\epsilon_1)}
  \nonumber\\
  \hspace*{-1cm}&&\hspace{1cm}\times
  \frac{\theta_1(v_i)\theta_1(v_i\!-\!\epsilon_1)\theta_1(v_i\!-\!2\epsilon_1)
  \cdot\theta_1(2v_i\!-\!4\epsilon_+\!-\!2\epsilon_1)
  \theta_1(2v_i\!-\!4\epsilon_+\!-\!3\epsilon_1)
  \theta_1(2v_i\!-\!4\epsilon_+\!-\!4\epsilon_1)}
  {\prod_{j(\neq i)}\theta_1(2\epsilon_++v_j)
  \theta_1(2\epsilon_++\epsilon_1+v_j)\theta_1(2\epsilon_++2\epsilon_1+v_j)}
  +(\epsilon_1\leftrightarrow\epsilon_2)\ .\nonumber
\end{eqnarray}
The first two lines are elliptic uplifts of the residue factors in
Nekrasov's $SU(3)$ partition function.
The residue sum from the $3$ poles on the second line is given by
\begin{eqnarray}\label{k=3-2}
  \hspace*{-1.5cm}&&\frac{\eta^6}{\theta_1(\epsilon_{1,2})^2
  \theta_1(2\epsilon_1\!-\!\epsilon_2)\theta_1(2\epsilon_2\!-\!\epsilon_1)}\sum_{i=1}^3
  \frac{\eta^{12}}{\prod_{j(\neq i)}\theta_1(v_{ij})\theta_1(v_{ij}\!-\!2\epsilon_+)
  \theta_1(v_{ij}\!-\!\epsilon_{1,2})\theta_1(v_{ij}\!-\!2\epsilon_+\!-\!\epsilon_{1,2})}
  \nonumber\\
  \hspace*{-1.5cm}&&\hspace{4cm}\times\frac{\theta_1(v_i)\theta_1(v_i-\epsilon_{1,2})
  \theta_1(2v_i-4\epsilon_+-2\epsilon_{1,2})\theta_1(2v_i-4\epsilon_+-\epsilon_1-\epsilon_2)}
  {\prod_{j(\neq i)}\theta_1(2\epsilon_++v_j)\theta_1(2\epsilon_++\epsilon_{1,2}+v_j)}\ .
\end{eqnarray}
Again, the first line is the elliptic uplift of the usual $SU(3)$ partition function.
The residue sum from the $18$ poles on the third line is given as
\begin{eqnarray}\label{k=3-3}
  \hspace*{-1cm}&&\frac{\eta^6}{\theta_1(\epsilon_{1,2})^2\theta_1(2\epsilon_1)
  \theta_1(\epsilon_2\!-\!\epsilon_1)}
  \sum_{i\neq j}\frac{\eta^{6}}
  {\theta_1(v_{ij})\theta_1(v_{ij}\!-\!2\epsilon_+)
  \theta_1(v_{ij}\!+\!\epsilon_1)\theta_1(v_{ij}\!-\!\epsilon_2)
  \theta_1(v_{ij}\!+\!\epsilon_2\!-\!\epsilon_1)\theta_1(v_{ij}\!-\!2\epsilon_1)}
  \nonumber\\
  \hspace*{-1cm}&&\times\frac{\eta^6}{\theta_1(v_{ik})\theta_1(v_{ik}-2\epsilon_+)
  \theta_1(v_{ik}-\epsilon_1)\theta_1(v_{ik}-2\epsilon_+-\epsilon_1)
  \theta_1(v_{jk})\theta_1(v_{jk}-2\epsilon_+)}\\
  \hspace*{-1cm}&&\times\frac{\theta_1(v_i)\theta_1(v_i\!-\!\epsilon_1)\theta_1(v_j)\cdot
  \theta_1(2v_i\!-\!4\epsilon_+\!-\!\epsilon_1)\theta_1(2v_i\!-\!4\epsilon_+\!-\!2\epsilon_1)
  \theta_1(2v_j\!-\!4\epsilon_+)\cdot\theta_1(v_k\!+\!4\epsilon_+\!+\!\epsilon_1)}
  {\theta_1(2\epsilon_+\!+\!v_i)\theta_1(2\epsilon_+\!+\!v_j)
  \theta_1(2\epsilon_+\!+\!\epsilon_1\!+\!v_j)\theta_1(v_k\!+\!2\epsilon_+)
  \theta_1(v_k\!+\!2\epsilon_+\!+\!\epsilon_{1,2})\theta_1(v_k\!+\!2\epsilon_+\!+\!2\epsilon_1)}
  +(\epsilon_1\leftrightarrow\epsilon_2)\nonumber
\end{eqnarray}
where $k$ is defined by $k\neq i,j$ for given $i,j$. The first two lines
are elliptic uplift of the usual residue. The final
residue from the fourth line is given by
\begin{equation}\label{k=3-4}
  \frac{\eta^6}{\theta_1(\epsilon_{1,2})^3}
  \prod_{i\neq j}\frac{\eta^{12}}{\theta_1(\epsilon_{1,2}+v_{ij})}\cdot\prod_{i=1}^3
  \frac{\theta_1(v_i)\theta_1(2v_i-4\epsilon_+)\cdot\theta_1(v_i+4\epsilon_+)}
  {\theta_1(v_i+2\epsilon_+)\theta_1(v_i+2\epsilon_++\epsilon_{1,2})}\ .
\end{equation}
The elliptic genus $Z_3$ is the sum of (\ref{k=3-1}), (\ref{k=3-2}), (\ref{k=3-3}),
(\ref{k=3-4}).

At the lowest $q^{-\frac{3}{2}}$ order in $q$ expansion,
we reproduce the $SU(3)$ $3$-instanton partition function. We checked this using
a computer.

We also consider the single particle index $f_3$, defined by
\begin{eqnarray}
  f_3&=&Z_3(\tau,\epsilon_{1,2},v_i)-Z_1(\tau,\epsilon_{1,2},v_i)
  \left[Z_2(\tau,\epsilon_{1,2},v_i)-\frac{Z_1(\tau,\epsilon_{1,2},v_i)^2
  +Z_1(2\tau,2\epsilon_{1,2}, 2v_i)}{2}\right]\nonumber\\
  &&-\frac{Z_1(\tau,\epsilon_{1,2},v_i)^3+3Z_1(\tau,\epsilon_{1,2},v_i)
  Z_1(2\tau,2\epsilon_{1,2},2v_i)+2Z_1(3\tau,3\epsilon_{1,2},3v_i)}{6}\ .
\end{eqnarray}
Again we study the genus $0$ limit:
\begin{equation}
  \lim_{\epsilon_{1,2}\rightarrow 0}\left(-4\pi^2\epsilon_1\epsilon_2f_3\right)
  =q^{-\frac{3}{2}}(y_{13})^3\sum_{d_1,d_2,d_3=0}^\infty
  (qy_{31})^{d_3}(y_{12})^{d_1}(y_{23})^{d_2}N^{(3)}_{d_1,d_2,d_3}\ .
\end{equation}
Some $N^{(3)}_{d_1,d_2,d_3}$'s are shown in Tables
\ref{k=3-p=0}, \ref{k=3-p=1}. Especially, the red numbers in Table \ref{k=3-p=0}
are computed in \cite{Haghighat:2014vxa} from the topological string calculus,
which completely agree with ours.

\hspace*{-.6cm}{\bf \underline{1 dimensional limit}:} For $k=1,2,3$, we have
illustrated or reported that the 1d reductions of our indices completely agree
with the $SU(3)$ instanton partition functions. We have also checked the agreements
of the 1d limits with $SU(3)$ instanton partition functions at $k\leq 5$.

We can also make a more general comparison of the two indices.
Note that we have collected considerable evidence that $U(1)_\phi$
decouples with our IR QFT on the first branch, and that the second branch does not
contribute to the BPS sector captured by the index. This implies that, even after
we turn on the chemical potential $m$ for $U(1)_\phi$ in (\ref{SU(3)-contour}),
$Z_k^{SU(3)}$ will not depend on $m$. To support the general arguments of
section 3.2, we have also checked it concretely at some low
$k$'s, by expanding in a few low orders in $q$.
Now converting the logic, we \textit{assume} that the index
(\ref{SU(3)-contour}) does not see $m$, and show using this fact that the 1d
limit of (\ref{SU(3)-contour}) is identical to the $SU(3)$ Nekrasov partition
function.

To show this, we first write the 1d limit of (\ref{SU(3)-contour}),
with the chemical potential $m$ turned on:
\begin{eqnarray}
  \hspace*{-.5cm}Z_k^{SU(3),{\rm 1d}}&=&\frac{(-1)^{\frac{k^2-k}{2}}}{k!}\oint \prod_{I=1}^k\left(2\pi du_I\right)\cdot
  \frac{{\rm sn}(2\epsilon_+)^k}{{\rm sn}(\epsilon_{1,2})^k}
  \prod_{I=1}^k\frac{{\rm sn}(2u_I\!-\!m){\rm sn}(2\epsilon_+\!+\!m\!-2u_I)
  {\rm sn}(u_I+\epsilon_++m)}{{\rm sn}(\epsilon_+\pm(u_I-v_{1,2,3}))
  {\rm sn}(\epsilon_++m-u_I-v_{1,2,3})}\nonumber\\
  &&
  \prod_{I<J}\frac{{\rm sn}(\pm u_{IJ}){\rm sn}(2\epsilon_+\!\mp \!u_{IJ})
  \cdot{\rm sn}(u_I\!+\!u_J\!-\!m){\rm sn}(2\epsilon_+\!+\!m\!-\!(u_I\!+\!u_J))}
  {{\rm sn}(\epsilon_{1,2}\pm u_{IJ})
  {\rm sn}(\epsilon_{1,2}+m-(u_I+u_J))}\ ,
\end{eqnarray}
where we replaced all $\frac{\theta_1(z)}{\eta}$ by $2\sin(\pi z)\equiv {\rm sn}(z)$.
Now we use the fact that this expression is independent of $m$ after contour integral.
Using this, we can take $m\rightarrow-i\infty$ and expand the integrand in $e^{-\pi i m}$
first, keeping the leading $\mathcal{O}(e^{-\pi im})^0$ term only.
For each ${\rm sn}$ factor containing $m$, one expands
${\rm sn}(m+z)=-ie^{\pi i(m+z)}+\cdots$ and ignore the subleading terms $\cdots$.
This way, one obtains
\begin{eqnarray}
  \hspace*{-.5cm}Z_k^{SU(3),{\rm 1d}}&=&\frac{(-1)^{\frac{k^2-k}{2}}}{(2\pi i)^kk!}\oint \prod_{I=1}^k\left(2\pi du_I\right)\cdot
  \frac{{\rm sn}(2\epsilon_+)^k}{{\rm sn}(\epsilon_{1,2})^k}
  \prod_{I=1}^k\frac{-e^{\pi i(m-2u_I)}e^{\pi i(m+2\epsilon_+-2u_I)}
  e^{\pi i(m+\epsilon_++u_I)}}
  {{\rm sn}(\epsilon_+\pm(u_I-v_{1,2,3}))
  e^{\pi i(m+\epsilon_+-u_I-v_{1,2,3})}}\nonumber\\
  &&\prod_{I<J}\frac{{\rm sn}(\pm u_{IJ}){\rm sn}(2\epsilon_+\!\mp \!u_{IJ})
  \cdot(-1)e^{\pi i(m-u_I-u_J)}e^{\pi i(m+2\epsilon_+-u_I-u_J)}}
  {{\rm sn}(\epsilon_{1,2}\pm u_{IJ})e^{\pi i(m+\epsilon_{1,2}-u_I-u_J)}}\\
  &=&\frac{(-1)^k}{(2\pi i)^kk!}\oint\prod_{I=1}^k\left(2\pi du_I\right)\cdot
  \frac{{\rm sn}(2\epsilon_+)^k}{{\rm sn}(\epsilon_{1,2})^k}
  \prod_{I=1}^k\frac{1}{{\rm sn}(\epsilon_+\pm(u_I-v_{1,2,3}))}\cdot
  \prod_{I<J}\frac{{\rm sn}(\pm u_{IJ}){\rm sn}(2\epsilon_+\!\mp \!u_{IJ})}
  {{\rm sn}(\epsilon_{1,2}\pm u_{IJ})}\ .\nonumber
\end{eqnarray}
The last expression is the contour integral for Nekrasov's $SU(3)$
instanton partition function \cite{Nekrasov:2002qd}. The overall sign $(-1)^k$
can be obtained by starting from the well-established correct
normalization summarized by eqn.(3.3) in \cite{Hwang:2014uwa}, and changing all
$2\sinh\frac{z_{\rm there}}{2}$ functions there by $2i\sin(\pi z_{\rm here})$
functions here in our convention, with the chemical potentials related by
$z_{\rm there}=2\pi iz_{\rm here}$. So we exactly prove, assuming that
$U(1)_\phi$ decouples in the index, that the Witten indices of our 1d gauge
theories are precisely the $SU(3)$ instanton partition functions.

\section{Strings of the $(E_6,E_6)$ conformal matter}\label{sec:E6xE6}

One application of our $SU(3)$ strings' gauge theories is that
one can now engineer self-dual string gauge theories for many other 6d CFTs
containing the $n=3$ atom. In this section, we study the self-dual strings of
the $(E_6,E_6)$ conformal matter \cite{DelZotto:2014hpa}, by constructing
the 2d gauge theories for them.

We start by briefly reviewing the conformal matter. The conformal matters are the 6d CFTs
of a single M5-brane probing an ADE singularity $\mathbb{C}^2/\Gamma_{ADE}\times\mathbb{R}$
in M-theory \cite{DelZotto:2014hpa}. The theories have $G\times G$ global symmetries where
$G$ is the algebra associated to the singularity.
When the singularity is of A-type, the theory is simply that of a standard bi-fundamental hypermultiplet of $G\times G$ symmetry. For the other types of singularity, the single
M5-brane breaks into several fractional M5-branes along the $\mathbb{R}$ direction, which
represent the tensor branch.
When $G=E_6$, the M5-brane splits into 4 fractional M5-branes. The IR 6d theory is a strongly interacting CFT with three tensor multiplets, whose three scalars parametrize the
relative distances between the four fractional M5-branes. The same CFT can be obtained by an
F-theory configuration of three curves of $1,3,1$ types, linearly intersecting in the given
order. So the middle $-3$ curve has the $SU(3)$ gauge symmetry and the left and the right
$-1$ curves support a $E_6$ global symmetry respectively.
This 6d CFT is called $(E_6,E_6)$ conformal matter theory.

In this section we shall study the 2d gauge theories for the self-dual strings of
the $(E_6,E_6)$ conformal matter theory. The 2d theories can be constructed by gluing
the $SU(3)$ string theory constructed in the previous sections with a pair of the
2d self-dual string theories in the E-string theory.

The 2d gauge theory describing the E-strings is engineered in \cite{Kim:2014dza}. It is an $\mathcal{N}=(0,4)$ supersymmetric gauge theory with $O(k)$ gauge group for $k$ strings.
This theory consists of the following $\mathcal{N}=(0,4)$ field contents:
\begin{eqnarray}\label{eq:E-string-mattercontent}
    (A_\mu,\lambda_{-\dot\alpha A}) \ &&: \ O(k) \ {\rm vector \ multiplet } \nonumber \\
    (a_{\alpha\dot\beta},\Psi_{+\alpha A}) \ &&: \ O(k) \ {\rm symmetric \ hypermultiplet} \nonumber \\
    \Xi_l \ &&: \ O(k)\times SO(16) \ {\rm Fermi \ multiplet} \ .
\end{eqnarray}
At low energy, the classical $SO(16)$ global symmetry of the Fermi multiplets $\Xi_l$
enhances to $E_8$ symmetry.
To construct the 2d gauge theories on the strings of the $(E_6,E_6)$ conformal matter theory, we will couple two copies of this theory to the $SU(3)$ minimal string theory. Thus the 2d gauge theory will become a $U(k_1)\times O(k_2)\times O(k_3)$ quiver gauge theory with appropriate interactions. Here $k_1$ is the number of $SU(3)$ strings and $k_2,k_3$ are
the number of E-strings on the left and right sides of the linear quiver, respectively.

Let us first focus on the couplings between the $U(k_1)$ gauge theory and the $O(k_2)$
theory at $k_3=0$, i.e. $k_1$ SU(3) strings and the $k_2$ E-strings.
This can also be understand as the strings of a 6d SCFT with rank $2$ tensor branch,
called `$13$' obtained by attaching the $SU(3)$ theory with an E-string theory
\cite{Heckman:2015bfa}. The two gauge nodes for $U(k_1)$ and $O(k_2)$ are connected by
bi-fundamental fields. In $\mathcal{N}=(0,2)$ SUSY notation, bi-fundamental fields are
organized as
\begin{align}
    \Phi + \tilde\Phi\ &: \ {\rm chiral\ multiplets \ in} \ ({\bf k}_1,{\bf k}_2)_{J=-1/2} \ + \ ({\bf \bar{k}}_1,{\bf k}_2)_{J=-1/2} \cr
    \Gamma + \tilde\Gamma\ &: \ {\rm Fermi\ multiplets \ in} \ ({\bf k}_1,{\bf k}_2)_{J_l=-1/2} \ + \ ({\bf \bar{k}}_1,{\bf k}_2)_{J_l=-1/2} \ ,
\end{align}
where the subscripts denote the charges under $U(1)_J$ and $U(1)_l \subset SU(2)_l$.
This is a guess, which is well inspired from 2d self-dual string quivers with D-brane origins
\cite{Haghighat:2013gba,Haghighat:2013tka,Haghighat:2014vxa}.
Then we choose 3 complex Fermi multiplets $\Xi_{i=1,2,3}$ among the $SO(16)$ fundamental Fermi multiplets and couple them to the bi-fundamental fields and the other fields in the $U(k_1)$ gauge node. We introduce the following superpotentials:
\begin{align}
    &J_{\Xi_i} = \tilde{q}^i\Phi \ , \quad E_{\Xi_i} = \tilde\Phi q_i \ , \cr
    &J_{\Gamma} = \tilde{a} \Phi - \Phi \tilde{a}' \ , \quad E_{\Gamma} = \tilde\Phi a - a'\tilde\Phi \ , \quad
    J_{\tilde\Gamma} =\tilde{a}'\tilde\Phi- \tilde\Phi \tilde{a}  \ , \quad E_{\tilde\Gamma} =  a \Phi - \Phi a' \ .
\end{align}
These superpotentials are not enough to satisfy the supersymmetry constraint
$\sum_i J_i E_i=0$.  To satisfy the SUSY constraint, we also need to modify the superpotentials for the adjoint Fermi multiplets $\lambda$ and $\lambda'$ in the $\mathcal{N}=(0,4)$ vectormultiplets as
\begin{equation}
    E_\lambda = -\Phi\tilde\Phi \ , \qquad E_{\lambda'} =
    (\tilde\Phi \Phi)_{A} \ .
\end{equation}
The superpotentials naturally break the $E_8$ global symmetry of the $O(k_2)$ theory into a subgroup $SO(10)\times U(1)\times SU(3)'$, and identify the $SU(3)'$ part with the $SU(3)$ global symmetry in the $U(k_1)$ gauge theory. Also, the $U(1)$ part rotating the three fermions $\Xi_{1,2,3}$ with charge $+1$ is identified with the $U(1)_g$ global symmetry in the $U(k_1)$ theory. The global symmetry of the combined system is therefore $SO(10)\times U(1)\times SU(3)$. This has to do with the fact that a $-1$ curve and $-3$ curve are
joined by taking the subgroup $SU(3)\times E_6\subset E_8$ of the E-string theory, and
gauging the $SU(3)$ in 6d. We naturally interpret
$SO(10)\times U(1)\subset SO(10)\times U(1)\times SU(3)$, which is a subgroup of $E_6$.
We expect this to enhance to $E_6$ at low
energy.\footnote{There is actually  one more anomaly free $U(1)$ flavor symmetry rotating all the bi-fundamental matters and the 3 Fermi multiplets with charges $Q[\Phi]=Q[\tilde{\Gamma}]=1, Q[\tilde\Phi] = Q[\Gamma] = Q[\Xi_i] = -1$. We find that the elliptic genus shows $E_6$ symmetry enhancement only when we turn off the mass parameter for this $U(1)$ symmetry. We will also see in the next section that the anomaly of the 2d gauge theory matches those for the self-dual string theory only if the presence of this $U(1)$ is
dismissed. So we will ignore this $U(1)$ symmetry in the following discussions, but we have
no clear explanation on why this is the case at the moment.}

To engineer the 2d gauge theories for the $(E_6,E_6)$ conformal matter strings,
we prepare one more $O(k_3)$ theory for the right E-strings and glue it to the $U(k_1)$
gauge node of this system in the similar manner. So we connect the $U(k_1)$ and $O(k_3)$
gauge theories similarly by the bi-fundamental fields and superpotentials described above.
Then the final theory becomes a $U(k_1)\times O(k_2)\times O(k_3)$ gauge theory with
$SO(10)^2\times U(1)\times SU(3)$ global symmetry.
We conjecture that this theory will flow in the infrared to the 2d CFT with the enhanced global symmetry $E_6\times E_6\times SU(3)$, and it will become the theory on self-dual strings in the  $(E_6,E_6)$ conformal matter theory. We will support this claim in this
section by computing the elliptic genus and confirming the symmetry enhancement.
Also in the next section, we show that the anomaly polynomials of the 2d gauge theories
completely agree with the anomalies on the self-dual strings, computed from
the anomaly inflow of the 6d bulk theory.

Note that, in this construction, two $E_8$ global symmetries of E-string theories
are broken to two copies of $SO(10)\times U(1)\times SU(3)'$ symmetry and both $U(1)$ symmetries are identified with the $U(1)_g$ symmetry of the $U(k_1)$ gauge theory via the superpotentials. Therefore, although we expect the global symmetry is enhanced to
$E_6\times E_6\times SU(3)$ in  the IR CFT, this UV gauge construction can see only the diagonal $U(1)$ component in $E_6\times E_6$.

Now we compute the elliptic genus of the 2d theory. The elliptic genus can be written as
\begin{align}
    Z^{(E_6,E_6)} =& \oint Z_{\rm 1-loop}^{SU(3)}(u,v)
    Z_{\rm 1-loop}^{\textrm{E-string}}(z,m,v) Z_{\rm 1-loop}^{\textrm{E-string}}(w,m',v) \\
    & \times
    \prod_{\rho_1\in{\rm fund}_1} \prod_{\rho_2\in{\rm fund}_2}\frac{\theta_1(-\epsilon_- \pm (\rho_1(u)-\rho_2(z))}{\theta_1(-\epsilon_+ \pm (\rho_1(u)-\rho_2(z))}
    \prod_{\rho_3\in{\rm fund}_3}\frac{\theta_1(-\epsilon_- \pm (\rho_1(u)-\rho_3(w))}{\theta_1(-\epsilon_+ \pm (\rho_1(u)-\rho_3(w))}  \nonumber \ ,
\end{align}
where the three integrands in the first line are the 1-loop contributions from the $SU(3)$ strings (explained in section 3.4), two E-string theories (explained in \cite{Kim:2013nva}),
and the second line comes from the 1-loop contributions of the bi-fundamental multiplets. Here, ${\rm fund}_{1,2,3}$ denote the fundamental representations of the $U(k_1),O(k_2),O(k_3)$ gauge groups respectively. $v=(v_1,v_2,v_3)$ is the
chemical potential for $SU(3)\times U(1)_g$. $u$ is the $U(k_1)$
holonomy on $T^2$, $z$ is that for $O(k_2)$, and $w$ is that for $O(k_3)$.
$m=(m_1,\cdots,m_5)$ are the chemical potentials for $SO(10)\times U(1)$ for the first
E-strings, where $\sum_{l=1}^5m_l$ is that for the $U(1)\sim U(1)_g$. Similarly,
$m^\prime=(m_1^\prime,\cdots,m_5^\prime)$ are chemical potentials for the second
$SO(10)\times U(1)_g$, where $U(1)_g$ is locked with the $U(1)_g$ for the first E-strings.
Thus, our chemical potentials are constrained by $\sum_{l=1}^lm_l=\sum_{l=1}^5m_l^\prime$,
which reflects the ignorance of our UV gauge theories on an $U(1)$ IR flavor symmetry.

Let us first consider the case with $(k_1,k_2,k_3) = (1,1,0)$. The elliptic genus
is given by
\begin{eqnarray}
    Z_{(1,1,0)} = &&\frac{\eta^4}{\theta_1(\epsilon_{1,2})^2} \oint du
    \frac{\eta^5\theta_1(2\epsilon_+)}{\prod_{i=1}^3\theta_1(\epsilon_+ \pm (u-v_i))} \cdot
    \frac{\theta_1(2u)\theta_1(2\epsilon_+-2u)\theta_1(u+\epsilon_+-v_1-v_2-v_3)}{\prod_{i=1}^3\theta_1(\epsilon_+-u-v_i)} \nonumber \\
    && \times \sum_{a=1}^4  \frac{\prod_{i=1}^3\theta_a(v_i)\prod_{l=1}^5\theta_a(m_l)}{2\eta^8} \cdot \frac{\theta_a(\epsilon_- \pm u)}{\theta_a(-\epsilon_+ \pm u)} \ .
\end{eqnarray}
The contour has JK poles at $\epsilon+u-v_i = 0$ and $-\epsilon_+ + u + p_a = 0$ where $p_a = 0,\frac{1}{2},\frac{\tau}{2},\frac{1+\tau}{2}$, but only the former three poles yield nonzero
residues. Thus the contour integral is evaluated as
\begin{eqnarray}
    Z_{(1,1,0)} =&& \frac{\eta^4}{\theta_1(\epsilon_{1,2})^2} \sum_{i=1}^3\frac{\eta^4\theta_1(v-v_i)\theta_1(2v_i-4\epsilon_+)}{\prod_{j\neq i}\theta_1(v_{ij})\theta_1(2\epsilon_+ -v_{ij})\theta_1(2\epsilon_+-v+v_j)}  \nonumber \\
    && \times \sum_{a=1}^4 \frac{\prod_{j=1}^3\theta_a(v_j)\prod_{l=1}^5\theta_a(m_l)}{2\eta^8} \frac{\theta_a(\epsilon_{1,2}-v_i)}{\theta_a(v_i-2\epsilon_+)\theta_a(v_i)} \ .
\end{eqnarray}
$Z_{(1,1,0)}$ can be expanded in the fugacities $q=e^{2\pi i \tau}$ and $t=e^{2\pi i\epsilon_+}$. Then the coefficients of the series organize themselves into characters of
the $SU(3)\times E_6$ flavor symmetry as
\begin{eqnarray}
    Z_{(1,1,0)} =&& \frac{q^{-2/3}}{\theta_1(\epsilon_{1,2})^2} \bigg[ t^2 + \chi^{SU(3)}_{\bf8}t^4 +\mathcal{O}(t^5)+ q\cdot \Big(t^{-1}\chi^{SU(2)}_{\bf 2}(u) -1 - t \chi^{SU(2)}_{\bf 2}(u) + \nonumber \\
    && +  t^2(1+ \chi^{SU(2)}_{\bf 3}(u) +\chi^{SU(3)}_{\bf 8} + \chi^{E_6}_{\bf 78})
    + t^3\chi^{SU(2)}_{\bf 2}(u)(\chi^{SU(3)}_{\bf \bar{3}}\chi^{E_6}_{\bf 27} + \chi^{SU(3)}_{\bf 3}\chi^{E_6}_{\bf \overline{27}}) \nonumber \\
    && + t^4 (1+ \chi^{SU(2)}_{\bf 3}(u)\chi^{SU(3)}_{\bf 8}+\chi^{SU(3)}_{\bf 27} +3\chi^{SU(3)}_{\bf 8} + \chi^{SU(3)}_{\bf 10} + \chi^{SU(3)}_{\bf \overline{10}}   + \chi^{SU(3)}_{\bf 8}\chi^{E_6}_{\bf 78} \nonumber \\
    && \qquad + \chi^{SU(3)}_{\bf 6}\chi^{E_6}_{\bf 27} + \chi^{SU(3)}_{\bf \bar{6}}\chi^{E_6}_{\bf \overline{27}}) +\mathcal{O}(t^5) \Big)   \nonumber \\
    && +q^2 \Big( t^{-3}\chi^{SU(2)}_{\bf 2}(u)\chi^{SU(3)}_{\bf 8} + t^{-2}(\chi^{SU(3)}_{\bf3}\chi^{E_6}_{\bf \overline{27}} + \chi^{SU(3)}_{\bf \bar{3}}\chi^{E_6}_{\bf 27})  \nonumber \\
    && + t^{-1} \chi^{SU(2)}_{\bf2}(u)(2+\chi^{E_6}_{\bf 78}+ \chi^{SU(3)}_{\bf 8})-(1+\chi^{SU(2)}_{\bf 3}(u) +\chi^{E_6}_{\bf 78} + \chi^{SU(3)}_{\bf 8})  \nonumber \\
    && - t \, \chi^{SU(2)}_{\bf2}(u)(2+\chi^{E_6}_{\bf 78}+ \chi^{SU(3)}_{\bf 8})+t^2\left(5+\chi^{SU(2)}_{\bf 3}(u)(2+\chi^{SU(3)}_{\bf 8}+\chi^{E_6}_{\bf78}) \right. \nonumber \\
    && \left.+ 3\chi^{SU(3)}_{\bf8}+\chi^{SU(3)}_{\bf27}+(2+\chi^{SU(3)}_{\bf8})\chi^{E_6}_{\bf27}+ \chi^{E_6}_{\bf 650}\right) +\mathcal{O}(t^3)\Big) + \mathcal{O}(q^3) \bigg]
\end{eqnarray}
where $u=e^{2\pi i \epsilon_-}$. $\chi_{\bf r}^{SU(2)}(u)$ is a $SU(2)$ character of ${\bf r}$ dimensional representation with its fugacity $u$. $\chi^{SU(3)}_{\bf r}$ is a $SU(3)$ character of ${\bf r}$ dimensional representation written in terms of $SU(3)$ fugacities $e^{2\pi i v_i}$, now restricted to be traceless, $\sum_iv_i=0$. The trace part
$v=\sum_{i=1}^3v_i$ for $U(1)_g$ is combined with the $SO(10)$ fugacities and sent
to the $E_6$ characters. $\chi^{E_6}_{\bf r}$ is a $E_6$ character of ${\bf r}$ dimensional representation with fugacities $e^{2\pi im_l}$ and $e^v$. Therefore, these results support
the $SO(10)\times U(1)_g\rightarrow E_6$ symmetry enhancement of the
2d gauge theory in IR.

Next, we turn to the case with $(k_1,k_2,k_3)=(1,1,1)$. The elliptic genus is given by
\begin{eqnarray}
    \hspace{-1cm}Z_{(1,1,1)} = \!\!\!\!\!\!&&\frac{\eta^6}{\theta_1(\epsilon_{1,2})^3}  \oint du
    \frac{\eta^5\theta_1(2\epsilon_+)}{\prod_{i=1}^3\theta_1(\epsilon_+ \pm (u-v_i))} \cdot
    \frac{\theta_1(2u)\theta_1(2\epsilon_+-2u)\theta_1(u+\epsilon_+-v_1-v_2-v_3)}{\prod_{i=1}^3\theta_1(\epsilon_+-u-v_i)}  \quad \\
   \!\!\!\!\!\! && \times \sum_{a=1}^4  \frac{\prod_{i=1}^3\theta_a(v_i)\prod_{l=1}^5\theta_a(m_l)}{2\eta^8}  \frac{\theta_a(\epsilon_- \pm u)}{\theta_a(-\epsilon_+ \pm u)} \cdot
    \sum_{b=1}^4  \frac{\prod_{i=1}^3\theta_b(v_i)\prod_{l=1}^5\theta_b(m'_l)}{2\eta^8}  \frac{\theta_b(\epsilon_- \pm u)}{\theta_b(-\epsilon_+ \pm u)} \ . \nonumber
\end{eqnarray}
Nonzero residues again come only from the poles at $\epsilon_++u-v_i = 0$ only.
The result takes the following form
\begin{eqnarray}
    Z_{(1,1,1)} =&& \frac{\eta^6}{\theta_1(\epsilon_{1,2})^3} \sum_{i=1}^3\frac{\eta^4\theta_1(v-v_i)\theta_1(2v_i-4\epsilon_+)}{\prod_{j\neq i}\theta_1(v_{ij})\theta_1(2\epsilon_+ -v_{ij})\theta_1(2\epsilon_+-v+v_j)}   \\
    && \times \sum_{a=1}^4 \frac{\prod_{j\neq i}\theta_a(v_j)\prod_{l=1}^5\theta_a(m_l)}{2\eta^8} \frac{\theta_a(\epsilon_{1,2}-v_i)}{\theta_a(v_i-2\epsilon_+)} \cdot \sum_{b=1}^4 \frac{\prod_{j\neq i}^3\theta_b(v_j)\prod_{l=1}^5\theta_b(m_l)}{2\eta^8} \frac{\theta_b(\epsilon_{1,2}-v_i)}{\theta_b(v_i-2\epsilon_+)}\ .\nonumber
\end{eqnarray}
The series expansion by $q$ and $t$ fugacities can be written in terms of the characters of the flavor symmetry as
\begin{eqnarray}
    Z_{(1,1,1)} =&& \frac{q^{-1}}{\theta_1(\epsilon_{1,2})^3}\bigg[ Z^{q=0}_{(1,0,0)} +q \Big( -t^{-2} + t^{-1}\cdot 2\chi^{SU(2)}_{\bf2}(u) -2 - t\cdot 2\chi^{SU(2)}_{\bf 2}(u) \nonumber \\
    && + t^2(1+2\chi^{SU(2)}_{\bf3}(u)+\chi^{SU(3)}_{\bf8}+\chi^{E_6}_{\bf 78}(m_l,v)+\chi^{E_6}_{\bf 78}(m'_l,v))  \nonumber \\
    &&+ t^3\chi^{SU(2)}_{\bf2}(u)\left(\chi^{SU(3)}_{\bf3}(\chi^{E_6}_{\bf\overline{27}}(m_l,v)+\chi^{E_6}_{\bf\overline{27}}(m_l',v))+\chi^{SU(3)}_{\bf \bar{3}}(\chi^{E_6}_{\bf27}(m_l,v)+\chi^{E_6}_{\bf27}(m_l',v)\right) \nonumber \\
    &&+t^4 \left(1+2\chi^{SU(2)}_{\bf3}(u)\chi^{SU(2)}_{\bf8}+\chi^{SU(3)}_{\bf27}+\chi^{SU(3)}_{\bf10}+\chi^{SU(3)}_{\bf\overline{10}} +\chi^{SU(3)}_{\bf 6}(\chi^{E_6}_{\bf27}(m_l,v)+\chi^{E_6}_{\bf27}(m_l',v)) \right. \nonumber \\
    &&+ \left. \chi^{SU(3)}_{\bf\bar{6}}(\chi^{E_6}_{\bf\overline{27}}(m_l,v)+\chi^{E_6}_{\bf\overline{27}}(m_l',v)) +\chi^{SU(3)}_{\bf8}(4+\chi^{E_6}_{\bf 78}(m_l,v)+\chi^{E_6}_{\bf 78}(m'_l,v))\right)+\mathcal{O}(t^5) \Big) \nonumber\\
    && + \mathcal{O}(q^2) \bigg] \ ,\quad \qquad
\end{eqnarray}
where $Z^{q=0}_{(1,0,0)}$ is the elliptic genus at $(k_1,k_2,k_3)=(1,0,0)$ in the
lowest order in $q$. Here $\chi^{E_6}(m_l,v)$ and $\chi^{E_6}(m'_l,v)$ are $E_6$ characters with chemical potentials $m_l,v$ and $m_l',v$ respectively. $v=v_1+v_2+v_3$ is the chemical
potential for the diagonal $U(1)_g$ of two $U(1)$'s in $E_6\times E_6$.
Although the ignorance of our gauge theory on the other $U(1)$ makes the symmetry enhancement analysis slightly incomplete, this result still provides a very nontrivial evidence
for the IR global symmetry enhancement to $E_6\times E_6\times SU(3)$.

The 2d CFT on E-strings has an alternative realization using the $U(k)$ gauge theory which was studied in \cite{Kim:2015fxa}. So one may be able to construct 2d gauge theories on self-dual strings in the $(E_6,E_6)$ conformal matter using this alternative realization. It would be interesting to see if this construction provides a better 2d theory which can
fully probe the $E_6\times E_6$ global symmetry in IR.

\section{Anomaly inflows from 6d to 2d}

The 2d CFTs on the self-dual strings have quadratic anomalies for the global symmetries.
When the 2d CFTs couple to the bulk 6d CFT by gauging some of the 2d global symmetries, the 2d quadratic anomaly causes a local gauge anomaly and makes the 2d/6d system inconsistent. To cure this inconsistency, there should be some mechanism within this theory which cancels the apparent anomaly coming from the 2d CFT. The 2d anomaly is canceled by an anomaly inflow from the bulk 6d CFT. This is analogous to the inflow mechanism discussed in~\cite{Callan:1984sa}.
In this section, we will study the anomaly inflow mechanism and explicitly demonstrate how the anomaly cancellation occurs in concrete examples.

Consider a 6d CFT and also suppose that the theory has a Green-Schwarz term required by the 6d anomaly cancellation
\begin{equation}
    S_{GS} = \Omega^{ij}\int B_i \wedge I_j \ ,
\end{equation}
where $\Omega^{ij}$ is a symmetric matrix and $I_i$ is a four-form comprised of the metric and gauge fields. For the class of theories we will discuss below, the matrix $\Omega$ is chosen such that its $(i,j)$-th element encodes $-1$ times the intersection number between $i$-th and $j$-th compact 2-cycles in the base of $CY_3$. With this term, the equation of motion for the two-form gauge field  $B_i$ becomes
\begin{equation}
    dH_i = d*H_i = I_i \ .
\end{equation}
The Green-Schwarz term contributes to the anomaly polynomial as follows
\begin{equation}\label{eq:anomaly-no-strings}
    I_{GS} = \frac{1}{2}\Omega^{ij}I_i I_j \ .
\end{equation}

In the presence of the self-dual strings, there could be additional contributions to anomalies, so to the anomaly polynomial, due to the singularity around the strings. It turns out that there are two sources for such anomalies. The first source comes from the chiral zero modes on the worldsheet of the strings. This first class of anomalies is essentially from the quadratic anomalies in the 2d CFT living on the self-dual strings. Secondly,
the Green-Schwarz contribution can receive additional  contributions in the self-dual string background. This second source leads to the anomaly inflow toward the self-dual string singularity. For the full system with the self-dual strings to be well-defined, these two classes of anomalies should cancel each other.

Let us first discuss the Green-Schwarz contribution in the presence of the self-dual strings.
Self-dual strings located at $y^{1,2,3,4}=0$ are sources for the two-form potential $B_{\mu\nu}$. So they change the equation of motion for the two-form potential as
\begin{equation}\label{4-form-defect}
    dH_i = I_i + k_i \prod_{a=1}^4\delta(y^a)dy^a \ ,
\end{equation}
for self-dual string numbers $k_i$.\footnote{On the right hand side of (\ref{4-form-defect}), $I_i$ includes ${\rm tr}(F\wedge F)$ of the 6d gauge fields. This can also account for
self-dual strings when their sizes are nonzero on $\mathbb{R}^4$. To avoid double counting, we always regard the self-dual strings as point-like objects on $\mathbb{R}^4$ in this
section, which are missed by ${\rm tr}(F\wedge F)$. We also note that the anomaly inflows
to instanton strings with nonzero sizes are studied in \cite{Blum:1993yd}.}
Equivalently, they shift the 4-form as $I_i \rightarrow I_i +k_i \prod_{a=1}^4\delta(y^a)dy^a$. Note that under this shift the anomaly 8-form in (\ref{eq:anomaly-no-strings}) will be deformed as
\begin{align}
    I_{GS} &= \frac{1}{2}\Omega^{ij}\left(I_i+k_i \prod_{a=1}^4\delta(y^a)dy^a\right)\left(I_j+k_j \prod_{a=1}^4\delta(y^a)dy^a\right) \cr
    &= I^0_{GS} + \Omega^{ij}I_i k_j \prod_{a=1}^4\delta(y^a)dy^a + \frac{1}{2}\Omega^{ij}k_ik_j\left(\prod_{a=1}^4\delta(y^a)dy^a\right)^2 \ ,
\end{align}
where $I^0_{GS}$ is the Green-Schwarz contribution without the self-dual string source.
This includes the additional contributions from the self-dual strings that depend on the self-dual string numbers $k_i$.

Let us integrate the $k_i$ dependent contributions over a four-manifold $M_4$ transverse to the self-dual strings. The contribution depending linearly on $k_i$ can be easily computed.  Due to the delta function, the result simply becomes
\begin{equation}
    I_4^{(1)} =  \int_{M_4} d^4 y \, \Omega^{ij} I_i k_j \delta^4(y) = \Omega^{ij} k_iI_j |_{M_2} \ .
\end{equation}
Here and below, we shall slightly abuse the notation $F|_{M_2}$,
for a differential $4$-form $F$ restricted to the self-dual string worldvolume $M_2$,
which means that
the 2-form derived from $I_j$ using the anomaly descent formalism is restricted to $M_2$.
This gives an anomaly 4-form on the worldsheet of the self-dual strings.

The contribution  with $k^2$ involves the square of delta functions that naively makes the computation ill-defined. In order to have a well-defined string source, we first need to smooth out the delta function source. Following~\cite{Freed:1998tg}, the delta function source can be written in terms of a regular source
\begin{equation}
    \prod_{a=1}^4\delta(y^a)dy^a = d(\rho e_3/2) \ ,
\end{equation}
where $\rho(r)$ is a smooth function of the radial coordinate $r$ for $y^a$, with $\rho(0)=-1$ and $\rho(r)=0$ for sufficiently large $r$.  $e_3$ is the global angular form on the three-sphere around the self-dual strings normalized as $\int_{S^3}e_3= 2$.
This differential form $d(\rho e_3/2)$ represents a cohomology class, so-called Thom class, of the $SO(4)$ tangent bundle $T_4$ of $M_4$~\cite{Bott}.

The explicit form of $e_3$ is given by~\cite{Harvey:1998bx,Becker:1999kh}
\begin{equation}
    e_3 = -\frac{1}{2\pi^2}\epsilon_{abcd}\left[\frac{1}{3} (D\hat{y})^a (D\hat{y})^b(D\hat{y})^c \hat{y}^d -\frac{1}{2} F^{ab}(D\hat{y})^c\hat{y}^d\right] \ ,
\end{equation}
with $\hat{y}^a\equiv y^a/r$. Using a globally defined connection $\Theta^{ab}$ for the $SO(4)$ bundle, we define the covariant derivative as
\begin{equation}
    (D\hat{y})^a \equiv d\hat{y}^a - \Theta^{ab}\hat{y}^b \ , \qquad F^{ab} = d\Theta^{ab} - \Theta^{ac}\wedge \Theta^{cb} \ .
\end{equation}
Then the global angular form $e_3$ satisfies, as analyzed in \cite{Harvey:1998bx,Becker:1999kh},
\begin{equation}
    de_3 = -\frac{1}{16\pi^2}\epsilon_{abcd}F^{ab}\wedge F^{cd} = -2\chi(T_4) \ .
\end{equation}
The four-form $\chi(T_4)$ represents the Euler class of the $SO(4)$ bundle $T_4$.

One can now easily compute the square of the delta function using the smoothed source. The result becomes
\begin{equation}
    \int_{M_4} \left(\prod_{a=1}^4\delta(y^a)dy^a\right)^2 = \frac{1}{2}d(\rho e_3)|_{M_2} = \chi(T_4) \ .
\end{equation}
This result is also in accordance with the fact that the pullback of the Thom class to a submanifold $M_2$ by the zero section is isomorphic to the Euler class, i.e. $\delta_{M_2}|_{M_2} = \chi(T_4)$ where $\delta_{M_2}$ is the delta function supported on $M_2$, as discussed in~\cite{Bott,Witten:1996hc}.
Therefore the terms quadratic in $k_i$ give rise to another contribution to the anomaly 4-form,
\begin{equation}
    I^{(2)}_4 = \frac{1}{2}\Omega^{ij}k_ik_j \chi(T_4) \ .
\end{equation}
Then we evaluate the total anomaly inflow from the bulk Green-Schwarz term as
\begin{equation}
    I_4^{\rm inflow} = I^{(1)}_4 + I^{(2)}_4 = \Omega^{ij} k_i\left[ I_j +\frac{1}{2}k_j \chi(T_4) \right] \ .
\end{equation}
This should be canceled by the anomaly of the 2d CFT living on the self-dual strings. Thus we conclude that the 2d CFT for the self-dual strings should have the anomaly polynomial of the form
\begin{equation}\label{eq:2d-anomaly-inflow}
    I^{2d}_4 = -I_4^{\rm inflow} \ .
\end{equation}
This puts a very strong constraint on 2d theories for the self-dual strings.

The anomaly polynomials for general 6d SCFTs were computed in~\cite{Ohmori:2014kda}. We shall use the results of \cite{Ohmori:2014kda} to compute the anomaly inflow to the 2d CFTs on self-dual strings. For later convenience, let us briefly summarize our conventions on group theoretical factors which basically follows from the conventions in~\cite{Ohmori:2014kda}. The trace `Tr' is defined as
\begin{equation}
    {\rm tr}_{\rm adj} F^2 = h^\vee_G {\rm Tr}F^2 \ , \quad {\rm tr}_{\rm fund} F^2 = s_G {\rm Tr}F^2 \ .
\end{equation}
The numbers $h^\vee_G$ and $s_G$ for the simply-laced Lie algebras are summarized in Table~\ref{tb:group-factors}.
\begin{table}[t]
\begin{center}
  \begin{tabular}{ | c || c | c | c | c | c | c | c | c | }
    \hline
    $G$ & $SU(N)$ & $SO(N)$ & $USp(2N)$ & $G_2$ & $F_4$ & $E_6$ & $E_7$ & $E_8$\\
    \hline
    $h^\vee_G$ & $N$ & $N-2$ & $N+1$ & $4$ & $9$ & $12$ & $18$ & $30$\\
    \hline
    $s_G$ & $\tfrac{1}{2}$ & $1$ & $\tfrac{1}{2}$ & $1$ & $3$ & $3$ & $6$ & $30$\\
    \hline
 \end{tabular}
 \caption{Dual Coxeter numbers $h^\vee_G$ and constants $s_G$ for Lie algebras. \label{tb:group-factors}}
\end{center}
\end{table}
We also use the notation for the anomaly polynomial of Weyl fermions in representation $R$ as
\begin{equation}
    \hat{A}(T)\, {\rm tr}_R\left(e^{iF}\right) \ ,
\end{equation}
where $\hat{A}(T)$ is the A-roof genus and the curvature $F$ is taken to be anti-Hermitian.

The anomalies in two dimensions are captured by a four-form anomaly polynomial.  For a left-moving chiral fermion in a representation $R$ of the gauge group $G$, the contribution to the 2d anomaly polynomial is
\begin{equation}
    I_4^L =- \frac{{\rm tr}_R (F^2)}{2} - \frac{p_1(T_2)}{24} \ ,
\end{equation}
In this expression, $F$ is the background curvature for the symmetry $G$ and $p_1(T_2)$ is the first Pontryagin class of the tangent bundle to two-manifold $M_2$. Here we used a convention that the curvature $F$ includes a factor $\frac{1}{2\pi}$, so the second Chern class is simply $c_2(R) = {\rm Tr}_R F^2/2$. We will use this convention in the discussion below.
A right-moving chiral fermion has the same contribution with overall minus sign, i.e $I_4^R = -I_4^L$. The anomaly polynomial of a 2d theory is given by the summation of all contributions from chiral fermions.

Let us now compute the anomalies for some 2d gauge theories which are proposed
to flow to the self-dual string CFTs, and check if they agree with (\ref{eq:2d-anomaly-inflow}) obtained by the anomaly inflow mechanism.

\paragraph{SCFT with $SU(N)$ gauge group and $2N$ flavors}
We first discuss 6d $\mathcal{N}=(1,0)$ SCFTs with $SU(N)$ gauge group and $N_f=2N$ fundamental hypermultiplets.
These theories describe the low energy dynamics of two parallel M5-branes probing the singularity $\mathbb{C}^2/\mathbb{Z}_N$. The gauge anomaly cancellation uniquely fix the Green-Schwarz interactions in these theories~\cite{Ohmori:2014kda}. We have
\begin{equation}
    \Omega^{ij} = \left(\begin{array}{cc} 2 & -1 \\ -1 & 2\end{array}\right) \ , \quad I_1 = \frac{1}{4}{\rm Tr} F_G^2 + \frac{N}{2} c_2 (R) \ , \quad I_2 = \frac{1}{4}{\rm Tr}F_F^2 \ ,
\end{equation}
where $i=1$ denotes the gauge node with gauge group $G=SU(N)$ and $i=2$ denotes the flavor node with flavor group $F=U(2N)$. $c_2(R) = \frac{1}{4}{\rm Tr}F_R^2$ is the second Chern class of the $SU(2)_R$ symmetry bundle. In the self-dual string background at $k_1=k$ (and $k_2=0$), we compute the anomaly inflow as
\begin{align}\label{eq:inflow-SQCD}
    I_4^{\rm inflow} &= \Omega^{ij} k_i \left[I_j + \frac{1}{2}k_j \chi_4(T_4)\right] \cr
    &= \frac{k}{2} {\rm Tr}F_G^2 -\frac{k}{4}{\rm Tr}F_F^2 +kN c_2(R) + k^2 \chi_4(T_4) \ .
\end{align}

Now let us check whether the anomaly inflowed from the 6d bulk CFT cancels the anomaly of the 2d gauge theory on  self-dual strings. The dynamics of the $k$ self-dual strings can be described by a 2d $\mathcal{N}=(0,4)$ supersymmetric $U(k)$ gauge theory with matters, as first discussed in~\cite{Haghighat:2013tka}. The 2d theory has the following $\mathcal{N}=(0,4)$ field contents:
\begin{align}
    (A_\mu,\lambda_{-\dot\alpha A}) \ & : \ U(k) \ {\rm vector\ multiplet}  \cr
    (a_{\alpha\dot\beta}, \Psi_{+\alpha A}) \ & : \ U(k) \ {\rm adjoint \ hypermultiplet} \cr
    (q_{\dot\alpha},\psi_{+A}) \ & : \ U(k)\times SU(N) \ {\rm bifundamental \ hypermultiplet} \cr
    (\eta_-) \ & : \ U(k)\times U(2N) \ {\rm bifundamental \ Fermi \ multiplet}  \ .
\end{align}
The theory with $N=1$ reduces to the M-string theory introduced in~\cite{Haghighat:2013gba}. One can also obtain this 2d theory from the $\mathbb{Z}_N$ orbifold of $k$ M-strings~\cite{Haghighat:2013tka}.

It is straightforward to evaluate the anomaly polynomial of the 2d gauge theory.
Each chiral fermion contributes to the anomaly polynomial as
\begin{align}
    \lambda_{-\dot\alpha A}+\Psi_{+\alpha A} \ \rightarrow & \    \frac{1}{2} \times 2\times k^2\times \left(\frac{{\rm tr}_{\rm fund}(F_{r}^2)}{2} -\frac{{\rm tr}_{\rm fund}(F_{l}^2)}{2}\right) = -k^2\left(c_2(l)-c_2(r)\right) \ , \cr
    \psi_{+A} +\eta_- \ \rightarrow& \ k \times \left( -2\times \frac{{\rm tr}_{\rm fund}(F_G^2)}{2}  - N \times \frac{{\rm tr}_{\rm fund}(F_R^2)}{2} +\frac{{\rm tr}_{\rm fund}(F_F^2)}{2} \right)\cr
    & = - \frac{k}{2} {\rm Tr} F_G^2 +\frac{k}{4} {\rm Tr}F_F^2 - kN c_2(R) \nonumber
\end{align}
where $c_2(l),\, c_2(r)$ are the second Chern classes of the $SU(2)_l, \,SU(2)_r$ bundles, respectively.
Here, since $\lambda_-$ and $\Psi_+$ are real multiplets, we have multiplied the overall factor $\frac{1}{2}$ to their contribution. The anomaly polynomial of the 2d theory is then given by
\begin{equation}\label{eq:anomaly-SQCD}
    I_4^{2d} = - \frac{k}{2} {\rm Tr}F_G^2 + \frac{k}{4} {\rm Tr}F_F^2 - kN c_2(R) - k^2(c_2(l)-c_2(r)) \ .
\end{equation}
Note that the last two terms are the Euler class for the $SO(4)=SU(2)_l\times SU(2)_r$ bundle $T_4$, i.e.  $\chi_4(T_4) = c_2(l)-c_2(r)$.

We can now compare this 2d result with the anomaly inflow computation in (\ref{eq:inflow-SQCD}). Clearly two results are the same but opposite overall signs. Therefore, the net 6d anomaly in the presence of 2d self-dual strings, which is given by the sum of  (\ref{eq:inflow-SQCD}) and (\ref{eq:anomaly-SQCD}), vanishes
\begin{equation}
    I^{2d}_4 + I^{\rm inflow}_4 = 0 \ .
\end{equation}
This shows that the 2d gauge theories discussed in this subsection can be consistently coupled to the 6d $SU(N)$ CFTs with $2N$ hypermultiplets.

\paragraph{E-string}
We now turn to the E-string theory, which is one of the most basic 6d theories preserving $\mathcal{N}=(1,0)$ superconformal symmetry. In M-theory setup, this theory arises as a low energy theory on an M5-brane embedded within an end-of-the world brane (or M9-brane) with $E_8$ symmetry~\cite{Ganor:1996mu,Seiberg:1996vs}.  In F-theory, the E-string theory is defined by the elliptically fibered Calabi-Yau threefold over the base $\mathcal{O}(-1)\rightarrow \mathbb{P}^1$~\cite{Witten:1996qb,Morrison:1996pp}. The theory has one dimensional Coulomb branch and $E_8$ global symmetry.

The anomaly polynomial of the E-string theory was computed in~\cite{Ohmori:2014pca} (See also~\cite{Ohmori:2014kda}). The Green-Schwarz contribution is
\begin{equation}
    I_{GS} = \frac{1}{2}\left(\frac{\chi_4(N)}{2} + \frac{{\rm Tr}F_{E_8}^2}{4} + \frac{p_1(T)}{4} + \frac{p_1(N)}{4}\right)^2 = \frac{1}{2}\left(\frac{{\rm Tr}F_{E_8}^2}{4} + \frac{p_1(T_6)}{4}-c_2(R)\right)^2 \ ,
\end{equation}
where $N$ denotes the $SO(4)$ normal bundle and $T_6$ denotes the tangent bundle. One can then easily compute the anomaly inflow toward the singularity of $k$ self-dual strings,
\begin{equation}
    I_4^{\rm inflow} = k \left(-\frac{{\rm Tr}F_{E_8}^2}{4} - \frac{p_1(T_6)}{4}+c_2(R) + \frac{k}{2}\chi_4(T_4)\right) \ .
\end{equation}
We again expect that this anomaly is canceled by the anomaly of the 2d self-dual string theory.

The self-dual strings called E-strings are M2-branes suspended between the M5-brane and the M9-brane. The E-strings at low energy admit weakly coupled $O(k)$ gauge theory description which was discussed in Section~\ref{sec:E6xE6}.
The left- and right-moving chiral fermions in (\ref{eq:E-string-mattercontent}) contribute to the anomaly polynomial as
\begin{align}
    \lambda_{-\dot\alpha A} \ \rightarrow & \ k(k-1) \times\left(\frac{1}{2}\left(c_2(r)+c_2(R)\right) +\frac{p_1(T_2)}{24} \right) \ , \cr
    \Psi_{+\alpha A} \ \rightarrow & \ -k(k+1) \times \left(\frac{1}{2}(c_2(l)+c_2(R)) + \frac{p_1(T_2)}{24}\right) \ , \cr
    \Xi_l \ \rightarrow& \ k\times \left( \frac{{\rm Tr}F_{SO(16)}^2}{4} +\frac{p_1(T_2)}{3} \right) \ . \nonumber
\end{align}
Collecting all the contributions, the anomaly polynomial of the 2d theory is given by
\begin{align}
    I_4^{2d} &= \frac{k}{4}{\rm Tr}F_{E_8}^2 -kc_2(R) + \frac{k}{4}\left(p_1(T_2) -2c_2(l)-2c_2(r)\right) - \frac{k^2}{2}(c_2(l)-c_2(r)) \cr
    &= \frac{k}{4}{\rm Tr}F_{E_8}^2 -kc_2(R) + \frac{k}{4}p_1(T_6) -\frac{k^2}{2}\chi_4(T_4)
\end{align}
where we used $p_1(T_6)  = p_1(T_2) - 2c_2(l) - 2c_2(r)$ for the bundle $T_6 = T_2\times T_4$, and $\chi_4(T_4) = c_2(l)-c_2(r)$.
Here we also identified the second Chern class for the $SO(16)$ bundle with the second Chern class of the $E_8$ bundle, assuming the $E_8$ enhancement.
Therefore, the total anomaly in the presence of the E-strings vanishes again, i.e. $I_4^{2d} + I_4^{\rm inflow}=0$.

\paragraph{$SO(8)$ minimal SCFT} Our next example is the 6d minimal SCFT with $SO(8)$ gauge group. In the F-theory construction, this theory corresponds to an elliptic fibration over a base $\mathcal{O}(-4) \rightarrow \mathbb{P}^1$~\cite{Bershadsky:1997sb}.

For the minimal SCFTs with gauge group $G=SU(3),SO(8),F_4,E_{6,7,8}$, the Green-Schwarz contribution to the anomaly polynomial takes the following simple expression\footnote{The equation (1.10) in~\cite{Ohmori:2014kda} misses the last term in the parenthesis, but this term is needed to cancel the terms proportional to $p_1(T)$ in the vector multiplet contribution given in (A.3) of~\cite{Ohmori:2014kda}.}~\cite{Ohmori:2014kda}:
\begin{equation}
    I_{GS} = \frac{w_G}{2}\left(\frac{1}{4}{\rm Tr}F^2 + \frac{h^\vee_G}{w_G}c_2(R)  + \frac{h^\vee_G}{w_G}\frac{p_1(T_6)}{12}\right)^2 \ ,
\end{equation}
where $w_G$ is the coefficient  in the relation ${\rm tr}_{\rm adj}F^4 = \frac{3}{4}w_G {\rm Tr}F^2$.
Then we find the anomaly inflow to the singularity in the $k$ self-dual string background as
\begin{equation}\label{eq:anomaly-inflow-minimalCFT}
    I_4^{\rm inflow} = k\times w_G\left(\frac{1}{4}{\rm Tr}F_G^2 + \frac{h_G^\vee}{w_G}c_2(R)+\frac{h^\vee_G}{w_G}\frac{p_1(T_6)}{12} + \frac{k}{2}\chi_4(T_4)\right) \ .
\end{equation}


\begin{table}[t]
\begin{center}
  \begin{tabular}{ | c || c | | c | c | c | c | c | c | c | }
    \hline
    $G$ & $SU(2)$ & $SU(3)$ & $SO(8)$ & $G_2$ & $F_4$ & $E_6$ & $E_7$ & $E_8$\\
    \hline
    $w_G$ & $\frac{8}{3}$ & 3 & 4 & $\frac{10}{3}$ & $5$ & $6$ & $8$ & $12$\\
    \hline
 \end{tabular}
\end{center}
\caption{Values of $w_G$ for various groups}
\end{table}

Let us compute the anomaly of the 2d theory living on the self-dual strings. The 2d theory has a gauge theory description as discussed in~\cite{Haghighat:2014vxa}. It is the $Sp(k)$ gauge theory with the field contents as follows:
\begin{align}
    (A_\mu,\lambda_{-\dot\alpha A}) \ & : \ Sp(k) \ {\rm vector \ multiplet } \cr
    (a_{\alpha\dot\beta}, \Psi_{+\alpha A}) \ & : \ Sp(k) \ {\rm antisymmetric \ hypermultiplet} \cr
    (q_{\dot\alpha},\psi_{+A}) \ & : \ Sp(k)\times SO(8) \ {\rm bifundamental \ hypermultiplet} \ .
\end{align}
The chiral fermions have the following contributions to the anomaly polynomial:
\begin{align}
    \lambda_{-\dot\alpha A} \ \rightarrow & \ 2k(2k+1) \times\left(\frac{1}{2}\left(c_2(r)+c_2(R)\right) +\frac{p_1(T_2)}{24} \right) \ , \cr
    \Psi_{+\alpha A} \ \rightarrow & \ -2k(2k-1) \times \left(\frac{1}{2}(c_2(l)+c_2(R)) + \frac{p_1(T_2)}{24}\right) \ , \cr
    \psi_{+A} \ \rightarrow& \ - 2k\times \left( \frac{{\rm Tr}F_{G}^2}{2} +4c_2(R) +\frac{p_1(T_2)}{3} \right) \ . \nonumber
\end{align}
The sum of these contributions yields the anomaly polynomial of the 2d theory on the $SO(8)$ minimal strings. The result is
\begin{align}
    I_4^{2d} = -k{\rm Tr}F_G^2 -6kc_2(R) -\frac{k}{2}p_1(T_6) - 2k^2\chi_4(T_4) \ .
\end{align}
This precisely cancels the anomaly inflow in (\ref{eq:anomaly-inflow-minimalCFT}) with $h^\vee_G=6,\, w_G = 4$ when $G=SO(8)$. Therefore the 2d/6d coupled system for the $SO(8)$ minimal SCFT is also consistent.

\paragraph{$SU(3)$ minimal SCFT}
We will now show that our 2d gauge theories proposed for the $SU(3)$ minimal strings
have the worldsheet anomalies which precisely cancel the anomaly inflow from the bulk CFT. This will provide another strong evidence for our 2d gauge theory being the correct theory on the $SU(3)$ strings.

The anomaly polynomial of our gauge theory can be read off from the field content given in (\ref{original-fields}) and (\ref{new-fields}). The contributions from the chiral fermions are
\begin{align}
    & \lambda_0 + \lambda + \Psi_+ + \tilde{\Psi}_+\ \rightarrow \ k^2\left(c_2(r) -c_2(l)\right) \ , \quad
    \psi_+ + \tilde\psi_+ \ \rightarrow \ -2k\left(\frac{F_G^2}{4} + \frac{3F_R^2}{8} +\frac{p_1(T_2)}{8}\right) \ , \cr
   & \xi + \tilde\xi \ \rightarrow \ -k(k-1)\left(\frac{c_2(l)}{2} + \frac{F_R^2}{8} +\frac{p_1(T_2)}{24}\right) \ , \quad
    \hat\lambda + \check{\lambda} \ \rightarrow \ k(k+1)\left(\frac{F_R^2}{8}+\frac{c_2(r)}{2} + \frac{p_1(T_2)}{24}\right) \ , \cr
   & \chi \ \rightarrow \ -k\left(\frac{F_G^2}{4} + \frac{3{\rm Tr}F_R^2}{8} + \frac{p_1(T_2)}{8} \right)  \ , \quad
    \zeta \ \rightarrow \ k\left(\frac{{\rm Tr}F_R^2}{8} + \frac{p_1(T_2)}{24}\right) \ , \nonumber
\end{align}
where we used the notation $J_R =R/2 ,\, J_r = -J-R/2$ and $F_R$ is the background curvature for the $J_R$.
Thus the anomaly four-form of the 2d gauge theory on the $SU(3)$ minimal strings is
\begin{equation}
    I_4^{2d} = -\frac{3k}{4}{\rm Tr}F_G^2 -3kc_2(R)- \frac{k}{4}p_1(T_6) - \frac{3k^2}{2}\chi_4(T_4) \ .
\end{equation}
This perfectly cancels the anomaly from the inflow mechanism in (\ref{eq:anomaly-inflow-minimalCFT}) with $h^\vee_G=3,\, w_G=3$. Therefore, our 2d theory can be consistently coupled to the bulk 6d $SU(3)$ minimal SCFT.

\paragraph{$(E_6,E_6)$ conformal matter}

Lastly, let us consider the anomaly of the 2d self-dual string theories in the 6d $(E_6,E_6)$ conformal matter theory.

The 6d anomaly polynomial of the conformal matter theory was calculated in \cite{Ohmori:2014kda}. There are three Green-Schwarz terms from three tensor multiplets of an $SU(3)$ theory and two E-string theories. For an E-string theory, the Green-Schwarz term is given by
\begin{equation}
    I_{GS}^{Estr} = \frac{1}{2}(I_4^{Estr})^2\ , \qquad I_4^{Estr} =-\frac{1}{4}{\rm Tr} F_{E_6}^2-\frac{1}{4}{\rm Tr} F_{SU(3)}^2 -\frac{p_1(T_6)}{4} +c_2(R) \ ,
\end{equation}
where we denote by $F_{E_6}$ and $F_{SU(3)}$ the background field strengths for the subgroup $E_6\times SU(3)\subset E_8$. The Green-Schwarz contribution coming from the $SU(3)$ gauge group is different from those of the $SU(3)$ minimal CFT because there is additional $SU(3)$ gauge anomaly from the E-string theory when we gauge the $SU(3)$ subgroup of the $E_8$ global symmetry. In \cite{Ohmori:2014kda}, the Green-Schwarz term from the $SU(3)$ part is given by
\begin{eqnarray}
    I_{GS}^{SU(3)} &&= \frac{1}{2}(\tilde{I}^{SU(3)}_4)^2 \ , \nonumber \\
    \tilde{I}_4^{SU(3)} &&= \frac{1}{4}{\rm Tr}F^2_{SU(3)} + 5c_2(R)-\frac{1}{4}p_1(T_6)-\frac{1}{4}{\rm Tr}F_{E_6^L}^2 -\frac{1}{4}{\rm Tr} F_{E_6^R}^2 \ ,
\end{eqnarray}
with the background field strengths $F_{E_6^{L,R}}$ for the $E^L_6\times E^R_6$ global symmetry. The total Green-Schwarz contribution to the conformal matter theory is the sum of these three terms.
\begin{equation}
    I_{GS}^{(E_6,E_6)} = I_{GS}^{Estr}(F_{E_6^L}) + I_{GS}^{SU(3)} +I_{GS}^{Estr}(F_{E_6^R}) \ .
\end{equation}

We find it more natural to write this Green-Schwarz contribution using the symmetric matrix $\Omega^{ij}$. It turns out that the above expression can be rewritten as
\begin{equation}
    I_{GS}^{(E_6,E_6)} =\frac{1}{2}(\Omega^{-1})_{ij} I_4^i I_4^j \ ,
\end{equation}
where $I^i_4\equiv \Omega^{ij}I_{4,j}$ are the 4-forms in the Bianchi identities of  tensor fields, which are given as follows:
\begin{eqnarray}
    I_{4}^1 &&= \frac{3}{4}{\rm Tr}F_{SU(3)}^2 + 3c_2(R) +\frac{1}{4}p_1(T_6) \ ,\nonumber \\
    I_{4}^2 &&= I_4^{Estr}(F_{E_6^L},F_{SU(3)}) \ , \quad I_{4}^3 = I_4^{Estr}(F_{E_6^R},F_{SU(3)}) \ .
\end{eqnarray}
Here the symmetric matrix is fixed by the intersection numbers of 2-cycles as
\begin{equation}
    \Omega = \left(\begin{array}{ccc}3 & -1 & -1 \\ -1 & 1 & 0 \\ -1 & 0 & 1\end{array}\right) \ .
\end{equation}
Note that $I^1_4$ is equal to the 4-form of the Green-Schwarz contribution in the $SU(3)$ minimal CFT.

From these Green-Schwarz terms, one can find the anomaly inflow toward the 2d string theory. The result takes the form of
\begin{eqnarray}\label{eq:anomal-4-form-conformal-matter}
    I_4^{\rm inflow} =&& \Omega^{ij}k_i\left[I_{4,j} + \frac{1}{2}k_j \chi_4(T_4) \right] \nonumber \\
    =&& \frac{1}{2}(3k_1^2+k_2^2+k_3^2-2k_1k_2-2k_1k_3)\chi(T_4)+k_1\left(\frac{3}{4}{\rm Tr}F_{SU(3)}^2 + 3c_2(R) +\frac{p_1(T_6)}{4} \right) \nonumber \\
    &&+k_2\left(-\frac{1}{4}{\rm Tr} F_{E_6^L}^2-\frac{1}{4}{\rm Tr} F_{SU(3)}^2 -\frac{p_1(T_6)}{4} +c_2(R)\right) \nonumber \\
    && +k_3\left(-\frac{1}{4}{\rm Tr} F_{E_6^R}^2-\frac{1}{4}{\rm Tr} F_{SU(3)}^2 -\frac{p_1(T_6)}{4} +c_2(R)\right) \ .
\end{eqnarray}

Let us then compare this anomaly inflow to the anomaly polynomial of the 2d gauge theory constructed in Section \ref{sec:E6xE6}. The computation of the 2d anomaly polynomial is almost straightforward.  We can simply add the anomaly polynomials of the $SU(3)$ minimal string theory and two E-string theories and additionally the contributions from the bi-fundamental fermions. The contributions from the bi-fundamental fermions are
\begin{equation}
    I_{4}^{(k_1,k_2)} = k_1k_2(c_2(l)-c_2(r)) = k_1k_2\chi_4(T_4) \ , \quad I_{4}^{(k_1,k_3)} = k_1k_3\chi_4(T_4) \ .
\end{equation}
Then the full anomaly polynomial of the 2d self-dual string theory in the $(E_6,E_6)$ conformal matter theory is given by
\begin{eqnarray}
    I_4^{2d} =&& k_1k_2\chi_4(T_4)+k_1k_3\chi_4(T_4)+I_4^{O(k_3)}+I_4^{U(k_1)} + I_4^{O(k_2)}  \nonumber \\
    =&& k_1(k_2+k_3)\chi_4(T_4)-k_1\left(\frac{3}{4}{\rm Tr}F_{SU(3)}^2 + 3c_2(R) + \frac{p_1(T_6)}{4} +\frac{3}{2}k_1\chi_4(T_4)\right) \nonumber \\
    &&+k_2\left(\frac{1}{4}{\rm Tr}F_{E_6^L}^2+\frac{1}{4}{\rm Tr}F_{SU(3)}^2-c_2(R)+\frac{p_1(T_6)}{4}-\frac{k_2}{2}\chi_4(T_4)\right)\nonumber \\
    &&+k_3\left(\frac{1}{4}{\rm Tr}F_{E_6^R}^2+\frac{1}{4}{\rm Tr}F_{SU(3)}^2-c_2(R)+\frac{p_1(T_6)}{4}-\frac{k_2}{2}\chi_4(T_4)\right)\ .
\end{eqnarray}
This anomaly  exactly cancels the anomaly inflow in (\ref{eq:anomal-4-form-conformal-matter}). This result obviously supports that our 2d gauge theory with $U(k_1)\times O(k_2)\times O(k_3)$ gauge group realizes the 2d self-dual string CFTs in the $(E_6,E_6)$ conformal matter theory at low energy.

\section{Conclusion and remarks}

In this paper, we constructed 2d gauge theories which we suggest to flow in IR
to the $\mathcal{N}=(0,4)$ superconformal field theories on the self-dual strings
of the 6d $SU(3)$ SCFT. We made various tests of this proposal.
We studied the  moduli spaces at classical and quantum levels, providing evidences for the
$\mathcal{N}=(0,4)$ SUSY enhancement in IR. The elliptic genus of our gauge
theories provides very powerful checks our proposal, which is compared with
the topological string calculus of \cite{Haghighat:2014vxa}. One can go
beyond testing our theories, and provide all-genus summations of the corresponding
topological string amplitudes from our approach. We also generalized our gauge
theory constructions to study the strings of the $E_6\times E_6$ conformal matter.
For all the new 2d gauge theories discovered in this paper, we provided another powerful
support for them by checking that their 2d anomalies agree with the anomalies on
the self-dual strings, computed from the anomaly inflow.

The studies made in this paper suggest various new directions and challenges.
In this section, we discuss some future directions. Some comments made in this
section are now being explored with substantial progress \cite{kkkp}.

We studied the classical moduli spaces of our gauge theory, and some 
1-loop corrections. The relevant branch of the classical moduli space appears 
to be consistent with our claim of $\mathcal{N}=(0,4)$ SUSY enhancements, 
within the approximations we made. However, it will be desirable to make 
more systematic studies, perhaps along the lines of \cite{Melnikov:2012nm}.

The studies of the elliptic genera illustrate
that our method is working very precisely. More amusing successes in this direction
will be reported in \cite{kkkp}, concerning 6d strings and exceptional instantons
without D-brane origins. As for the studies made in this paper, it will be
interesting to further study the strings of the
$(E_6,E_6)$ conformal matter, using the new 2d QFTs we constructed in section 4.
In this paper, we have only checked $E_6\times E_6$ symmetry enhancement
and the 2d anomalies. Beyond these, for instance, one may make more concrete
studies of the 6d CFT compactified on a circle, by
comparing it with suitable 5d super Yang-Mills theory systems.

We find that one exciting application of our findings is
the study of other exceptional instantons and instanton strings from gauge theory
approaches. As emphasized in the introduction and section 2, this is closely related
to the fact that instanton strings in non-Higgsable 6d $SU(3)$ gauge theories should
be regarded as the simplest exceptional instantons. The fact that the standard ADHM
formulation of $SU(3)$ instantons fails in 2d is one sign of this.
The non-Higgsable 6d $SU(3)$ theory is related to a sequence of exceptional gauge
theories with matters via unHiggsings shown in (\ref{Higgsing-SU3}).
We find that one can study the strings of the $G_2$ SCFT with a
hypermultiplet in ${\bf 7}$, and $SO(7)$ SCFT with two spinor matters, by
straightforwardly extending the QFTs of section 3.1.
We shall explain the 2d gauge theories and their elliptic genera
in a follow up work \cite{kkkp}.
With these two kinds of 2d gauge theories ($G_2$ instanton strings,
and $SO(7)$ instanton strings with exotic matters),
one can also construct natural 2d quivers for the strings of other `atomic'
CFTs,  $32$, $322$, $232$ of Table \ref{other}, with 2d/3d tensor branches.
In particular, gluing the $232$ atom with two E-string theories,
we can study the strings of the $(E_7,E_7)$ conformal matter $12321$
\cite{DelZotto:2014hpa}.

Another related application is a new ADHM-like description of $G_2$
instanton particles in 5d
gauge theories. Namely, we find 1d gauge theories for multi-instantons of the
5d $G_2$ gauge theories coupled to some numbers of matters in fundamental
representation ${\bf 7}$, and compute their Nekrasov partition functions
\cite{kkkp}. It would be interesting to see if our methods allow us to
study other exceptional instantons, with gauge groups $F_4,E_6,E_7,E_8$.

Perhaps in section 3.1, the reader might have wondered how one can guess
the complicated list (\ref{new-fields}) of extra fields.
It is in fact much easier to understand (\ref{new-fields}) by starting
from $SO(7)$ or $G_2$ instanton theories, and Higgs them to $SU(3)$.
It will be explained in \cite{kkkp}.

In recent years, Yang-Mills instantons played important roles in
understanding 5d/6d SCFTs. For instance, the instanton partition function of
\cite{Nekrasov:2002qd,Nekrasov:2003rj} played important
roles in studying curved space partition functions,
such as the 5d indices on $S^4\times S^1$
\cite{Kim:2012gu}, or the 6d indices on $S^5\times S^1$
\cite{Kim:2012ava,Lockhart:2012vp,Kim:2012qf,Kim:2013nva}. Also,
there have been substantial developments in the calculation of the
instanton partition functions
\cite{Kim:2011mv,Hwang:2014uwa,Cremonesi:2014xha}. Yang-Mills instanton strings
also play interesting roles in 6d SCFTs, exhibiting surprisingly subtle features.
For instance, it will be interesting to see if the elliptic genera computed
from our 2d gauge theories can be used to better understand other 6d CFT observables
in the symmetric phase \cite{Kim:2012ava,Lockhart:2012vp,Kim:2012qf,Kim:2013nva}.

One curious aspect that we find in section 5 about the 2d anomalies
is that, for $k$ strings (or instantons), some anomalies
are proportional to $k^2$. At large $k$, this is much larger than what
one expects from the instanton moduli spaces,
which only show quantities proportional to $k$. So this implies that many
light degrees of freedom of these 2d SCFTs are supported at small instanton
singularities. Note also that the central charges of these CFTs are all proportional
to $k$, which can be computed from the asymptotic regions of the nonlinear sigma
models using free field theory techniques. So there should be many degrees of
freedom in this CFT which are not captured by the central charges.
Just to compare, consider the 2d SCFT living on the D1-D5 system,
which is the 2d CFT living on the instanton strings of 6d maximal super-Yang-Mills
theory. The theory preserves $\mathcal{N}=(4,4)$ supersymmetry, and it is well
known that the small instanton singularity on the classical moduli space is replaced
by a `throat' region \cite{Witten:1997yu}. The throat region induces a novel UV continuum,
but does not host novel $k^2$ degrees of freedom for $k$ D-strings.
Incidently, the anomalies of this $\mathcal{N}=(4,4)$ CFT are all proportional to
$k$, not $k^2$. So the $k^2$ light
degrees of freedom captured by our anomaly analysis implies that the physics of
the small instanton singularities will be very different from that of the
6d maximal super Yang-Mills.

2d $\mathcal{N}=(0,4)$ SCFTs for self-dual strings
are related to 4d $\mathcal{N}=2$ SCFTs engineered by D3-branes probing 7-brane
singularities, by compactifying the latter on $S^2$ with R-symmetry twists.
Especially, the strings of the 6d minimal SCFTs are related to interesting
4d SCFTs compactified on $S^2$. For single strings,
the associated 4d theories are very well known. For $G=E_6,E_7,E_8$
(i.e. at $n=6,8,12$) in Table \ref{minimal}, the corresponding
4d theories are the Minahan-Nemeschansky theories
\cite{Minahan:1996fg}. At $n=5$ with $G=F_4$, the 2d theory is a compactification
of the 4d $E_6$ Minahan-Nemeschansky theory with a defect on $S^2$.
At $n=4$, the 2d theory is obtained by compactifying the 4d $\mathcal{N}=2$
$SU(2)$ gauge thoeries with $N_f=4$ flavors, which is a Lagrangian QFT. The
self-dual string gauge theories at $n=4$ studied in \cite{Haghighat:2014vxa}
were essentially engineered this way from 4d. Finally, at $n=3$, the 2d SCFT for
single $SU(3)$ self-dual string is an $S^2$ reduction of the 4d Argyres-Douglas
theory \cite{Argyres:1995jj} of type $(A_1,D_4)$ or $H_2$. Recent studies show
that some classes of Argyres-Douglas theories can be studied
from Lagrangian approaches \cite{Maruyoshi:2016tqk}. It will
be interesting to see if such 4d approaches can be applied to the $H_2$ model,
and compactified on $S^2$ to address the physics that we studied in
this paper.

\vskip 0.5cm

\hspace*{-0.8cm} {\bf\large Acknowledgements}
\vskip 0.2cm

\hspace*{-0.75cm} We thank Michele Del Zotto, Koji Hashimoto, Albrecht Klemm,
Kimyeong Lee, Guglielmo Lockhart, David Tong and especially Joonho Kim
for helpful discussions. HK is supported by the Perimeter Institute for
Theoretical Physics. Research at Perimeter Institute is supported by the Government
of Canada through Industry Canada and by the Province of Ontario through the Ministry
of Economic Development and Innovation. SK is supported in part by the National
Research Foundation of Korea (NRF) Grant 2015R1A2A2A01003124. JP is supported in part
by the NRF Grant 2015R1A2A2A01007058.

\end{document}